\documentstyle[11pt,a4,cite,epsf,axodraw]{article}

\newcommand{\be}{\begin{equation}}
\newcommand{\ee}{\end{equation}}
\newcommand{\bea}{\begin{eqnarray}}
\newcommand{\eea}{\end{eqnarray}}

\newcommand{\nn}{\nonumber}

\newcommand{\cO}{{\cal O}}
 
\newcommand{\eq}[1]{Eq.~(\ref{#1})}

\newcommand{\NPB}[3]{Nucl.\ Phys.\ {\bf B{#1}} (19{#2}) {#3}}
\newcommand{\PRD}[3]{Phys.\ Rev.\ {\bf D{#1}} (19{#2}) {#3}}
\newcommand{\PLB}[3]{Phys.\ Lett.\ {\bf B{#1}} (19{#2}) {#3}}
\newcommand{\PRL}[3]{Phys.\ Rev.\ Lett.\ {\bf {#1}} (19{#2}) {#3}}

\newcommand{\AP}[3]{Ann.\ Phys.\ {\bf {#1}} (19{#2}) {#3}}
\newcommand{\ZPC}[3]{Z.\ Phys.\ {\bf C{#1}} (19{#2}) {#3}}
\newcommand{\EPJC}[3]{Eur.\ Phys.\ J.\ {\bf C{#1}} (19{#2}) {#3}}
 
\newcommand{\ks}{{\mbox k \!\!\! /}}
\newcommand{\ps}{{\mbox p \!\!\! /}}
\newcommand{\qs}{{\mbox q \!\!\! /}}

\begin{document}
 
\begin{titlepage}
\begin{flushright} 
RAL-TR-1998-078 \\
November 1998 
\end{flushright}
\vspace*{2cm}
 
\begin{center}
 
{\LARGE {\bf The Pinch Technique Beyond One Loop:}}
\vspace{5pt}

{\LARGE {\bf The Gauge-Independent Two-Loop Quark}}
\vspace{5pt}

{\LARGE {\bf Self-Energy}}
\\[1.5cm]

{\large {\bf N.J. Watson}}\footnote{
Present address: Rutherford Appleton Laboratory, Chilton, Didcot,
Oxon., OX11 0QX, U.K. \\
Email: jay.watson@rl.ac.uk }
\\[1cm]
 
\noindent
Division de Physique Th\'eorique, \\
Institut de Physique Nucl\'eaire, \\
Universit\'e de Paris-Sud, \\
F-91406 Orsay Cedex, France.

\end{center}
 
\vspace*{1.5cm}
\begin{abstract}

It is shown how the pinch technique algorithm may be 
consistently extended beyond the one-loop level to obtain the 
gauge-independent two-loop fermion self-energy 
$-i\hat{\Sigma}^{(2)}(p)$
in QCD in the pinch technique approach. 
The starting point for the construction is the general
diagrammatic representation of the two-loop quark self-energy 
in terms of renormalized one-loop two- and three-point function
and tree level Bethe-Salpeter-type quark-gluon scattering kernel 
insertions in the one-loop quark self-energy.
Using factors of longitudinal gluon four-momentum only
from lowest order gauge field
propagators and triple gauge vertices to trigger the
relevant Ward identities, the function 
$-i\hat{\Sigma}^{(2)}(p)$ 
is explicitly constructed
from the consideration of the two-loop QCD corrections to
the Compton scattering of a photon off a quark.
It is shown that the resulting pinch technique self-energy
$-i\hat{\Sigma}^{(2)}(p)$ 
is gauge-independent at all momenta, 
does not shift the position of the propagator pole and 
is multiplicatively renormalizable by local counterterms.
The demonstration of the gauge independence is based on
an efficient diagrammatic method to deal with the several
dozen two-loop diagrams involved. 
It is explicitly shown by this example that the 
general correspondence 
between the pinch technique $n$-point functions and those obtained
in the background field method in the Feynman quantum gauge
$\xi_{Q} = 1$ does not persist beyond one loop.

\end{abstract}
 
\vfill
 
 
\end{titlepage}
 

\section{Introduction}

The pinch technique (PT) 
\cite{PT0,PT1,PTquark,PT2pt1,PT2pt2,PTqcdeffch,PTeweffch}
is a well-defined algorithm for the
{\em rearrangement}\, of contributions to conventional one-loop
$n$-point functions in gauge theories to obtain one-loop
``effective'' $n$-point functions which, most notably, 
are entirely independent of the particular gauge fixing procedure used
(covariant, non-covariant, background field etc).
This rearrangement of one-loop perturbation theory is based on the
systematic use of the tree level Ward identities of the theory
to cancel among Feynman integrands all factors of longitudinal
four-momentum associated with gauge fields propagating in loops.
In addition to being gauge-independent, the PT $n$-point functions
display a wide range of further
desirable theoretical properties.
In particular, they satisfy simple tree-level-like Ward identities
corresponding to the gauge invariance of the classical lagrangian.
As a result of these properties, the PT has been advocated as the 
appropriate theoretical framework for a broad range of applications 
in which one is forced to go beyond the strictly order-by-order 
computation of $S$-matrix elements, or to consider amplitudes for 
explicitly off-shell processes 
\cite{PT2pt1,PT2pt2,PTqcdeffch,PTeweffch,PTapps}.

A fundamental criticism of the PT approach, however, is that, to date,
it has remained restricted to the one-loop level: it has yet
to be shown how the PT algorithm may be consistently extended beyond
the one-loop level to obtain one-particle-irreducible (1PI)
multi-loop $n$-point functions with the same desirable properties
as at one loop. From a theoretical point of view, 
this extension is clearly essential if it is to be shown that the PT
is more than just an artefact of the one-loop approximation.
In particular, one would like to know whether the correspondence
\cite{BFMPT}
between the PT gauge-independent $n$-point functions
and those obtained in the background field method (BFM) 
\cite{abbott,capper}
in the Feynman quantum gauge $\xi_{Q} = 1$ 
persists beyond one loop.\footnote{
For explanations of why the PT one-loop
$n$-point functions are distinguished
on physical grounds from those obtained in the BFM for arbitrary
values of the quantum gauge fixing parameter $\xi_{Q}$, see
Refs.~\cite{PT2pt2,PTeweffch}.}
{}From a phenomenological point of view,
many of the applications of the PT increasingly
demand accuracy beyond the one-loop level.

In attempting to extend the PT beyond the one-loop approximation,
two basic problems arise:

(1) {\em How to deal consistently with triple gauge vertices all 
three legs
of which are associated with gauge fields propagating in loops?}\,
In the PT at the one-loop level, 
the factors of longitudinal four-momentum associated with gauge fields 
propagating in loops originate from tree level gauge 
field propagators and triple gauge vertices. In particular, 
the triple gauge vertices which occur in one-loop diagrams
always have one external leg $A_{\mu}^{a}(q)$ and
two internal legs $A_{\rho}^{r}(k_{1})$, $A_{\sigma}^{s}(k_{2})$.
It is then possible to decompose such vertices so as to isolate
unambiguously the longitudinal factors $k_{1\rho}$, $k_{2\sigma}$
associated with the internal gauge fields.
Beyond one loop, however, there occur triple gauge vertices for
which all three legs are internal.
It is thus unclear how
to decompose such vertices
in order to identify the associated longitudinal factors
which then trigger the PT rearrangement.

(2) {\em How to deal consistently with the ``induced''
factors of longitudinal internal gauge field four-momentum 
which originate from internal loop corrections?}
Beyond the one-loop level, in addition to
the factors of longitudinal internal gauge field four-momentum
from tree level gauge field propagators and triple gauge vertices,
there occur further
such factors originating from the invariant tensor structure of
internal loop corrections. A simple example are the longitudinal
factors occurring in the transverse structure
of the gluon self-energy in QCD.
It is thus unclear whether or not such ``induced'' factors
should also be used to trigger the PT rearrangement;
and, if so, how this may be done consistently.

In addition to these two problems of principle, 
one is also faced at the diagrammatic level with the
algebraic complexity of attempting to 
implement the PT rearrangement among
multi-loop $n$-point functions starting from arbitrary gauges.

The purpose of this paper is to take the first step towards
solving the first of the above problems. 
At the two-loop level to be considered here, 
the two problems are uncoupled
and so can be investigated separately and successively:
one has first of all to solve (1)
in order to identify the one-loop internal corrections 
needed before one can address (2).
By investigating here only the first problem, 
we effectively assume
that the correct approach will turn out to
be {\em not}\, to use the ``induced''
longitudinal factors to trigger further the PT rearrangement.
In particular,
throughout this paper, ``the PT'' will be understood to mean
the PT algorithm implemented using
only the longitudinal factors from lowest order 
gauge field propagators and triple gauge vertices.
The possibility that one should in fact then go on to
use the ``induced'' longitudinal factors to trigger further
the PT rearrangement will be briefly discussed in the conclusions. 

We consider the construction in the PT approach
of the two-loop fermion self-energy in QCD.
On the one hand, this example is non-trivial in that it
involves diagrams with triple gauge vertices all three legs
of which are internal. On the other hand, the quark self-energy
is simpler than, e.g., the two-loop 
gluon self-energy in the PT approach, since 
(i) the construction of the corresponding PT one-loop function
is almost trivial, and (ii) the PT two-loop quark self-energy
may be obtained from two-loop processes
involving no diagrams with tree level four-point functions
(quadruple gauge vertices or, possibly, gauge-ghost vertices).
The starting point for 
the construction is the general
diagrammatic representation of the two-loop quark self-energy 
in terms of renormalized one-loop 
two- and three-point function
and tree level Bethe-Salpeter-type quark-gluon scattering
kernel insertions in the one-loop quark self-energy.
In terms of this diagrammatic representation,
the first problem above may be recast as: 
how to reorganize consistently the contributions to 
the two-loop Feynman integrands involved
so as to isolate, first, the PT one-loop
$n$-point functions occurring as internal corrections, 
and thence the corresponding PT tree level 
Bethe-Salpeter-type kernel insertion?
It will be shown here how, for the case of the 
two-loop fermion self-energy, this first problem
may be consistently solved---and that the solution turns
out to be remarkably simple.

The paper is organized as follows. 
In section~2, the construction of the PT gauge-independent
one-loop quark self-energy 
$-i\hat{\Sigma}^{(1)}(p)$ is reviewed.
In section~3, the general diagrammatic representation
which provides the basis for the construction of the
PT two-loop quark self-energy
$-i\hat{\Sigma}^{(2)}(p)$ is described.
In section~4, it is shown how the problem of
purely internal triple gauge vertices in the PT
is solved for the case in hand,
enabling the consistent construction of
$-i\hat{\Sigma}^{(2)}(p)$ 
starting from the ordinary Feynman gauge.
In section~5, the construction is generalized to 
the class of arbitrary linear covariant gauges
in order to demonstrate
explicitly the gauge independence of 
$-i\hat{\Sigma}^{(2)}(p)$.
In section~6, it is shown
that the resulting two-loop function is renormalizable
via the explicit calculation of the required two-loop
counterterms.
It is also shown that the
PT two-loop quark self-energy 
leaves unchanged the position of the
pole in the Dyson-summed fermion propagator (to two-loop order), 
and does not coincide with the self-energy obtained
in the BFM at $\xi_{Q} = 1$.
A discussion and conclusions are given in section~7.

 
\setcounter{equation}{0}

\section{The pinch technique one-loop quark self-energy }

We begin by reviewing the construction of the PT quark self-energy
at the one-loop level \cite{PTquark}.
This is worthwhile since the technique
used here for the rather trivial one-loop case will be extended
to the non-trivial two-loop case in the subsequent sections.

We work in the class of ordinary linear covariant gauges with QCD
gauge parameter $\xi$ ($\xi = 0$ is the Landau gauge).
The free gluon and quark propagators are thus given by
\bea
iD_{\mu\nu}(q,\xi)
&=&
\frac{i}{q^{2}+i\epsilon}\biggl( 
-g_{\mu\nu} + (1-\xi)\frac{q_{\mu}q_{\nu}}{q^{2}} 
\biggr)\,\,,
\label{D} \\
iS(q)
&=&
\frac{i}{\qs-m+i\epsilon}\,\,,
\label{S}
\eea
respectively, where $m$ is the renormalized mass of the given 
quark.\footnote{Throughout, our approach is purely perturbative; all
issues of confinement are ignored.}

It will be convenient to introduce transverse and longitudinal
projection operators as follows: 
\be
t_{\mu\nu}(q) = g_{\mu\nu} - \frac{q_{\mu}q_{\nu}}{q^{2}}\,\,,
\qquad\qquad
l_{\mu\nu}(q) = \frac{q_{\mu}q_{\nu}}{q^{2}}\,\,.
\ee
Also, we use as abbreviation for the volume element
appearing in the Feynman integrals
\be
[dk]
=
\mu^{2\epsilon}\frac{d^{d}k}{(2\pi)^{d}}
\ee
(we work always in $d = 4-2\epsilon$ dimensions with
`t~Hooft mass scale $\mu$). 

In the conventional perturbation theory approach,
the renormalized one-loop fermion self-energy (two-point function) 
in QCD is specified by the diagrams shown in Fig.~1:
\bea
{\rm Fig.\,1a}
\,\,=\,\, 
-i\Sigma^{(1)}(p,\xi)
&=&
C_{F}g^{2}\int[dk]D^{\rho\rho'\!}(k,\xi)
\gamma_{\rho}S(p-k)\gamma_{\rho'}  \nn \\
& &
+ (Z_{2}-1)_{\xi}^{(1)}i(\ps - m) - (Z_{2}(Z_{m}-1))_{\xi}^{(1)}im 
\qquad\phantom{\biggl|}
\label{fig1}  \\
&=&
-i\Bigl(\Sigma_{1}^{(1)}(p^{2},\xi)
\,+\,(\ps - m)\Sigma_{2}^{(1)}(p^{2},\xi) \Bigr)\,\,, 
\eea
where $C_{F}$ is the quadratic Casimir coefficient for the
fundamental representation 
($C_{F} = (N^{2}-1)/2N$ \hfill for \hfill SU($N$)) \hfill
and \hfill $g$ \hfill is \hfill the \hfill renormalized \hfill
QCD \hfill gauge \hfill coupling \hfill
($\alpha_{s} = g^{2}/4\pi$). 
\linebreak
In \hfill \eq{fig1}, \hfill $Z_{2}$ \hfill and \hfill $Z_{m}$ \hfill
are \hfill the \hfill quark \hfill wavefunction \hfill and \hfill 
mass \hfill renormalization \hfill constants,

\pagebreak

\begin{center}
\begin{picture}(430,110)(-90,185)
 

\ArrowLine(  0,280)( 24,280)
\GCirc( 35,280){11}{0.8}
\put( 35,280){\makebox(0,0)[c]{\Large{\bf 1}}}
\ArrowLine( 46,280)( 70,280)
\put( 35,250){\makebox(0,0)[c]{(a)}}
 
\put( 80,280){\makebox(0,0)[c]{$=$}}
 
\ArrowLine( 90,280)(160,280)
\GlueArc(125,280)(15,  0,180){2}{7}
\put(125,250){\makebox(0,0)[c]{(b)}}
 
\put(170,280){\makebox(0,0)[c]{$+$}}
 
\ArrowLine(180,280)(215,280)
\ArrowLine(215,280)(250,280)
\GCirc(215,280){4}{0}
\put(215,250){\makebox(0,0)[c]{(c)}}

\put(-95,225){\makebox(0,0)[l]{\small
Fig.\ 1. The Feynman diagrams specifying the conventional
renormalized one-loop fermion self-energy }}

\put(-95,210){\makebox(0,0)[l]{\small
$-i\Sigma^{(1)}(p,\xi)$ in QCD.
The straight (curly) line represents the fermion
(gluon) propagator. The black }}

\put(-95,195){\makebox(0,0)[l]{\small
blob represents the sum of the fermion wavefunction
and fermion mass counterterms. }}

\end{picture}
\end{center}

\noindent
respectively
($\psi_{0} = Z_{2}^{1/2}\psi$ and
$m_{0} = Z_{m}m$, where the subscript 0 denotes
bare quantities);
the subscript $\xi$ indictes the class of linear covariant gauges,
while the superscript $(1)$ denotes the $\cO(\alpha_{s})$
term in the perturbative expansion.
Carrying out the integration, the one-loop functions 
$\Sigma_{1}^{(1)}$ and $\Sigma_{2}^{(1)}$ are given by 
(cf.\ e.g.\ Ref.~\cite{coq})
\bea
\Sigma_{1}^{(1)}(p^{2},\xi)
&=&
\frac{\alpha_{s}}{4\pi}C_{F}m\biggl\{ 
3C_{\rm UV} -3\ln\biggl(\frac{m^{2}-p^{2}}{\mu^{2}}\biggr) + 1 
+ \biggl( 3  + \xi - \xi\frac{m^{2}}{p^{2}}\biggr)L
\biggr\} + (Z_{m}-1)_{\xi}^{(1)}m\,\,, \nn \\
\label{Sigma11}
\\
\Sigma_{2}^{(1)}(p^{2},\xi)
&=&
\frac{\alpha_{s}}{4\pi}C_{F}\xi\biggl\{ 
-C_{\rm UV} + \ln\biggl(\frac{m^{2}-p^{2}}{\mu^{2}}\biggr) - 1
- \frac{m^{2}}{p^{2}}L
\biggr\}  - (Z_{2}-1)_{\xi}^{(1)}\,\,, \label{Sigma12}
\eea
where $C_{\rm UV} = \epsilon^{-1} + \ln(4\pi) - \gamma_{E}$,
with $\gamma_{E}$ Euler's constant, 
and $L = 1 + (m^{2}/p^{2})\ln[1 -(p^{2}/m^{2})]$.

In general, and as exemplified by Eqs.~(\ref{Sigma11}) and (\ref{Sigma12}),
the conventional self-energy $\Sigma(p,\xi)$
depends on the particular choice of gauge for 
all values of the fermion four-momentum $p$
except at $\ps = M$ where $M$ is the pole mass, defined from
the solution of the transcendental equation
\be
\ps - m - \Sigma(p,\xi)\Bigl|_{p\!\!\!/ = M}
=
0\,\,.
\ee
The gauge-independent \cite{polegind} (and infrared-finite \cite{poleir})
pole mass $M$ is related in perturbation theory to the 
gauge-independent renormalized mass $m$ by
\be
M
=
m\biggl\{ 1 +   
  \biggl(\frac{\alpha_{s}}{4\pi}\biggr)c_{1} 
+ \biggl(\frac{\alpha_{s}}{4\pi}\biggr)^{2}c_{2} 
+ \ldots
\biggr\}\,\,,
\ee
where the coefficients $c_{i}$ are renormalization 
scheme- and scale-dependent. For example, in the modified minimal
subtraction ($\overline{\rm MS}$) scheme, from \eq{Sigma11},
$c_{1} = C_{F} [4 + 3\ln(\mu^{2}/M^{2})]$,
while $c_{2}$ may be found in Refs.~\cite{grayetal,fleischeretal}.

In the PT approach, the conventional self-energy (\ref{fig1})
is considered as just one particular correction to a given
tree level process $1+2 \rightarrow 3 + 4$ involving the fermion
as virtual intermediate state.
The construction of the PT fermion self-energy 
is then based on the recognition that the {\em integrands\,} 
for the conventional vertex and box
corrections to this process
have well-defined components with exactly
the same structure as that of the conventional self-energy.
These components occur as a result of the presence in the integrands
of factors of longitudinal four-momentum $k_{\rho}$ associated
with the gluons $A_{\rho}^{r}(k)$ propagating in loops.
When contracted with the adjacent tree level vertices, 
these longitudinal factors trigger the associated elementary Ward identities. 
The triggering of these Ward identities results in the
rearrangement of the 
\hfill propagator \hfill and \hfill vertex \hfill structure \hfill of 
\hfill the \hfill integrand. \hfill 
After \hfill systematically \hfill contracting \hfill all 

\pagebreak

\begin{center}
\begin{picture}(440,145)(-20, 0)
 

\GCirc(200,100){14.2}{0.7}
\Photon(170,70)(190,90){3}{3}
\Photon(230,70)(210,90){-3}{3}
\ArrowLine(170,130)(190,110)
\ArrowLine(210,110)(230,130)
\put(171,83){\vector(1, 1){10}}
\put(220,93){\vector(1,-1){10}}
 
\put(160, 60){\makebox(0,0)[c]{$p_{1},\nu$}}
\put(160,140){\makebox(0,0)[c]{$p_{2}$}}
\put(240, 60){\makebox(0,0)[c]{$p_{3},\mu$}}
\put(240,140){\makebox(0,0)[c]{$p_{4}$}}

\put(-20,25){\makebox(0,0)[l]{\small
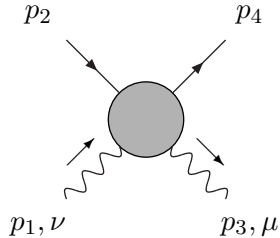
Fig.\ 2. The kinematics of the QCD Compton scattering process
$q\gamma \rightarrow q\gamma$. The wavy lines represent}}
 
\put(-20,10){\makebox(0,0)[l]{\small
the photons.}}
 
\end{picture}
\end{center}
 
\noindent
such longitudinal factors, the contributions to
the PT fermion self-energy (``effective'' two-point function)
are identified from the resulting integrands
by their naive self-energy-like propagator structure.

In order to construct the PT one-loop quark self-energy, we 
consider the ${\cal O}(\alpha_{s})$ QCD corrections to the lowest order
Compton scattering $q\gamma\rightarrow q\gamma$ 
of a photon off a quark with electromagnetic charge $Q$.
The kinematics of this process are shown in Fig.~2.
Given that the gluons only couple directly with the quark,
this is the very simplest process from which to construct the PT
quark self-energy. Indeed, the fact that the non-abelian character
of the gluons plays essentially no role at the one-loop level
in this process will make the construction almost trivial
(one could use more complicated one-loop process involving also purely
non-abelian interactions, e.g.\ the Compton-like
scattering $qg\rightarrow qg$ 
of a gluon off a quark as in \cite{PTquark}).
For convenience, the quarks are taken to be on-shell, although this
need not be so \cite{PTqcdeffch}.

The set of six diagrams specifying the one-loop QCD corrections to 
the tree level scattering process are shown in Figs.~3a--f 
(the corresponding set of diagrams with crossed external photon 
legs are not shown, nor are the counterterm insertion diagrams).
The QCD contribution to the PT renormalized one-loop quark self-energy
(``effective'' two-point function) $-i\hat{\Sigma}^{(1)}(p)$
is defined from the coefficient $\hat{\Sigma}^{\prime(1)}(p,k)$ of the
component of the integrands for the diagrams in Figs.~3a--f which,
after the systematic contraction of all factors of longitudinal 
internal gluon four-momentum,
has the following self-energy-like structure:
\be
\int[dk]\,
iQ\gamma_{\mu}\,iS(p) 
\Bigl( -i\hat{\Sigma}^{\prime(1)}(p,k) \Bigr)
iS(p)\,iQ\gamma_{\nu}
\label{def1}
\ee
where 
\be
p = p_{1}+p_{2} = p_{3} + p_{4}
\ee
and $iQ\gamma_{\mu}$, $iQ\gamma_{\nu}$ are the tree level
$\gamma q\overline{q}$ vertices. Then
\be
-i\hat{\Sigma}^{(1)}(p)
=
-i\int[dk]\,\hat{\Sigma}^{\prime(1)}(p,k)
\,+\, (Z_{2}-1)_{\rm PT}^{(1)}i(\ps - m) 
\,-\, (Z_{2}(Z_{m}-1))_{\rm PT}^{(1)}im 
\,\,.
\label{def2}
\ee
In the above expression, $Z_{2}$ and $Z_{m}$ are the quark wavefunction
and mass renormalization constants in the PT framework,
determined at $\cO(\alpha_{s})$ from the divergence structure
of the integral in \eq{def2}
as $d \rightarrow 4$.
It is emphasized that the definition of the PT ``effective'' two-point
function is in terms of the Feynman {\em integrands}\, corresponding
to the diagrams for the interactions:
all rearrangements in the PT are carried out under the integral sign(s).

\pagebreak

\begin{center}
\begin{picture}(430,480)(0,-50)
 
 
\ArrowLine(  0,430)(  0,400)
\ArrowLine(  0,400)( 60,400)
\ArrowLine( 60,400)( 60,430)
\GlueArc( 30,400)(15,  0,180){2}{7}
\Photon( 0,380)( 0,400){-2}{3}
\Photon(60,380)(60,400){2}{3}
\put( 30,370){\makebox(0,0)[c]{(a)}}
 
\Line(  0,355)(  0,325)
\ArrowLine(  0,325)( 60,325)
\ArrowLine( 60,325)( 60,355)
\GlueArc(  0,325)(17,  0, 90){2}{4}
\Photon( 0,305)( 0,325){-2}{3}
\Photon(60,305)(60,325){2}{3}
\put( 30,295){\makebox(0,0)[c]{(b)}}
 
\ArrowLine(  0,280)(  0,250)
\ArrowLine(  0,250)( 60,250)
\Line( 60,250)( 60,280)
\GlueArc( 60,250)(17, 90,180){2}{4}
\Photon( 0,230)( 0,250){-2}{3}
\Photon(60,230)(60,250){2}{3}
\put( 30,220){\makebox(0,0)[c]{(c)}}
 
\Line(  0,205)(  0,175)
\ArrowLine(  0,175)( 60,175)
\Line( 60,175)( 60,205)
\Gluon(0,190)(60,190){2}{10}
\Photon( 0,155)( 0,175){-2}{3}
\Photon(60,155)(60,175){2}{3}
\put( 30,145){\makebox(0,0)[c]{(d)}}

\ArrowLine(  0,130)(  0, 90)
\ArrowLine(  0, 90)( 60, 90)
\ArrowLine( 60, 90)( 60,130)
\GlueArc(  0,110)(10,-90, 90){2}{4}
\Photon( 0, 80)( 0, 90){2}{1.5}
\Photon(60, 80)(60, 90){-2}{1.5}
\put( 30, 70){\makebox(0,0)[c]{(e)}}
 
\ArrowLine(  0, 55)(  0, 15)
\ArrowLine(  0, 15)( 60, 15)
\ArrowLine( 60, 15)( 60, 55)
\GlueArc( 60,35)(10, 90,270){2}{4}
\Photon( 0,  5)( 0, 15){2}{1.5}
\Photon(60,  5)(60, 15){-2}{1.5}
\put( 30, -5){\makebox(0,0)[c]{(f)}}
 
 
\put( 80,292){\makebox(0,0)[c]
{$ \left. \begin{array}{c}  \\ \\ \\ \\ \\ \\ \\ \\ \\ 
\\ \\ \\ \\ \\ \\ \\ \\ \\ \\
\end{array} \right\}$}}
 
\put( 120,292){\makebox(0,0)[c]{
\Large $=$}}
 
\put(160,292){\makebox(0,0)[c]
{$ \left\{ \begin{array}{c}  \\ \\ \\ \\ \\ \\ \\ \\ \\ 
\end{array} \right.$}}
 
 
\ArrowLine(180,355)(180,325)
\Line(180,325)(210,325)
\GlueArc(210,325)(17, 90, 180){2}{4}
\Photon(180,305)(180,325){-2}{3}
\put(195,295){\makebox(0,0)[c]{(g)}}
 
\Line(180,280)(180,250)
\ArrowLine(180,250)(210,250)
\Gluon(180,265)(210,265){2}{4}
\Photon(180,230)(180,250){-2}{3}
\put(195,220){\makebox(0,0)[c]{(h)}}
 
 
\put(230,292){\makebox(0,0)[c]
{$ \left. \begin{array}{c}  \\ \\ \\ \\ \\ \\ \\ \\ \\ 
\end{array} \right\}$}}

\put(305,320){\makebox(0,0)[c]{$k$}}
\put(300,310){\vector(1, 0){10}}
 
\Gluon(288,300)(322,300){2}{5}
\ArrowLine(290,284)(320,284)
 
\put(305,270){\makebox(0,0)[c]{$p-k$}}
 
\put(380,292){\makebox(0,0)[c]
{$ \left\{ \begin{array}{c}  \\ \\ \\ \\ \\ \\ \\ \\ \\ 
\end{array} \right.$}}
 
 
\Line(400,325)(430,325)
\ArrowLine(430,325)(430,355)
\GlueArc(400,325)(17,  0, 90){2}{4}
\Photon(430,305)(430,325){2}{3}
\put(415,295){\makebox(0,0)[c]{(i)}}
 
\ArrowLine(400,250)(430,250)
\Line(430,250)(430,280)
\Gluon(400,265)(430,265){2}{4}
\Photon(430,230)(430,250){2}{3}
\put(415,220){\makebox(0,0)[c]{(j)}}

\put(-5,-25){\makebox(0,0)[l]{\small
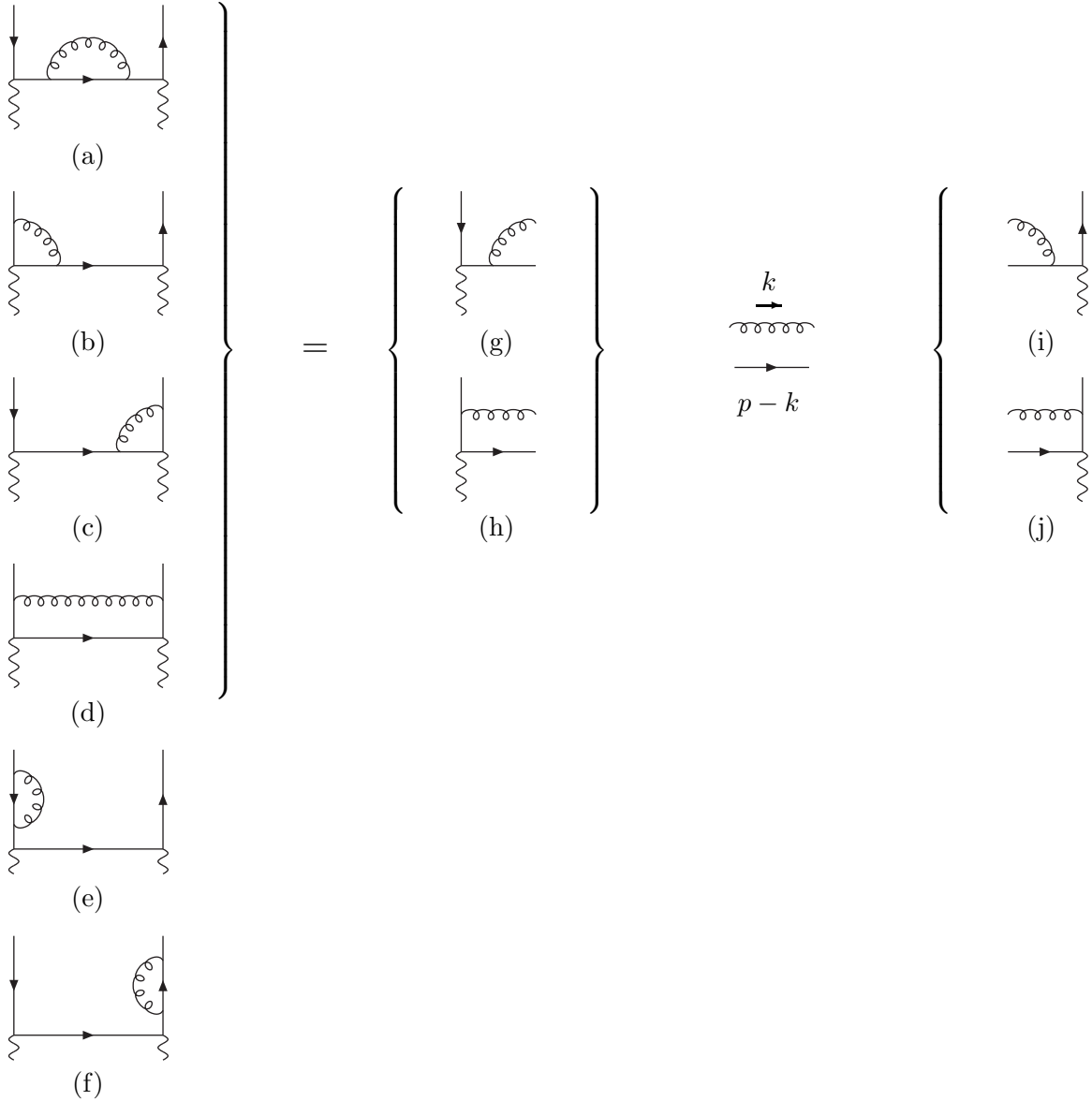
Fig.\ 3. a--f: The Feynman diagrams specifying the
one-loop QCD corrections to the Compton scat-}}

\put(-5,-40){\makebox(0,0)[l]{\small
tering process $q\gamma \rightarrow q\gamma$. }}

\end{picture}
\end{center}

We first consider the four diagrams shown in Figs.~3a--d.
In order to deal efficiently with these one-loop diagrams 
(and, in later sections, with various two-loop diagrams too)
it is convenient to define the connected four-point function
$G_{\rho\mu}^{r}(q_{1},q_{2},q_{3},q_{4})$
specifying the tree level coupling of a gluon
$A_{\rho}^{r}(q_{1})$ and a photon $A_{\mu}(q_{2})$
to a quark with electromagnetic charge $Q$.
The two relevant diagrams are shown in Fig.~4:
\bea
{\rm Figs.\,4a+4b}
&=&
iQg\,G_{\rho\mu}^{r}(q_{1},q_{2},q_{3},q_{4}) \\
&=&
iQg\Bigl( 
-\gamma_{\mu}S(q_{1}+q_{3})\gamma_{\rho}T^{r} 
-\gamma_{\rho}T^{r}S(q_{4}-q_{1})\gamma_{\mu}
\Bigr)
\label{G4}
\eea

\pagebreak

\begin{center}
\begin{picture}(440,145)(-20, 0)
 

\ArrowLine(145,100)(145,125)
\ArrowLine(100,100)(145,100)
\ArrowLine(100, 75)(100,100)
\Gluon(100,100)(100,125){4}{2}
\Photon(145,75)(145,100){3}{3}
\put( 90,120){\vector(0,-1){10}}
\put(155, 80){\vector(0, 1){10}}
 
\put( 90,140){\makebox(0,0)[c]{$q_{1},\rho,r $}}
\put(158, 60){\makebox(0,0)[c]{$q_{2},\mu $}}
\put( 90, 60){\makebox(0,0)[c]{$q_{3}$}}
\put(158,140){\makebox(0,0)[c]{$q_{4}$}}

\put(125, 40){\makebox(0,0)[c]{$(a)$}}

\ArrowLine(250, 75)(275,75)
\ArrowLine(275, 75)(275,120)
\ArrowLine(275,120)(300,120)
 
\Gluon(250,120)(275,120){4}{2}
\Photon(275,75)(300, 75){-3}{3}

\put(275, 40){\makebox(0,0)[c]{$(b)$}}
 
\put(-20,10){\makebox(0,0)[l]{\small
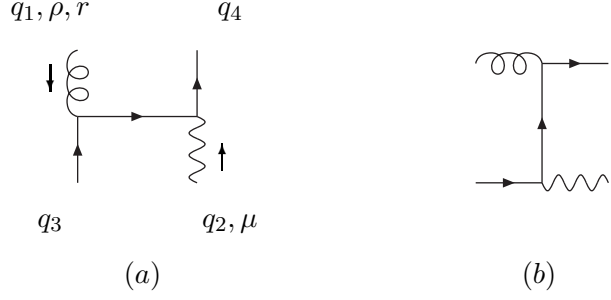
Fig.\ 4. The Feynman diagrams specifying
the tree level four-point function
$G_{\rho\mu}^{r}(q_{1},q_{2},q_{3},q_{4})$. }}

\end{picture}
\end{center}

\noindent
with $q_{1} + q_{2} + q_{3} = q_{4}$. 
In \eq{G4}, the hermitian matrices $T^{r}$ are the generators
for the fermion representation, satisfying
$[T^{a},T^{b}] = if^{abc}T^{c}$.
Using the elementary Ward identity
\be
\qs
=
S^{-1}(q + q') - S^{-1}(q')\,\,,
\label{wid}
\ee
this four-point function is easily shown to obey the Ward identity
\be
q_{1}^{\rho} G_{\rho\mu}^{r}(q_{1},q_{2},q_{3},q_{4})
=
\gamma_{\mu}S(q_{1}+q_{3})T^{r}S^{-1}(q_{3})
-S^{-1}(q_{4})T^{r}S(q_{4}-q_{1})\gamma_{\mu}\,\,.
\label{G4wid}
\ee

As indicated in Fig.~3, 
the sum of the four diagrams 3a--d may be expressed as the
sum of all possible contractions of the diagrams 3g and 3h with
the diagrams 3i and 3j, 
where the straight (curly) line between these diagrams
represents the propagator for the fermion (gluon) propagating in
the loops. The sums of the diagrams 3g+3h and 3i+3j are each
just the connected four-point function defined in \eq{G4}.
Thus, the sum of the four one-loop diagrams 3a--d 
for arbitrary $\xi$ may be written as
\be
{\rm Figs.\,3a\!-\!d}
=
Q^{2}g^{2}\int[dk]\,D^{\rho\rho'\!}(k,\xi)
G_{\rho\mu}^{r}(k,-p_{3},p-k,p_{4})
S(p-k)
G_{\rho'\!\nu}^{r}(-k,p_{1},p_{2},p-k)
\label{fig3ad}
\ee
(spinors for the on-shell external fermions have been omitted).
In order to identify the PT ``effective'' two-point component of
these diagrams, it is then necessary to contract systematically
the factors of longitudinal gluon four-momentum occurring in
the integrand in \eq{fig3ad}, triggering the tree level Ward identities.
Using the Ward identity (\ref{G4wid}), together with the fact that
$S^{-1}(p_{2}) = S^{-1}(p_{4}) = 0$ for the on-shell external fermions, 
we easily obtain
\bea
{\rm Figs.\,3a\!-\!d}
&=&
-Q^{2}g^{2}\int[dk]\,\frac{1}{k^{2}}
\biggl\{
G_{\rho\mu}^{r}(k,-p_{3},p-k,p_{4})
S(p-k)
G_{\phantom{\rho}\nu}^{\rho,r}(-k,p_{1},p_{2},p-k) \nn \\
& & -
C_{F}(1-\xi)\frac{1}{k^{2}}
\gamma_{\mu}S(p)\gamma_{\nu} \biggl\}\,\,.
\label{fig3ad2}
\eea

It remains to consider the two external leg corrections in 
Figs.~3e and 3f. These diagrams are given by
\bea
{\rm Figs.\,3e + 3f}
&=&
\frac{1}{2}Q^{2}g^{2}
\int[dk]\,D^{\rho\rho'\!}(k,\xi) \biggl\{
\gamma_{\mu}S(p)\gamma_{\nu} S(p_{2})
\gamma_{\rho}T^{r}S(p_{2}-k)\gamma_{\rho'\!}T^{r} \nn \\
& & +\,
\gamma_{\rho}T^{r}S(p_{4}-k)\gamma_{\rho'\!}T^{r}S(p_{4})
\gamma_{\mu}S(p)\gamma_{\nu} 
\biggr\}\,\,.
\label{fig3ef}
\eea

\begin{center}
\begin{picture}(400,195)(25,-40)
 

\put( 70,145){\makebox(0,0)[c]{$k^{\rho}$}}
\put( 80,144){\vector(1, 0){10}}
 
\Line(100,125)(130,125)
\ArrowLine(130,125)(130,155)
\GlueArc(100,125)(17,  0, 90){2}{4}
\Photon(130,105)(130,125){2}{3}
\put(100, 95){\makebox(0,0)[c]{(a)}}
 
\put(215,135){\makebox(0,0)[c]{pinch}}
\put(200,125){\vector(1, 0){30}}
 
\put(280,125){\makebox(0,0)[c]{\large{$+$}}}
 
\Line(300,125)(330,125)
\ArrowLine(330,125)(330,155)
\Gluon(300,140)(330,125){2}{5}
\Photon(330,105)(330,125){2}{3}
\put(330, 95){\makebox(0,0)[c]{(b)}}
 
\put(360,125){\makebox(0,0)[c]{\large{$+$}}}
\put(390,125){\makebox(0,0)[c]{$\cdots$}}
 
 
\put( 70, 67){\makebox(0,0)[c]{$k^{\rho}$}}
\put( 80, 66){\vector(1, 0){10}}
 
\ArrowLine(100, 50)(130, 50)
\Line(130, 50)(130, 80)
\Gluon(100, 65)(130, 65){2}{4}
\Photon(130, 30)(130, 50){2}{3}
\put(100, 20){\makebox(0,0)[c]{(c)}}
 
\put(215, 60){\makebox(0,0)[c]{pinch}}
\put(200, 50){\vector(1, 0){30}}
 
\put(280, 50){\makebox(0,0)[c]{\large{$-$}}}
 
\ArrowLine(300, 50)(330, 50)
\Line(330, 50)(330, 80)
\Gluon(300, 65)(330,50){2}{5}
\Photon(330, 30)(330, 50){2}{3}
\put(330, 20){\makebox(0,0)[c]{(d)}}
 
 
\put(5,-15){\makebox(0,0)[l]{\small
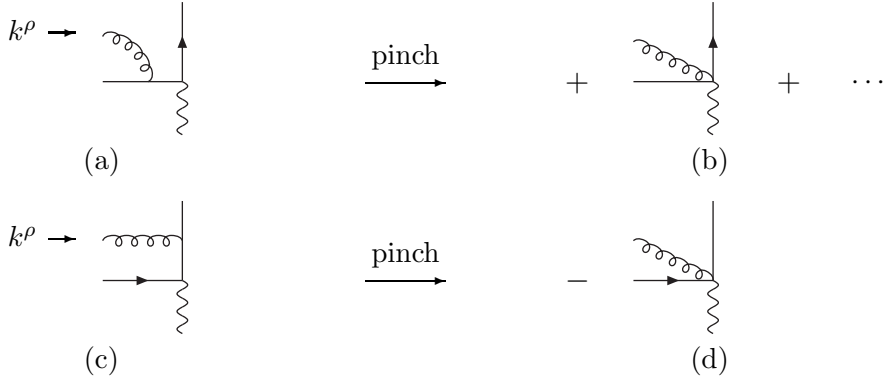
Fig.\ 5. The elementary cancellation occurring among 
the diagrams i and j in Fig.~3 when contracted
}}

\put(5,-30){\makebox(0,0)[l]{\small
with the gluon longitudinal four-momentum $k^{\rho}$, 
and expressed in the Ward identity Eq.~(\ref{G4wid}).}}
 
\end{picture}
\end{center}

\noindent
Using the elementary Ward identity (\ref{wid}) for the longitudinal
factors $k^{\rho}$, $k^{\rho'}$ occurring 
in \eq{fig3ef}, together with
$S^{-1}(p_{2}) = S^{-1}(p_{4}) = 0$ for the on-shell external fermions, 
we obtain
\bea
{\rm Figs.\,3e + 3f}
&=&
-\frac{1}{2}C_{F}Q^{2}g^{2}
\int[dk]\,\frac{1}{k^{2}} \biggl\{
\gamma_{\mu}S(p)\gamma_{\nu} S(p_{2})
\gamma_{\rho}S(p_{2}-k)\gamma_{\rho'\!} \nn \\
& & +\,
\gamma_{\rho}S(p_{4}-k)\gamma_{\rho'\!}S(p_{4})
\gamma_{\mu}S(p)\gamma_{\nu} 
\,+\,
2(1-\xi)\frac{1}{k^{2}}\gamma_{\mu}S(p)\gamma_{\nu}
\biggr\}\,\,.
\label{fig3ef2}
\eea

Combining the expressions (\ref{fig3ad2}) and (\ref{fig3ef2}), 
the last terms cancel, giving
\bea
\lefteqn { {\rm Figs.\,3a\!-\!f}
\,\,=\,\,
-Q^{2}g^{2}\int[dk]\,\frac{1}{k^{2}}
\biggl\{
G_{\rho\mu}^{r}(k,-p_{3},p-k,p_{4})
S(p-k)
G_{\phantom{\rho}\nu}^{\rho,r}(-k,p_{1},p_{2},p-k)  } \nn \\
& &
+ \frac{1}{2}C_{F}\Bigl(
\gamma_{\mu}S(p)\gamma_{\nu} S(p_{2})
\gamma_{\rho}S(p_{2}-k)\gamma_{\rho'\!}
+
\gamma_{\rho}S(p_{4}-k)\gamma_{\rho'\!}S(p_{4})
\gamma_{\mu}S(p)\gamma_{\nu} \Bigr) \biggl\}\,\,. \qquad\qquad
\label{fig3af}
\eea
Having thus contracted all longitudinal factors appearing
in the integrands in Eqs.~(\ref{fig3ad}) and (\ref{fig3ef}),
the expressions for the two four-point functions may now be
substituted into the resulting \eq{fig3af}.
The PT ``effective'' two-point component of the
integrand may then be isolated according to the definition (\ref{def1}):
\be
{\rm Figs.\,3a\!-\!f}|_{2-{\rm pt}}
=
-C_{F}Q^{2}g^{2}\int[dk]\,\frac{1}{k^{2}}
\gamma_{\mu}S(p)\gamma_{\rho}
S(p-k)
\gamma^{\rho}S(p)\gamma_{\nu}\,\,.
\label{fig3af2pt}
\ee
Then, from the definitions (\ref{def1}) and (\ref{def2}),
\be
-i\hat{\Sigma}^{(1)}(p)
=
-C_{F}g^{2}\int[dk]\,\frac{1}{k^{2}+i\epsilon}
\gamma_{\rho}S(p-k)\gamma^{\rho}
\,+\, (Z_{2}-1)_{\rm PT}^{(1)}i(\ps - m) 
\,-\, (Z_{2}(Z_{m}-1))_{\rm PT}^{(1)}im 
\label{Sigmahat1}
\ee
where the $+i\epsilon$ prescription has been restored in the gluon
propagator.

Several remarks are in order:

(1) In the contribution to the PT self-energy \eq{Sigmahat1}
from the diagrams 3a--d, an elementary cancellation has occurred between
(i) the component of the diagram 3i in which the internal 
fermion propagator $S(p)$ has been cancelled (pinched)
by the action of the $\xi$-dependent
longitudinal factor $k^{\rho}$ from the gluon propagator, and (ii)
the entire contribution of the diagram 3j in which the internal fermion
propagator $S(p_{4}-k)$ has been cancelled (pinched) by the
action of the same factor. This cancellation is illustrated in Fig.~5,
and is encoded in the Ward identity (\ref{G4wid}) for the
four-point function $G_{\mu\rho}^{r}$ in \eq{fig3ad}.
An identical cancellation takes place for the diagrams 3g and 3h
when contracted with the factor $k^{\rho'}$.
By dealing with the set of diagrams 3a--d, as opposed to individual
diagrams, we have been able to make these cancellations among
the conventional self-energy, vertex and box functions simply
and immediately. These cancellations leave just the tadpole-like term
proportional to $(1-\xi)k^{-4}$ in \eq{fig3ad2}, which
is then cancelled algebraically by a
similar term in \eq{fig3ef2} for the 
external leg corrections.\footnote{Note that we do not 
set these integrals to zero in dimensional regularization.}
While elementary in the one-loop case, 
this approach, involving the simultaneous 
consideration of subsets of diagrams,
when extended to the two-loop case
will greatly facilitate the construction 
of the PT two-loop quark self-energy in the following sections,
in particular the demonstration of its gauge independence.

(2) Here we have chosen to work in the class of ordinary
linear covariant gauges. However, we could have started in 
{\em any}\, gauge, i.e.\ with gluon propagator
\be
iD_{\mu\nu}(q,a,b)
=
\frac{i}{q^{2}+i\epsilon}
\biggl( -g_{\mu\nu} + a_{\mu}(q)q_{\nu} + q_{\mu}a_{\nu}(q)
+ b(q)q_{\mu}q_{\nu} \biggr)\,\,.
\label{Dqab}
\ee
For example, in the class of non-covariant gauges defined
by adding to the classical lagrangian the gauge-fixing term
${\cal L}_{\rm gf} = -(n\cdot A^{a})^{2}/2\lambda$
for some four-vector $n_{\mu}$ and constant $\lambda$,
one has $a_{\mu}(q) = n_{\mu}/n\cdot q$ and
$b(q) = -(n^{2} + \lambda q^{2})/(n\cdot q)^{2}$.
In all cases, an exactly similar algebraic cancellation takes place
among contributions to the PT ``effective'' two-point component
of the interaction.
In particular, all integrands involving terms
$(n\cdot k)^{-1}$, $(n\cdot k)^{-2}$ 
cancel algebraically, and so present no difficulties.
Thus, the PT self-energy \eq{Sigmahat1} is fully gauge-independent.

(3) The PT gauge-independent one-loop quark self-energy coincides
with that obtained in the conventional Feynman gauge $\xi = 1$, 
which in turn coincides with that obtained in the BFM in the 
Feynman quantum gauge $\xi_{Q} = 1$ \cite{PTquark}:
\be
-i\hat{\Sigma}^{(1)}(p)
=
-i\Sigma^{(1)}(p,\xi=1)
=
-i\Sigma_{\rm BFM}^{(1)}(p,\xi_{Q}=1)\,\,.
\ee
This is in fact obvious: in the integrands for the
one-loop corrections to the process
$q\gamma \rightarrow q\gamma$, the only sources of factors of
longitudinal internal gluon four-momentum are the
gluon propagators. Starting from the gauge $\xi = 1$,
these factors are not present. Thus, 
for this particular process in this particular gauge, no
Ward identities are triggered (there is ``no pinching''),
and one obtains the PT one-loop quark self-energy directly
from the conventional self-energy.

(4) The above construction of $\hat{\Sigma}^{(1)}(p)$
involved the process $q\gamma \rightarrow q\gamma$.
However, we could have used {\em any}\, one-loop process
involving the quark as intermediate virtual state,
i.e.\ the self-energy \eq{Sigmahat1} is universal.

\pagebreak

\begin{center}
\begin{picture}(440,270)(0,65)
 
 
\Photon(  0,325)( 14,325){2}{3}
\GCirc( 25,325){11}{0.8}
\put( 25,326){\makebox(0,0)[c]{\Large{\bf 2}}}
\Photon( 36,325)( 50,325){2}{3}
\put( 25,295){\makebox(0,0)[c]{(a)}}
 
\put( 57.5,325){\makebox(0,0)[c]{$=$}}

\Photon( 65,325)( 79,325){2}{3}
\ArrowArc( 90,325)(11,-90,270)
\Photon(101,325)(115,325){2}{3}
\GCirc( 90,317){6}{0.8}
\put( 91,318){\makebox(0,0)[c]{{\bf 1}}}
\put( 90,295){\makebox(0,0)[c]{(b)}}
 
\put(122.5,325){\makebox(0,0)[c]{$+$}}
 
\Photon(130,325)(144,325){2}{3}
\ArrowArc(155,325)(11, 90,450)
\Photon(166,325)(180,325){2}{3}
\GCirc(155,333){6}{0.8}
\put(156,334){\makebox(0,0)[c]{{\bf 1}}}
\put(155,295){\makebox(0,0)[c]{(c)}}
 
\put(187.5,325){\makebox(0,0)[c]{$+$}}
 
\Photon(195,325)(209,325){2}{3}
\ArrowArc(220,325)(11,-90,270)
\Photon(231,325)(245,325){2}{3}
\GCirc(229,325){6}{0.8}
\put(230,326){\makebox(0,0)[c]{{\bf 1}}}
\put(220,295){\makebox(0,0)[c]{(d)}}

\put(252.5,325){\makebox(0,0)[c]{$+$}}
 
\Photon(260,325)(274,325){2}{3}
\ArrowArc(285,325)(11,-90,270)
\Photon(296,325)(310,325){2}{3}
\GCirc(276,325){6}{0.8}
\put(276.5,325){\makebox(0,0)[c]{{\bf 1}}}
\put(285,295){\makebox(0,0)[c]{(e)}}
 
\put(317.5,325){\makebox(0,0)[c]{$-$}}
 
\Photon(325,325)(337.5,325){2}{2.5}
\ArrowArc(347.5,325)(10,100,180)
\CArc(347.5,325)(10,180,270)
\ArrowArc(352.5,325)(10,-80, 0)
\CArc(352.5,325)(10, 0, 90)
\Photon(362.5,325)(375,325){2}{2.5}
\GBox(345,311)(355,338){0.8}
\put(351,325){\makebox(0,0)[c]{{\bf 0}}}
\put(350,295){\makebox(0,0)[c]{(f)}}
 
\put(382.5,325){\makebox(0,0)[c]{$+$}}
 
\Photon(390,325)(440,325){2}{11}
\GCirc(415,325){2.5}{0}
\put(415,295){\makebox(0,0)[c]{(g)}}
 

\put( 57.5,250){\makebox(0,0)[c]{$=$}}
 
\Photon( 65,250)(79,250){2}{3}
\ArrowArc(90,250)(11,-90,270)
\Photon(101,250)(115,250){2}{3}
\PhotonArc(90,238)(9,30,150){2}{3.5}
\put( 90,220){\makebox(0,0)[c]{(h)}}
 
\put(122.5,250){\makebox(0,0)[c]{$+$}}
 
\Photon(130,250)(144,250){2}{3}
\ArrowArc(155,250)(11, 90,450)
\Photon(166,250)(180,250){2}{3}
\PhotonArc(155,262)(9,-150,-30){2}{3.5}
\put(155,220){\makebox(0,0)[c]{(i)}}
 
\put(187.5,250){\makebox(0,0)[c]{$+$}}
 
\Photon(195,250)(209,250){2}{3}
\ArrowArc(220,250)(11,-45,315)
\Photon(231,250)(245,250){2}{3}
\Photon(220,239)(220,261){2}{4}
\put(220,220){\makebox(0,0)[c]{(j)}}
 

\put( 57.5,175){\makebox(0,0)[c]{$+$}}
 
\Photon( 65,175)(79,175){2}{3}
\ArrowArc( 90,175)(11,-90,270)
\Photon(101,175)(115,175){2}{3}
\GCirc( 90,164){2.5}{0}
\put( 90,145){\makebox(0,0)[c]{(k)}}
 
\put(122.5,175){\makebox(0,0)[c]{$+$}}
 
\Photon(130,175)(144,175){2}{3}
\ArrowArc(155,175)(11, 90,450)
\Photon(166,175)(180,175){2}{3}
\GCirc(155,186){2.5}{0}
\put(155,145){\makebox(0,0)[c]{(l)}}
 
\put(187.5,175){\makebox(0,0)[c]{$+$}}
 
\Photon(195,175)(209,175){2}{3}
\ArrowArc(220,175)(11,-90,270)
\Photon(231,175)(245,175){2}{3}
\GCirc(231,175){2.5}{0}
\put(220,145){\makebox(0,0)[c]{(m)}}
 
\put(252.5,175){\makebox(0,0)[c]{$+$}}
 
\Photon(260,175)(274,175){2}{3}
\ArrowArc(285,175)(11,-90,270)
\Photon(296,175)(310,175){2}{3}
\GCirc(274,175){2.5}{0}
\put(285,145){\makebox(0,0)[c]{(n)}}
 
\put(317.5,175){\makebox(0,0)[c]{$+$}}
 
\Photon(325,175)(375,175){2}{11}
\GCirc(350,175){2.5}{0}
\put(350,145){\makebox(0,0)[c]{(o)}}


\put(0,120){\makebox(0,0)[l]{\small
Fig.\ 6. The Feynman diagrams specifying the
renormalized two-loop vacuum polarization in QED.}}
 
\put(0,105){\makebox(0,0)[l]{\small
The grey blobs marked ``1'' represent the 
renormalized one-loop internal corrections. 
The grey box }}

\put(0, 90){\makebox(0,0)[l]{\small
marked ``0'' represents the lowest order 
contribution to the electron-positron scattering kernel.
The }}

\put(0, 75){\makebox(0,0)[l]{\small
black blobs represent the counterterm insertions. }}

\end{picture}
\end{center}

 
\setcounter{equation}{0}

\section{The pinch technique two-loop quark self-energy}


We now turn to the 1PI two-loop quark self-energy in the PT approach.
In this section, the general diagrammatic representation of 
the two-loop quark self-energy
in terms of one-loop two- and three-point function and tree level 
Bethe-Salpeter-type 
kernel insertions in the one-loop quark self-energy is described.
It is this diagrammatic representation which 
will provide the basis for the
explicit construction of the PT two-loop quark self-energy 
$-i\hat{\Sigma}^{(2)}(p)$ in the subsequent sections.

In order to gain insight, it is instructive to consider first
the renormalized two-loop photon self-energy (vacuum polarization) 
$i\Pi_{\mu\nu}^{(2)}(p)$ in QED.
The diagrams which contribute to this function are shown in Fig.~6.
In Figs.~6b--g, the two-loop vacuum polarization has been written
in terms of renormalized one-loop two- and three-point
functions, represented by the grey blobs,
inserted in all possible ways
in the diagram for the one-loop vacuum polarization 
$i\Pi_{\mu\nu}^{(1)}(p)$, 
{\em minus}\, a contribution involving the lowest
order term in the perturbative expansion of the 
{\em electron-positron scattering kernel} $K$, 
represented by the grey box,
inserted in the diagram for $i\Pi_{\mu\nu}^{(1)}(p)$.
The kernel $K$ is characterized by the fact that 
it has no contributions involving 
annihilation into a one-particle (photon) intermediate state 
and is 2PI with respect to the electron-positron 
pair of lines 
(cf.\ e.g.\ chapter 19 of Ref.~\cite{bjdr}).
The contribution in Fig.~6g is the two-loop 
counterterm required
finally to renormalize the two-loop function.
The representation in Figs.~6b--g involving the electron-positron
scattering kernel follows
from the Dyson-Schwinger equations for QED 
(cf.\ in particular Fig.~19.45 of Ref.~\cite{bjdr}; 
for a review of the applications of Dyson-Schwinger equations, 
see Ref.~\cite{DS}).
The diagrams which contribute \hfill to \hfill the \hfill one-loop
\hfill internal \hfill corrections \hfill and \hfill at \hfill 
lowest \hfill order \hfill to \hfill the \hfill
scattering \hfill kernel \hfill are \hfill in \hfill turn 

\pagebreak

\begin{center}
\begin{picture}(440,305)(-85,-95)


\ArrowLine(  0,200)( 24,200)
\GCirc( 35,200){11}{0.8}
\put( 35,200){\makebox(0,0)[c]{\Large{\bf 1}}}
\ArrowLine( 46,200)( 70,200)
\put( 35,170){\makebox(0,0)[c]{(a)}}
 
\put( 85,200){\makebox(0,0)[c]{$=$}}
 
\ArrowLine(100,200)(170,200)
\PhotonArc(135,200)(15,  0,180){2}{7.5}
\put(135,170){\makebox(0,0)[c]{(b)}}
 
\put(185,200){\makebox(0,0)[c]{$+$}}
 
\ArrowLine(200,200)(235,200)
\ArrowLine(235,200)(270,200)
\GCirc(235,200){4}{0}
\put(235,170){\makebox(0,0)[c]{(c)}}
 

\ArrowLine(  0,120)( 24,120)
\Photon( 35,120)( 35, 85){2}{5}
\GCirc( 35,120){11}{0.8}
\put( 35,120){\makebox(0,0)[c]{\Large{\bf 1}}}
\ArrowLine( 46,120)( 70,120)
\put( 35, 70){\makebox(0,0)[c]{(d)}}
 
\put( 85,120){\makebox(0,0)[c]{$=$}}
 
\ArrowLine(100,120)(120,120)
\Line(120,120)(150,120)
\ArrowLine(150,120)(170,120)
\PhotonArc(135,120)(15,  0,180){2}{7.5}
\Photon(135,120)(135, 85){2}{5}
\put(135, 70){\makebox(0,0)[c]{(e)}}
 
\put(185,120){\makebox(0,0)[c]{$+$}}
 
\ArrowLine(200,120)(235,120)
\ArrowLine(235,120)(270,120)
\Photon(235,120)(235, 85){2}{5}
\GCirc(235,120){4}{0}
\put(235, 70){\makebox(0,0)[c]{(f)}}
 

\ArrowLine(  5,-5)( 25, 0)
\ArrowLine( 45, 0)( 65,-5)
\ArrowLine( 65,30)( 45,25)
\ArrowLine( 25,25)(  5,30)
\GBox(25,-10)(45,35){0.8}
\put( 35,13){\makebox(0,0)[c]{\Large{\bf 0}}}
\put( 35,-30){\makebox(0,0)[c]{(g)}}
 
\put( 85, 12.5){\makebox(0,0)[c]{$=$}}
 
\ArrowLine(105,-5)(135,2.5)
\ArrowLine(135,2.5)(165,-5)
\ArrowLine(165,30)(135,22.5)
\ArrowLine(135,22.5)(105,30)
\Photon(135,2.5)(135,22.5){2}{3}
\put(135,-30){\makebox(0,0)[c]{(h)}}

\put(-85, -55){\makebox(0,0)[l]{\small
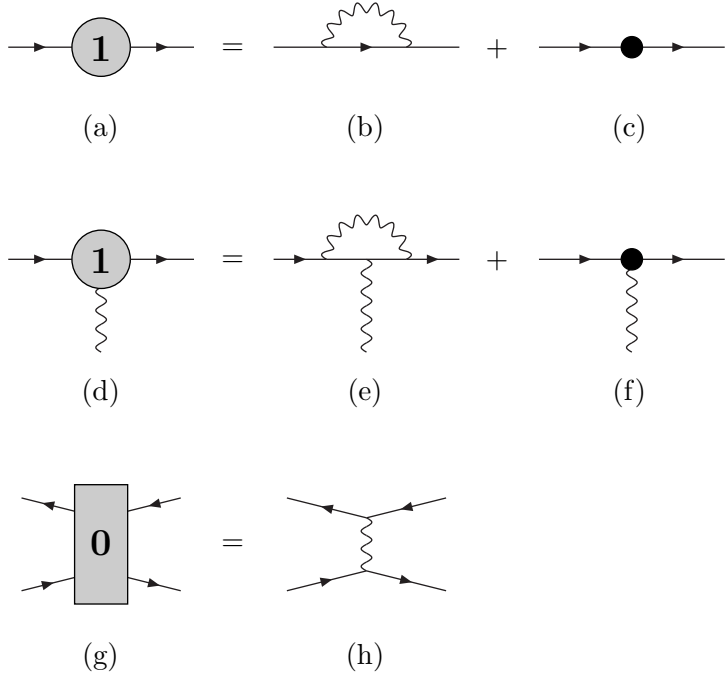
Fig.\ 7. The Feynman diagrams specifying the renormalized
one-loop fermion self-energy (7a) and }}

\put(-85,-70){\makebox(0,0)[l]{\small
photon-fermion vertex (7d), and the lowest order 
contribution to the electron-positron scattering}}
 
\put(-85,-85){\makebox(0,0)[l]{\small
kernel (7g) in QED.}}

\end{picture}
\end{center}

\noindent
shown in Fig.~7.
Substituting the diagrams of Fig.~7 into the representation
in Figs.~6b--g, we recover the usual two-loop perturbation theory
diagrams for the vacuum polarization as shown in Figs.~6h--o.
In particular, in the language of individual two-loop Feynman diagrams,
the effect of the kernel insertion contribution of Fig.~6f
is to compensate for the overcounting which would otherwise occur
due to the two
vertex insertion contributions of Figs.~6d and 6e.

For the case of the conventional renormalized two-loop fermion self-energy 
$-i\Sigma^{(2)}(p,\xi)$ in QCD,
the diagrams which contribute to this function are shown in Fig.~8.
In Figs.~8b--g, the two-loop fermion self-energy has been written
in terms of renormalized one-loop two- and three-point
functions, represented by the grey blobs,
inserted in all possible ways 
in the diagram for the one-loop fermion self-energy
$-i\Sigma^{(1)}(p,\xi)$,
{\em minus}\, the contribution shown in Fig.~8f
involving the grey box inserted in the diagram for
$-i\Sigma^{(1)}(p,\xi)$.
The representation of the two-loop QCD fermion self-energy 
in Figs.~8b--g is the exact analogue
of the representation of the two-loop QED 
photon self-energy in Figs.~6b--g.
In particular, the component represented by the grey box
in Fig.~8f consists of all 
possible ways of coupling an incoming and an outgoing
quark-gluon pair (with external legs truncated) at tree level,
{\em except} for the contribution involving annihilation
into a one-particle (quark) intermediate state.
The box in Fig.~8f is thus the lowest order contribution to a 
{\em Bethe-Salpeter-type kernel}\, $V$, 
the analogue for the quark-gluon pair 
of the lowest order QED electron-positron kernel 
appearing in Fig.~6f.
The contribution in Fig.~8g represents the two-loop 
counterterms required
finally to renormalize the two-loop function.
The diagrams which contribute to the 
one-loop internal corrections
in Figs.~8b--e and the tree level Bethe-Salpeter-type 
quark-gluon kernel in Fig.~8f

\pagebreak 

\begin{center}

\begin{picture}(440,275)(0,145)

 
\ArrowLine(  0,400)( 14,400)
\GCirc( 25,400){11}{0.8}
\put( 25,401){\makebox(0,0)[c]{\Large{\bf 2}}}
\ArrowLine( 36,400)( 50,400)
\put( 25,375){\makebox(0,0)[c]{(a)}}
 
\put( 57.5,400){\makebox(0,0)[c]{$=$}}
 
\Line( 65,400)(115,400)
\GlueArc( 90,400)(15,  0,180){1.8}{8}
\GCirc( 90,400){6}{0.8}
\put( 91,401){\makebox(0,0)[c]{{\bf 1}}}
\put( 90,375){\makebox(0,0)[c]{(b)}}
 
\put(122.5,400){\makebox(0,0)[c]{$+$}}
 
\ArrowLine(130,400)(180,400)
\GlueArc(155,400)(15,  0, 90){1.8}{4}
\GlueArc(155,400)(15, 90,180){1.8}{4}
\GCirc(155,415){6}{0.8}
\put(156,416){\makebox(0,0)[c]{{\bf 1}}}
\put(155,375){\makebox(0,0)[c]{(c)}}
 
\put(187.5,400){\makebox(0,0)[c]{$+$}}
 
\ArrowLine(195,400)(245,400)
\GlueArc(220,400)(15,  0,180){1.8}{8}
\GCirc(235,400){6}{0.8}
\put(236,401){\makebox(0,0)[c]{{\bf 1}}}
\put(220,375){\makebox(0,0)[c]{(d)}}
 
\put(252.5,400){\makebox(0,0)[c]{$+$}}
 
\ArrowLine(260,400)(310,400)
\GlueArc(285,400)(15,  0,180){1.8}{8}
\GCirc(270,400){6}{0.8}
\put(271,401){\makebox(0,0)[c]{{\bf 1}}}
\put(285,375){\makebox(0,0)[c]{(e)}}
 
\put(317.5,400){\makebox(0,0)[c]{$-$}}
 
\ArrowLine(325,400)(375,400)
\GlueArc(350,400)(15,  0,180){1.8}{8}
\GBox(345,395)(355,420){0.8}
\put(351,408){\makebox(0,0)[c]{{\bf 0}}}
\put(350,375){\makebox(0,0)[c]{(f)}}
 
\put(382.5,400){\makebox(0,0)[c]{$+$}}
 
\ArrowLine(390,400)(415,400)
\ArrowLine(415,400)(440,400)
\GCirc(415,400){2.5}{0}
\put(415,375){\makebox(0,0)[c]{(g)}}


\put( 57.5,325){\makebox(0,0)[c]{$=$}}
 
\ArrowLine( 65,325)(115,325)
\GlueArc( 90,325)(15,  0,180){1.8}{8}
\GlueArc( 90,325)(8,  0,180){1.8}{4}
\put( 90,300){\makebox(0,0)[c]{(h)}}
 
\put(122.5,325){\makebox(0,0)[c]{$+$}}
 
\ArrowLine(130,325)(180,325)
\GlueArc(155,325)(15,  0, 60){1.8}{2}
\GlueArc(155,325)(15,120,180){1.8}{2}
\GlueArc(155,340)( 7,-15,195){1.8}{4}
\GlueArc(155,340)( 7,-165,-15){1.8}{2}
\put(155,300){\makebox(0,0)[c]{(i)}}
 
\put(187.5,325){\makebox(0,0)[c]{$+$}}
 
\ArrowLine(195,325)(245,325)
\GlueArc(220,325)(15,  0, 60){1.8}{2}
\GlueArc(220,325)(15,120,180){1.8}{2}
\DashArrowArc(220,340)( 7.5,-90,270){2}
\put(220,300){\makebox(0,0)[c]{(j)}}
 
\put(252.5,325){\makebox(0,0)[c]{$+$}}
 
\ArrowLine(260,325)(310,325)
\GlueArc(285,325)(15,  0, 60){1.8}{2}
\GlueArc(285,325)(15,120,180){1.8}{2}
\ArrowArc(285,340)(7.5,-90,270)
\put(285,300){\makebox(0,0)[c]{(k)}}

\put(317.5,325){\makebox(0,0)[c]{$+$}}
 
\ArrowLine(325,325)(375,325)
\GlueArc(344.5,325)(11,  0,180){1.8}{6}
\GlueArc(355.5,325)(11,180,360){1.8}{6}
\put(350,300){\makebox(0,0)[c]{(l)}}
 
\put(382.5,325){\makebox(0,0)[c]{$+$}}
 
\Line(390,325)(440,325)
\GlueArc(415,325)(15,  0, 90){1.8}{4}
\GlueArc(415,325)(15, 90,180){1.8}{4}
\Gluon(415,325)(415,340){1.8}{2}
\put(415,300){\makebox(0,0)[c]{(m)}}


\put( 57.5,250){\makebox(0,0)[c]{$+$}}
 
\Line( 65,250)(115,250)
\GlueArc( 90,250)(15,  0,180){1.8}{8}
\GCirc( 90,250){2.5}{0}
\put( 90,225){\makebox(0,0)[c]{(n)}}
 
\put(122.5,250){\makebox(0,0)[c]{$+$}}
 
\ArrowLine(130,250)(180,250)
\GlueArc(155,250)(15,  0, 85){1.8}{4}
\GlueArc(155,250)(15, 95,180){1.8}{4}
\GCirc(155,265){2.5}{0}
\put(155,225){\makebox(0,0)[c]{(o)}}
 
\put(187.5,250){\makebox(0,0)[c]{$+$}}
 
\ArrowLine(195,250)(245,250)
\GlueArc(220,250)(15,  0,180){1.8}{8}
\GCirc(235,250){2.5}{0}
\put(220,225){\makebox(0,0)[c]{(p)}}
 
\put(252.5,250){\makebox(0,0)[c]{$+$}}
 
\ArrowLine(260,250)(310,250)
\GlueArc(285,250)(15,  0,180){1.8}{8}
\GCirc(270,250){2.5}{0}
\put(285,225){\makebox(0,0)[c]{(q)}}
 
\put(317.5,250){\makebox(0,0)[c]{$+$}}
 
\ArrowLine(325,250)(350,250)
\ArrowLine(350,250)(375,250)
\GCirc(350,250){2.5}{0}
\put(350,225){\makebox(0,0)[c]{(r)}}
 

\put(0,200){\makebox(0,0)[l]{\small
Fig.\ 8. The Feynman diagrams specifying the
conventional renormalized two-loop fermion self-energy}} 
 
\put(0,185){\makebox(0,0)[l]{\small
in QCD. The grey blobs marked ``1'' represent
the renormalized conventional one-loop internal 
correc-
}}
 
\put(0,170){\makebox(0,0)[l]{\small
tions.
The grey box marked ``0''
represents the conventional
lowest order contribution to the Bethe-
}}

\put(0,155){\makebox(0,0)[l]{\small
Salpeter-type quark-gluon kernel. 
The black blobs
represent the conventional counterterm insertions. 
}}

\end{picture}
\end{center}

\noindent
are in turn shown in  
Fig.~9.\footnote{For the one-loop gluon self-energy in Fig.~9d,
the diagram involving the quadruple gauge vertex has been omitted
since it vanishes in dimensional regularization.} 
Substituting the diagrams of Fig.~9 into the representation
in Figs.~8b--g, we recover the usual two-loop perturbation theory
diagrams for the conventional
quark self-energy as shown in Figs.~8h--r.
Just as in the case of the two-loop vacuum polarization,
when the representation in Figs.~8b--g
is written in terms of individual two-loop Feynman diagrams,
the effect of the kernel insertion contribution in Fig.~8f is to
compensate for the overcounting which would otherwise occur
due to the two
vertex insertion contributions in Figs.~8d and 8e.

{}From the point of view of the PT,
the significance of the diagrammatic representations 
shown in Figs.~6b--g and 8b--g 
lies in the fact that they provide an entirely general
representation of the respective two-loop two-point
functions which is:
(i) {\em explicitly in terms of 
renormalized one-loop corrections}\, to the
tree level propagators and vertices occurring
in the corresponding one-loop two-point function,
these corrections being just those obtained in perturbation
theory at the one-loop level; and
(ii) {\em symmetric}, in the sense that the internal
corrections appear inserted in all possible ways
in the corresponding one-loop two-point function,
so that no orientation of the overall set of diagrams 
is preferred.\footnote{It should be pointed out that
the symmetric organisations shown in Figs.~6b--g 
and 8b--g of the contributions to the two-loop
photon and quark self-energies differ from
the asymmetric organisations obtained {\em directly}\, 
from the Dyson-Schwinger equation for the 
respective self-energies, expanded out to two-loop order
in terms of renormalized one-loop internal corrections
(cf.\ e.g.\ Ref.~\cite{DS}).
It is precisely the introduction of the kernel insertion
contributions which enables the self-energies
to be represented in the symmetric way 
shown in Figs.~6b--g and 8b--g.}
In the case of the two-loop \hfill
fermion \hfill self-energy \hfill in \hfill the \hfill 
PT \hfill framework, \hfill it \hfill is \hfill this \hfill
symmetric\hfill  property \hfill of \hfill the

\pagebreak

\begin{center}
\begin{picture}(430,390)(0, -95)
 

\ArrowLine(  0,280)( 24,280)
\GCirc( 35,280){11}{0.8}
\put( 35,280){\makebox(0,0)[c]{\Large{\bf 1}}}
\ArrowLine( 46,280)( 70,280)
\put( 35,250){\makebox(0,0)[c]{(a)}}
 
\put( 80,280){\makebox(0,0)[c]{$=$}}
 
\ArrowLine( 90,280)(160,280)
\GlueArc(125,280)(15,  0,180){2}{7}
\put(125,250){\makebox(0,0)[c]{(b)}}
 
\put(170,280){\makebox(0,0)[c]{$+$}}
 
\ArrowLine(180,280)(215,280)
\ArrowLine(215,280)(250,280)
\GCirc(215,280){4}{0}
\put(215,250){\makebox(0,0)[c]{(c)}}
 

\Gluon(  0,200)( 24,200){2}{3}
\GCirc( 35,200){11}{0.8}
\put( 35,200){\makebox(0,0)[c]{\Large{\bf 1}}}
\Gluon( 46,200)( 70,200){2}{3}
\put( 35,170){\makebox(0,0)[c]{(d)}}
 
\put( 80,200){\makebox(0,0)[c]{$=$}}
 
\Gluon( 90,200)(114,200){2}{3}
\GlueArc(125,200)(11,  0,180){2}{5}
\GlueArc(125,200)(11,180,360){2}{5}
\Gluon(136,200)(160,200){2}{3}
\put(125,170){\makebox(0,0)[c]{(e)}}
 
\put(170,200){\makebox(0,0)[c]{$+$}}
 
\Gluon(180,200)(204,200){2}{3}
\DashArrowArc(215,200)(11,-90,270){2}
\Gluon(226,200)(250,200){2}{3}
\put(215,170){\makebox(0,0)[c]{(f)}}
 
\put(260,200){\makebox(0,0)[c]{$+$}}
 
\Gluon(270,200)(294,200){2}{3}
\ArrowArc(305,200)(11,-90,270)
\Gluon(316,200)(340,200){2}{3}
\put(305,170){\makebox(0,0)[c]{(g)}}
 
\put(350,200){\makebox(0,0)[c]{$+$}}
 
\Gluon(360,200)(393,200){2}{4}
\Gluon(397,200)(430,200){2}{4}
\GCirc(395,200){4}{0}
\put(395,170){\makebox(0,0)[c]{(h)}}

 
\ArrowLine(  0,120)( 24,120)
\Gluon( 35,124)( 35, 85){2}{5}
\GCirc( 35,120){11}{0.8}
\put( 35,120){\makebox(0,0)[c]{\Large{\bf 1}}}
\ArrowLine( 46,120)( 70,120)
\put( 35, 70){\makebox(0,0)[c]{(i)}}
 
\put( 80,120){\makebox(0,0)[c]{$=$}}
 
\ArrowLine( 90,120)(110,120)
\Line(110,120)(140,120)
\ArrowLine(140,120)(160,120)
\GlueArc(125,120)(15,  0,180){2}{7}
\Gluon(125,120)(125, 85){2}{5}
\put(125, 70){\makebox(0,0)[c]{(j)}}
 
\put(170,120){\makebox(0,0)[c]{$+$}}
 
\ArrowLine(180,120)(250,120)
\GlueArc(215,120)(15,180,270){2}{3}
\GlueArc(215,120)(15,270,360){2}{3}
\Gluon(215,105)(215, 85){2}{3}
\put(215, 70){\makebox(0,0)[c]{(k)}}
 
\put(260,120){\makebox(0,0)[c]{$+$}}
 
\ArrowLine(270,120)(305,120)
\ArrowLine(305,120)(340,120)
\Gluon(305,118)(305, 85){2}{5}
\GCirc(305,120){4}{0}
\put(305, 70){\makebox(0,0)[c]{(l)}}
 
 
\ArrowLine(  5,-5)( 25, 0)
\ArrowLine( 45, 0)( 65,-5)
\Gluon( 65,29)( 45,24){-2}{3}
\Gluon( 25,24)(  5,29){-2}{3}
\GBox(25,-10)(45,35){0.8}
\put( 35,13){\makebox(0,0)[c]{\Large{\bf 0}}}
\put( 35,-30){\makebox(0,0)[c]{(m)}}
 
\put( 80, 12.5){\makebox(0,0)[c]{$=$}}
 
\ArrowLine( 95,-5)(125,2.5)
\ArrowLine(125,2.5)(155,-5)
\Gluon(154,30)(125,22.5){-2}{4}
\Gluon(125,22.5)( 96,30){-2}{4}
\Gluon(125,2.5)(125,22.5){2}{3}
\put(125,-30){\makebox(0,0)[c]{(n)}}
 
\put(170, 12.5){\makebox(0,0)[c]{$+$}}
 
\ArrowLine(184,-5)(246,-5)
\Gluon(187,30)(234,-5){2}{10}
\Gluon(196,-5)(209, 5){2}{2}
\Gluon(222,14)(243,30){2}{4}
\put(215,-30){\makebox(0,0)[c]{(o)}}

\put(-5,-55){\makebox(0,0)[l]{\small
Fig.\ 9. The Feynman diagrams specifying the conventional
renormalized one-loop quark self-energy }}

\put(-5,-70){\makebox(0,0)[l]{\small
(9a), gluon self-energy (9d) and quark-gluon vertex (9i), and the 
lowest order contributions to the }}

\put(-5,-85){\makebox(0,0)[l]{\small

conventional Bethe-Salpeter-type quark-gluon kernel
(9m) in QCD. The dashed lines represent ghosts.}}

\end{picture}
\end{center}

\noindent
representation which will enable the consistent solution 
of the first problem described in the introduction.

In the above examples, the renormalized
one-loop corrections and
the tree level kernels, the one-loop diagrams into which
they are inserted and the two-loop counterterms are each
the gauge-dependent functions obtained in conventional
perturbation theory. 
If the extension of the PT beyond one loop
to construct the PT 1PI two-loop fermion self-energy
$-i\hat{\Sigma}^{(2)}(p)$ in QCD 
is to be {\em consistent}, then clearly 
the one-loop corrections to the tree level
propagators and vertices appearing in the contributions to
$-i\hat{\Sigma}^{(2)}(p)$ analogous to Figs.~8b--e 
must consist of the PT renormalized 
gauge-independent one-loop corrections 
inserted in the diagram for
the PT gauge-independent one-loop fermion self-energy 
$-i\hat{\Sigma}^{(1)}(p)$.\footnote{
In the case of the vacuum polarization in QED, the
corresponding requirement is trivial to satisfy: the PT one-loop
$n$-point functions in QED coincide with those
obtained in the ordinary Feynman gauge, so that one has
simply to set $\xi = 1$ in Figs.~6b--f
(recall that the vacuum polarization in QED is gauge-independent
to all orders).}
It then remains to determine the PT tree level 
quark-gluon kernel $\hat{V}^{(0)}$
which appears in the contribution 
analogous to that in Fig.~8e; and also the 
required PT two-loop counterterm contributions, 
analogous to those

\pagebreak

\begin{center}
\begin{picture}(440,140)(0, 55)

 
\ArrowLine(  0,175)( 14,175)
\GCirc( 25,175){11}{0.8}
\put( 25,175){\makebox(0,0)[c]{\Large{\bf 2}}}
\ArrowLine( 36,175)( 50,175)
\put( 25,183){\makebox(0,0)[c]{\Huge $\hat{\phantom{X}}$}}
\put( 25,150){\makebox(0,0)[c]{(a)}}
 
\put( 57.5,175){\makebox(0,0)[c]{$=$}}

\Line( 65,175)(115,175)
\GlueArc( 90,175)(15,  0,180){1.8}{8}
\GCirc( 90,175){6}{0.8}
\put( 90,175){\makebox(0,0)[c]{{\bf 1}}}
\put( 90,178){\makebox(0,0)[c]{\huge $\hat{\phantom{X}}$}}
\put( 90,150){\makebox(0,0)[c]{(b)}}
 
\put(122.5,175){\makebox(0,0)[c]{$+$}}
 
\ArrowLine(130,175)(180,175)
\GlueArc(155,175)(15,  0, 90){1.8}{4}
\GlueArc(155,175)(15, 90,180){1.8}{4}
\GCirc(155,190){6}{0.8}
\put(155.5,190){\makebox(0,0)[c]{{\bf 1}}}
\put(155.5,193){\makebox(0,0)[c]{\huge $\hat{\phantom{X}}$}}
\put(155,150){\makebox(0,0)[c]{(c)}}
 
\put(187.5,175){\makebox(0,0)[c]{$+$}}
 
\ArrowLine(195,175)(245,175)
\GlueArc(219,175)(14, -5,180){1.8}{8}
\GCirc(235,175){6}{0.8}
\put(236,175){\makebox(0,0)[c]{{\bf 1}}}
\put(236,178){\makebox(0,0)[c]{\huge $\hat{\phantom{X}}$}}
\put(220,150){\makebox(0,0)[c]{(d)}}
 
\put(252.5,175){\makebox(0,0)[c]{$+$}}
 
\ArrowLine(260,175)(310,175)
\GlueArc(286,175)(14,  0,185){1.8}{8}
\GCirc(270,175){6}{0.8}
\put(270.5,175){\makebox(0,0)[c]{{\bf 1}}}
\put(270,178){\makebox(0,0)[c]{\huge $\hat{\phantom{X}}$}}
\put(285,150){\makebox(0,0)[c]{(e)}}
 
\put(317.5,175){\makebox(0,0)[c]{$-$}}
 
\ArrowLine(325,175)(375,175)
\GlueArc(350,175)(15,  0,180){1.8}{8}
\GBox(345,170)(355,195){0.8}
\put(351,183){\makebox(0,0)[c]{{\bf 0}}}
\put(351,193){\makebox(0,0)[c]{\huge $\hat{\phantom{X}}$}}
\put(350,150){\makebox(0,0)[c]{(f)}}
 
\put(382.5,175){\makebox(0,0)[c]{$+$}}
 
\ArrowLine(390,175)(415,175)
\ArrowLine(415,175)(440,175)
\GCirc(415,175){2.5}{0}
\put(416,175){\makebox(0,0)[c]{\huge $\hat{\phantom{X}}$}}
\put(415,150){\makebox(0,0)[c]{(g)}}


\put(0,125){\makebox(0,0)[l]{\small
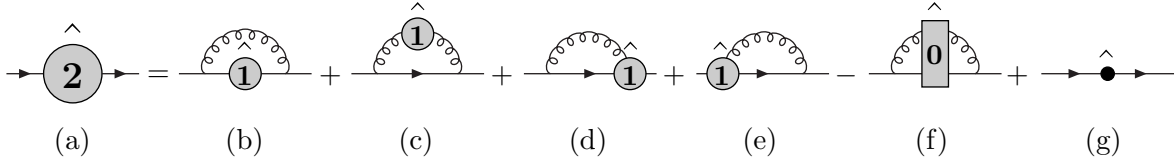
Fig.\ 10. The diagrammatic representation of the PT 
renormalized two-loop fermion self-energy in }}
 
\put(0,110){\makebox(0,0)[l]{\small
QCD. 
The grey, hatted blobs marked ``1'' 
represent the PT renormalized
one-loop internal corrections.}}
 
\put(0, 95){\makebox(0,0)[l]{\small
The grey, hatted box marked ``0'' represents the
as-yet-undetermined PT tree level quark-gluon kernel.}}

\put(0, 80){\makebox(0,0)[l]{\small
The black, hatted blob represents the
as-yet-undetermined PT two-loop
counterterm contributions.}}

\put(0, 65){\makebox(0,0)[l]{\small
The gluon propagators are all as in the Feynman gauge. }}

\end{picture}
\end{center}

\noindent
in Fig.~8f.
The contributions to $-i\hat{\Sigma}^{(2)}(p)$ are
thus as illustrated in Fig.~10, in which the hats
denote the PT $n$-point functions, 
the as-yet-undetermined PT quark-gluon kernel
and the as-yet-undetermined PT two-loop counterterms.

It is at this point that we
encounter the first problem described in the introduction,
concerning purely internal
triple gauge vertices in the PT approach.
In the PT at the one-loop level, the factors of longitudinal
four-momentum $k_{1\rho}$, $k_{2\sigma}$ originating from
the tree level interaction of an external
gluon  $A_{\mu}^{a}(q)$ and
two internal gluons $A_{\rho}^{r}(k_{1})$, $A_{\sigma}^{s}(k_{2})$
are isolated using the now-familiar decomposition of the
triple gauge vertex:
\be
\Gamma_{\mu\rho\sigma}(q,k_{1},k_{2})
=
  \Gamma_{\mu\rho\sigma}^{F}(q;k_{1},k_{2})
+ \Gamma_{\mu\rho\sigma}^{P}(q;k_{1},k_{2})\,\,,
\label{Gamma}
\ee
where
\bea
\Gamma_{\mu\rho\sigma}^{F}(q;k_{1},k_{2})
&=&
(k_{1}-k_{2})_{\mu}g_{\rho\sigma} 
-2q_{\rho}g_{\sigma\mu} + 2q_{\sigma}g_{\rho\mu}\,\,, 
\label{GammaF} \\
\Gamma_{\mu\rho\sigma}^{P}(q;k_{1},k_{2})
&=&
-k_{1\rho}g_{\sigma\mu} + k_{2\sigma}g_{\rho\mu}\,\,,
\label{GammaP}
\eea
with $q+k_{1}+k_{2} = 0$. The component $\Gamma^{F}$
by definition
contributes no factors of longitudinal internal four-momentum
$k_{1\rho}$, $k_{2\sigma}$,
and obeys a Ward identity
$q^{\mu}\Gamma_{\mu\rho\sigma}^{F}(q;k_{1},k_{2})
= (k_{2}^{2} - k_{1}^{2})g_{\rho\sigma}$
involving the difference of two inverse gluon propagators in
the Feynman gauge.
The terms in $\Gamma^{F}$ have simple interpretations:
the term $(k_{1}-k_{2})_{\mu}g_{\rho\sigma}$ is a convection
term, independent of the spin of the fields to which the
gluon $A_{\mu}^{a}(q)$ couples, while the term
$-2q_{\rho}g_{\sigma\mu} +2q_{\sigma}g_{\rho\mu}$
is a spin-1 magnetic interaction.
The vertex $\Gamma^{F}$ {\em coincides}\, with that
specifying the tree level interaction of a background gluon
$A_{\mu}^{a}(q)$ and two quantum gluons
$Q_{\rho}^{r}(k_{1})$, $Q_{\sigma}^{s}(k_{2})$
in the BFM in the Feynman quantum gauge $\xi_{Q} = 1$.
{\em It is this component $\Gamma^{F}$ which appears as the triple
gauge vertex in the PT one-loop $n$-point functions.}\footnote{
For a discussion of the role played by the vertex \eq{GammaF} 
in the decomposition of QCD amplitudes into supersymmetric and
non-supersymmetric parts, see Ref.~\cite{bernetal}.}
The component $\Gamma^{P}$ by definition
involves only the factors of longitudinal internal four-momentum
$k_{1\rho}$, $k_{2\sigma}$.
{\em It is this component $\Gamma^{P}$ which 
generates the ``pinch parts'' of the corresponding diagrams.}

The decomposition (\ref{Gamma})--(\ref{GammaP}) is represented
diagrammatically in Fig.~11.
Using the representation of Fig.~11, the diagrams contributing
to the PT one-loop quark self-energy, gluon self-energy
and quark-gluon vertex are shown in Fig.~12
(all gluon propagators in Fig.~12 are as in the Feynman gauge).

\pagebreak

\begin{center}

\begin{picture}(440,140)(-20,0)
 

\Gluon( 80,100)(110,100){3}{3}
\Gluon(110,100)(125,126){-3}{3}
\Gluon(110,100)(125, 74){-3}{3}
\put( 60,100){\makebox(0,0)[c]{$q,\mu $}}
\put(126, 60){\makebox(0,0)[c]{$k_{1},\rho $}}
\put(126,140){\makebox(0,0)[c]{$k_{2},\sigma $}}
\put(105, 40){\makebox(0,0)[c]{$\Gamma_{\mu\rho\sigma}(q,k_{1},k_{2})$}}
 
\put(155,100){\makebox(0,0)[c]{
\Large $=$}}
 
\Gluon(180,100)(210,100){3}{3}
\Gluon(210,100)(225,126){-3}{3}
\Gluon(210,100)(225, 74){-3}{3}
\Line(220,100)(230,117)
\Line(220,100)(230, 83)
\put(205, 40){\makebox(0,0)[c]{$\Gamma_{\mu\rho\sigma}^{F}(q;k_{1},k_{2})$}}
 
\put(255,100){\makebox(0,0)[c]{
\Large $+$}}
 
\Gluon(280,100)(310,100){3}{3}
\Gluon(310,100)(325,126){-3}{3}
\Gluon(310,100)(325, 74){-3}{3}
\put(320,100){\vector(2, 3){10}}
\put(320,100){\vector(2,-3){10}}
\put(305, 40){\makebox(0,0)[c]{$\Gamma_{\mu\rho\sigma}^{P}(q;k_{1},k_{2})$}}
 
\put(-20, 10){\makebox(0,0)[l]{\small
Fig.\ 11. The diagrammatic representation of the
decomposition Eq.~(\ref{Gamma}) of the triple gauge vertex. }}
 
\end{picture}
\end{center}

\begin{center}
\begin{picture}(430,315)(0,5)
 

\ArrowLine(  0,280)( 24,280)
\GCirc( 35,280){11}{0.8}
\put( 35,280){\makebox(0,0)[c]{\Large{\bf 1}}}
\put( 35,289){\makebox(0,0)[c]{\Huge{\bf $\hat{\phantom{X}}$}}}
\ArrowLine( 46,280)( 70,280)
\put( 35,250){\makebox(0,0)[c]{(a)}}
 
\put( 80,280){\makebox(0,0)[c]{$=$}}
 
\ArrowLine( 90,280)(160,280)
\GlueArc(125,280)(15,  0,180){2}{7}
\put(125,250){\makebox(0,0)[c]{(b)}}
 
\put(170,280){\makebox(0,0)[c]{$+$}}
 
\ArrowLine(180,280)(215,280)
\ArrowLine(215,280)(250,280)
\GCirc(215,280){4}{0}
\put(216,281){\makebox(0,0)[c]{\Huge{\bf $\hat{\phantom{X}}$}}}
\put(215,250){\makebox(0,0)[c]{(c)}}
 

\Gluon(  0,200)( 24,200){2}{3}
\GCirc( 35,200){11}{0.8}
\put( 35,200){\makebox(0,0)[c]{\Large{\bf 1}}}
\put( 35,209){\makebox(0,0)[c]{\Huge{\bf $\hat{\phantom{X}}$}}}
\Gluon( 46,200)( 70,200){2}{3}
\put( 35,170){\makebox(0,0)[c]{(d)}}
 
\put( 80,200){\makebox(0,0)[c]{$=$}}
 
\Gluon( 90,200)(114,200){2}{3}
\GlueArc(125,200)(11,  0,180){2}{5}
\GlueArc(125,200)(11,180,360){2}{5}
\Gluon(136,200)(160,200){2}{3}
\Line(117,200)(123,206)
\Line(117,200)(123,194)
\Line(133,200)(127,206)
\Line(133,200)(127,194)
\put(125,170){\makebox(0,0)[c]{(e)}}
 
\put(170,200){\makebox(0,0)[c]{$+$}}
 
\Gluon(180,200)(204,200){2}{3}
\DashArrowArc(215,200)(11,-90,270){2}
\Gluon(226,200)(250,200){2}{3}
\put(215,170){\makebox(0,0)[c]{(f)}}
 
\put(260,200){\makebox(0,0)[c]{$+$}}
 
\Gluon(270,200)(294,200){2}{3}
\ArrowArc(305,200)(11,-90,270)
\Gluon(316,200)(340,200){2}{3}
\put(305,170){\makebox(0,0)[c]{(g)}}
 
\put(350,200){\makebox(0,0)[c]{$+$}}
 
\Gluon(360,200)(393,200){2}{4}
\Gluon(397,200)(430,200){2}{4}
\GCirc(395,200){4}{0}
\put(396,201){\makebox(0,0)[c]{\Huge{\bf $\hat{\phantom{X}}$}}}
\put(395,170){\makebox(0,0)[c]{(h)}}

 
\ArrowLine(  0,120)( 24,120)
\Gluon( 35,124)( 35, 85){2}{5}
\GCirc( 35,120){11}{0.8}
\put( 35,120){\makebox(0,0)[c]{\Large{\bf 1}}}
\put( 35,129){\makebox(0,0)[c]{\Huge{\bf $\hat{\phantom{X}}$}}}
\ArrowLine( 46,120)( 70,120)
\put( 35, 70){\makebox(0,0)[c]{(i)}}
 
\put( 80,120){\makebox(0,0)[c]{$=$}}
 
\ArrowLine( 90,120)(110,120)
\Line(110,120)(140,120)
\ArrowLine(140,120)(160,120)
\GlueArc(125,120)(15,  0,180){2}{7}
\Gluon(125,120)(125, 85){2}{5}
\put(125, 70){\makebox(0,0)[c]{(j)}}
 
\put(170,120){\makebox(0,0)[c]{$+$}}
 
\ArrowLine(180,120)(250,120)
\Gluon(215,101)(196,120){2}{4}
\Gluon(215,101)(234,120){-2}{4}
\Gluon(215,101)(215, 85){2}{2}
\Line(215,109)(223,116)
\Line(215,109)(208,116)
\put(215, 70){\makebox(0,0)[c]{(k)}}
 
\put(260,120){\makebox(0,0)[c]{$+$}}
 
\ArrowLine(270,120)(305,120)
\ArrowLine(305,120)(340,120)
\Gluon(305,118)(305, 85){2}{5}
\GCirc(305,120){4}{0}
\put(306,121){\makebox(0,0)[c]{\Huge{\bf $\hat{\phantom{X}}$}}}
\put(305, 70){\makebox(0,0)[c]{(l)}}

\put(-5, 45){\makebox(0,0)[l]{\small
Fig.\ 12. The Feynman diagrams specifying the PT
renormalized one-loop quark self-energy (12a), gluon}}

\put(-5, 30){\makebox(0,0)[l]{\small
self-energy (12d) and quark-gluon vertex (12i). 
The gluon propagators are all as in the
Feynman gauge.}}

\put(-5, 15){\makebox(0,0)[l]{\small
These Feynman diagrams coincide with those of the BFM
in the Feynman quantum gauge $\xi_{Q} = 1$. }}

\end{picture}
\end{center}

In the case of the PT two-loop quark self-energy 
$-i\hat{\Sigma}^{(2)}(p)$,
the first problem of section~1 
therefore becomes that of how to decompose the 
triple gauge vertices involved
so as to obtain, first of all, 
and starting from an arbitrary gauge,
the {\em internal}\, one-loop pinch parts
required to construct
the PT gauge-independent one-loop self-energy and vertex
functions of Fig.~12 as the internal corrections
shown in Figs.~10b--e.
The key to the required rearrangement will turn out to be
a second, new decomposition of the triple gauge vertex:
\bea
\Gamma_{\rho\sigma\tau}(k_{1},k_{2},k_{3})
&\equiv&
- 2\Gamma_{\tau\rho\sigma}^{P}(k_{3};k_{1},k_{2}) \nn \\
& &
+\, \Gamma_{\rho\sigma\tau}^{F}(k_{1};k_{2},k_{3})
+ \Gamma_{\sigma\tau\rho}^{F}(k_{2};k_{3},k_{1}) 
- \Gamma_{\tau\rho\sigma}^{F}(k_{3};k_{1},k_{2}) \,\,.
\label{GammaFFFP}
\eea
Using the definitions (\ref{GammaF}) and (\ref{GammaP}), it is easily
verified that the above identity indeed holds.
This \hfill identity \hfill is \hfill to \hfill be \hfill 
compared \hfill with \hfill the \hfill trivial \hfill identity \hfill  
relating \hfill the \hfill diagram \hfill in \hfill Fig.~8m \hfill to

\pagebreak

\begin{center}

\begin{picture}(440,125)(-95, -80)


\put(-20, 43){\makebox(0,0)[c]{\small $k_{1},\rho,r $}}
\put( 50, 43){\makebox(0,0)[c]{\small $k_{2},\sigma,s $}}
\put(-20,-15){\makebox(0,0)[c]{\small $p-k_{1}$}}
\put( 50,-15){\makebox(0,0)[c]{\small $p+k_{2}$}}
\ArrowLine(-15,-5)(  5, 0)
\ArrowLine( 25, 0)( 45,-5)
\Gluon( 45,29)( 25,24){-2}{3}
\Gluon(  5,24)(-15,29){-2}{3}
\GBox( 5,-10)(25,35){0.8}
\put( 15,33){\makebox(0,0)[c]{\Huge $\hat{\phantom{X}}$}}
\put( 15,13){\makebox(0,0)[c]{\Large{\bf 0}}}
\put( 15,-30){\makebox(0,0)[c]{(a)}}

\put( 80, 12.5){\makebox(0,0)[c]{$=$}}

\ArrowLine( 95,-5)(125,2.5)
\ArrowLine(125,2.5)(155,-5)
\Gluon(154,30)(125,22.5){-2}{4}
\Gluon(125,22.5)( 96,30){-2}{4}
\Gluon(125,2.5)(125,22.5){2}{3}
\Line(110,33.8)(125,30)
\Line(125,30)(140,33.8)
\put(125,-30){\makebox(0,0)[c]{(b)}}
 
\put(170, 12.5){\makebox(0,0)[c]{$+$}}
 
\ArrowLine(184,-5)(246,-5)
\Gluon(187,30)(234,-5){2}{10}
\Gluon(196,-5)(209, 5){2}{2}
\Gluon(222,14)(243,30){2}{4}
\put(215,-30){\makebox(0,0)[c]{(c)}}
 
\put(-95,-55){\makebox(0,0)[l]{\small
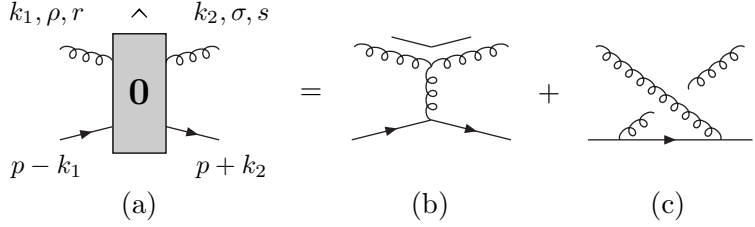
Fig.\ 13. The Feynman diagrams specifying the
contributions to the PT tree level Bethe-Salpeter- }}

\put(-95,-70){\makebox(0,0)[l]{\small
type quark-gluon kernel. The gluon propagator
in (b) is as in the Feynman gauge.}}

\end{picture}
\end{center}

\noindent
the triple gauge vertex contributions
to the one-loop vertex corrections in Figs.~8d and 8e
and the kernel in Fig.~8f
for the case of the conventional two-loop quark self-energy:
\be
\Gamma_{\rho\sigma\tau}(k_{1},k_{2},k_{3})
\,\,\equiv\,\,
\Gamma_{\rho\sigma\tau}(k_{1},k_{2},k_{3})
+ \Gamma_{\sigma\tau\rho}(k_{2},k_{3},k_{1}) 
- \Gamma_{\tau\rho\sigma}(k_{3},k_{1},k_{2}) \,\,.
\label{Gammatriv}
\ee

In the context of the process $q\gamma \rightarrow q\gamma$,
the construction of the PT renormalized
two-loop quark self-energy $-i\hat{\Sigma}^{(2)}(p)$ 
may thus be broken down into three steps:

\begin{enumerate}

\item
The rearrangement of the integrands for the
conventional two-loop QCD corrections to 
$q\gamma \rightarrow q\gamma$
to obtain as self-energy-like (``effective'' two-point)
components the integrands for the contributions to 
$-i\hat{\Sigma}^{(2)}(p)$ 
involving the PT one-loop $n$-point functions
as internal corrections shown in Figs.~10b--e.

\item
The isolation of the remaining self-energy-like
component of the two-loop corrections to
$q\gamma \rightarrow q\gamma$
in order to determine the PT tree level quark-gluon kernel 
$\hat{V}^{(0)}$ appearing in the contribution to
$-i\hat{\Sigma}^{(2)}(p)$ shown in Fig.~10f.

\item
The calculation of the ultraviolet-divergent part of the 
contributions in Figs.~10b--f in order to determine the
PT $\cO(\alpha_{s}^{2})$ renormalization constants 
$(Z_{2}-1)_{\rm PT}^{(2)}$ and $(Z_{m}-1)_{\rm PT}^{(2)}$
required finally to renormalize 
$-i\hat{\Sigma}^{(2)}(p)$, and shown in Fig.~10g.

\end{enumerate}

\noindent
We see that, assuming the first step can be successfully made,
the construction of the PT two-loop quark self-energy
reduces to obtaining the PT tree level quark-gluon
kernel and the required PT two-loop counterterms.

At this point, it is important to remark that it is
by no means clear that,
having made the first of the above steps,
the second and third steps can be 
consistently carried out too.
Thus, in the second step, 
it is not clear that the resulting PT kernel 
$\hat{V}^{(0)}$ appearing in the diagram 10f will,
for example, indeed have no
contributions involving annihilation into a 
one-particle (quark) intermediate state.
And in the third step, it is not clear that
the rearrangement of the two-loop contributions
required in the first two steps
will result in no divergences of the form, for example,
$\epsilon^{-1}\ln(-p^{2}/\mu^{2})$ 
which then cannot be renormalized by local counterterms.

In the remainder of this paper, it is shown how
the construction outlined above can indeed be 
consistently carried out to obtain the two-loop quark self-energy 
$-i\hat{\Sigma}^{(2)}(p)$ in the PT approach.
In particular, it is demonstrated that the 
contributions to the PT tree level quark-gluon kernel 
$\hat{V}^{(0)}$ are as shown in Fig.~13,
i.e.\ exactly those of the conventional 
Feynman gauge quark-gluon kernel except that the
triple gauge vertex involves just the component
$\Gamma^{F}$. Thus, when inserted in some diagram, 
e.g.\ as in Fig.~10f, {\em the PT kernel
provides no factors of longitudinal gluon four-momentum
to trigger the Ward identities for the adjacent vertices}.
It is explicitly shown that the PT self-energy $-i\hat{\Sigma}^{(2)}(p)$ 
thus obtained is gauge-independent at all momenta, is
multiplicatively renormalizable by local counterterms
and does not shift the propagator pole position
(to $\cO(\alpha_{s}^{2})$). 
Furthermore, it is shown that 
the PT two-loop quark self-energy 
{\em differs} from that obtained in the BFM at $\xi_{Q} = 1$.

 
\setcounter{equation}{0}

\section{Construction of $-i\hat{\Sigma}^{(2)}(p)$}


In this section, we consider the first two steps in the construction of  
$-i\hat{\Sigma}^{(2)}(p)$ just outlined. 
To this end, we consider the $\cO(\alpha_{s}^{2})$ 
corrections to $q\gamma\rightarrow q\gamma$ consisting of 
``genuine'' two-loop diagrams, i.e.\ involving two $d$-dimensional
loop momentum integrals. In order to make the required
rearrangements, it will be convenient to divide these 
two-loop diagrams into three distinct classes.
Furthermore, for simplicity, we choose here to start from
the Feynman gauge $\xi = 1$; the generalization  of the
construction to arbitrary $\xi$ will be presented in section 5.
We note immediately that in the Feynman gauge
the conventional self-energy diagrams
8b and 8h provide directly the required contribution
to $-i\hat{\Sigma}^{(2)}(p)$ of the diagram 10b,
involving the renormalized PT one-loop quark self-energy 
as an internal correction to, in turn, the PT one-loop
quark self-energy.

\subsection{Two-loop corrections involving one-loop 
gluon self-energy insertions}

We first consider the two-loop QCD corrections to the
process $q\gamma\rightarrow q\gamma$ which consist of one-loop
gluon self-energy insertions in the one-loop QCD corrections.
The conventional renormalized
one-loop covariant gauge gluon self-energy
is specified by the diagrams shown in Figs.~9d--h:
\be
{\rm Fig.\,9d}
=
i\Pi_{\mu\nu}^{(1)}(q,\xi)
=
i\int[dk]\,\Pi_{\mu\nu}^{\prime(1)}(q,k,\xi)
\,-\,
(Z_{3}-1)_{\xi}^{(1)}iq^{2}t_{\mu\nu}(q) \,\,.
\label{Pixi}
\ee
In the above expression,
$\Pi_{\mu\nu}^{\prime(1)}(q,k,\xi)$ is the gauge-dependent
integrand, given directly by the usual covariant gauge Feynman rules
for the diagrams 9e--g, while the last term is the
$\cO(\alpha_{s})$
contribution of the covariant gauge
counterterm diagram 9h, where $Z_{3}$ is the 
gluon wavefunction renormalization constant
($A_{0\mu}^{a} = Z_{3}^{1/2}A_{\mu}^{a}$).
The integrand in \eq{Pixi} 
may be written as the difference between the integrand
$\hat{\Pi}_{\mu\nu}^{\prime(1)}(q,k)$
for the PT gauge-independent one-loop
gluon self-energy and the integrand 
$\Delta\Pi_{\mu\nu}^{\prime(1)}(q,k,\xi)$
for the gauge-dependent ``pinch part'' of the conventional 
self-energy:
\be
\Pi_{\mu\nu}^{\prime(1)}(q,k,\xi) 
=
 \hat{\Pi}_{\mu\nu}^{\prime(1)}(q,k) 
- \Delta\Pi_{\mu\nu}^{\prime(1)}(q,k,\xi)\,\,.
\label{Pidecomp}
\ee
The pinch part $-\Delta\Pi_{\mu\nu}^{\prime(1)}(q,k,\xi)$
of the conventional self-energy is that which
is cancelled by the gauge-dependent
self-energy-like pinch part
$+\Delta\Pi_{\mu\nu}^{\prime(1)}(q,k,\xi)$
of the integrands for one-loop vertex and box diagrams
in the PT at the one-loop level. This cancellation leaves 
$\hat{\Pi}_{\mu\nu}^{\prime(1)}(q,k)$
as the integrand for the unrenormalized
PT gauge-independent one-loop gluon
``effective'' two-point function. The PT integrand
$\hat{\Pi}_{\mu\nu}^{\prime(1)}(q,k)$
coincides \cite{BFMPT} 
with that obtained using the Feynman rules
of the BFM \cite{abbott} with $\xi_{Q} = 1$.

In the particular case of the Feynman gauge $\xi = 1$, 
the pinch part of the self-energy integrand (\ref{Pixi}) 
is given by\footnote{In \eq{Piprimemunupinch0},
the conventional covariant gauge ghost term has been symmetrized.
In \eq{Piprimemunupinch1},
the dimensional regularization rule $\int[dk]\,k^{-2} = 0$
has been used to drop terms which vanish upon integration.}
\bea
i\Delta\Pi_{\mu\nu}^{\prime(1)}(q,k,1)
&=&
\frac{1}{2}C_{\!A}g^{2}\frac{1}{k^{2}(k+q)^{2}}\biggl\{ 
-\Gamma_{\mu\rho\sigma}^{F}(q;k,-k-q)
\Gamma_{\nu}^{F,\rho\sigma}(-q;-k,k+q) \nn \\
& &
-\, 2(2k+q)_{\mu}(2k+q)_{\nu} 
\,+\,\Gamma_{\mu\rho\sigma}(q,k,-k-q)
\Gamma_{\nu}^{\phantom{\nu}\rho\sigma}(-q,-k,k+q) \nn \\
& &
\,+\, k_{\mu}(k+q)_{\nu} 
\,+\, (k+q)_{\mu}k_{\nu} \biggr\} 
\label{Piprimemunupinch0} \\
&=&
iq^{2}t_{\mu\nu}(q)\Delta\Pi^{\prime(1)}(q,k,1)
\label{Piprimemunupinch1}
\eea
where
\be
\Delta\Pi^{\prime(1)}(q,k,1)
=
-2iC_{\!A}g^{2}\frac{1}{k^{2}(k+q)^{2}}\,\,.
\label{Piprimepinch1}
\ee
In the above expressions,
$C_{\!A}$ is the quadratic Casimir coefficient for the
adjoint representation ($C_{\!A} = N$ for SU($N$)).
We note that the integrand $\Delta\Pi_{\mu\nu}^{\prime(1)}(q,k,1)$
is explicitly transverse.

The set of two-loop diagrams for the 
$\cO(\alpha_{s}^{2})$ QCD corrections to
$q\gamma \rightarrow q\gamma$
which involve the {\em unrenormalized}\, 
one-loop gluon self-energy as
internal corrections are shown in Figs.~14a--f
(the corresponding set of diagrams with
crossed photon legs are not shown).
Just as in the case of the one-loop QCD corrections to
$q\gamma \rightarrow q\gamma$ discussed in section 2,
and as indicated in Fig.~14,
the sum of the four diagrams 14a--d may be expressed as the
sum of all possible contractions of the diagrams 14g and 14h with
the diagrams 14i and 14j,
where the curly line with the grey blob between these diagrams
represents the unrenormalized
one-loop-corrected covariant gauge gluon propagator.
The sums of the diagrams 14g+14h and 14i+14j are again each
just the connected four-point function defined in \eq{G4}.
Thus, the sum of the four two-loop diagrams 14a--d 
for arbitrary $\xi$ may be written
\bea
{\rm Figs.\,14a\!-\!d}
&=&
-Q^{2}g^{2}\int[dk_{1}][dk_{2}]\,
\frac{1}{k_{1}^{4}}
\Pi^{\prime(1)\rho\rho'}(k_{1},k_{2},\xi) 
\nn \\
& &
\times 
G_{\rho\mu}^{r}(k_{1},-p_{3},p-k_{1},p_{4})
S(p-k_{1})
G_{\rho'\!\nu}^{r}(-k_{1},p_{1},p_{2},p-k_{1})\,\,.
\label{fig14ad}
\eea
\eq{fig14ad} is precisely the two-loop analogue of \eq{fig3ad}; the two
expressions differ only by the fact that, instead of the tree level
gluon propagator in (\ref{fig3ad}), there appears the one-loop-corrected
gluon propagator in (\ref{fig14ad}).

Substituting in (\ref{fig14ad}) 
the decomposition (\ref{Pidecomp}) of the gluon self-energy integrand,
and then using the expression
(\ref{Piprimemunupinch1}) 
for the case of the Feynman gauge gives
\bea
{\rm Figs.\,14a\!-\!d}|_{\xi = 1}
&=&
-Q^{2}g^{2}\int[dk_{1}][dk_{2}]\,
\frac{1}{k_{1}^{4}}
\Bigl(  \hat{\Pi}^{\prime(1)\rho\rho'}(k_{1},k_{2})
-k_{1}^{2}t^{\rho\rho'\!}(k_{1}) 
\Delta\Pi^{\prime(1)}(k_{1},k_{2},1)  \Bigr)
\nn\\
& &
\times 
G_{\rho\mu}^{r}(k_{1},-p_{3},p-k_{1},p_{4})
S(p-k_{1})
G_{\rho'\!\nu}^{r}(-k_{1},p_{1},p_{2},p-k_{1}) \,\,.
\label{fig14ad2}
\eea

\pagebreak

\begin{center}
\begin{picture}(430,480)(0,-50)


\ArrowLine(  0,430)(  0,400)
\ArrowLine(  0,400)( 60,400)
\ArrowLine( 60,400)( 60,430)
\GlueArc( 30,400)(15,  0,180){2}{6}
\GCirc( 30,415){5}{0.7}
\Photon( 0,380)( 0,400){-2}{3}
\Photon(60,380)(60,400){2}{3}
\put( 30,370){\makebox(0,0)[c]{(a)}}

\Line(  0,355)(  0,325)
\ArrowLine(  0,325)( 60,325)
\ArrowLine( 60,325)( 60,355)
\GlueArc(  0,325)(17,  0, 90){2}{5}
\GCirc( 12,337){5}{0.7}
\Photon( 0,305)( 0,325){-2}{3}
\Photon(60,305)(60,325){2}{3}
\put( 30,295){\makebox(0,0)[c]{(b)}}

\ArrowLine(  0,280)(  0,250)
\ArrowLine(  0,250)( 60,250)
\Line( 60,250)( 60,280)
\GlueArc( 60,250)(17, 90,180){2}{5}
\GCirc( 48,262){5}{0.7}
\Photon( 0,230)( 0,250){-2}{3}
\Photon(60,230)(60,250){2}{3}
\put( 30,220){\makebox(0,0)[c]{(c)}}
 
\Line(  0,205)(  0,175)
\ArrowLine(  0,175)( 60,175)
\Line( 60,175)( 60,205)
\Gluon(0,190)(60,190){2}{10}
\GCirc( 30,190){5}{0.7}
\Photon( 0,155)( 0,175){-2}{3}
\Photon(60,155)(60,175){2}{3}
\put( 30,145){\makebox(0,0)[c]{(d)}}

\ArrowLine(  0,130)(  0, 90)
\ArrowLine(  0, 90)( 60, 90)
\ArrowLine( 60, 90)( 60,130)
\GlueArc(  0,110)(10,-90, 90){2}{4}
\GCirc( 11,110){5}{0.7}
\Photon( 0, 80)( 0, 90){2}{1.5}
\Photon(60, 80)(60, 90){-2}{1.5}
\put( 30, 70){\makebox(0,0)[c]{(e)}}

\ArrowLine(  0, 55)(  0, 15)
\ArrowLine(  0, 15)( 60, 15)
\ArrowLine( 60, 15)( 60, 55)
\GlueArc( 60,35)(10, 90,270){2}{4}
\GCirc( 49, 35){5}{0.7}
\Photon( 0,  5)( 0, 15){2}{1.5}
\Photon(60,  5)(60, 15){-2}{1.5}
\put( 30, -5){\makebox(0,0)[c]{(f)}}
 

\put( 80,292){\makebox(0,0)[c]
{$ \left. \begin{array}{c}  \\ \\ \\ \\ \\ \\ \\ \\ \\ 
\\ \\ \\ \\ \\ \\ \\ \\ \\ \\
\end{array} \right\}$}}
 
\put( 120,292){\makebox(0,0)[c]{
\Large $=$}}
 
\put(160,292){\makebox(0,0)[c]
{$ \left\{ \begin{array}{c}  \\ \\ \\ \\ \\ \\ \\ \\ \\ 
\end{array} \right.$}}
 
 
\ArrowLine(180,355)(180,325)
\Line(180,325)(210,325)
\GlueArc(210,325)(17, 90, 180){2}{4}
\Photon(180,305)(180,325){-2}{3}
\put(195,295){\makebox(0,0)[c]{(g)}}
 
\Line(180,280)(180,250)
\ArrowLine(180,250)(210,250)
\Gluon(180,265)(210,265){2}{4}
\Photon(180,230)(180,250){-2}{3}
\put(195,220){\makebox(0,0)[c]{(h)}}
 
 
\put(230,292){\makebox(0,0)[c]
{$ \left. \begin{array}{c}  \\ \\ \\ \\ \\ \\ \\ \\ \\ 
\end{array} \right\}$}}
 
\put(305,325){\makebox(0,0)[c]{$k_{1}$}}
\put(300,315){\vector(1, 0){10}}
 
\Gluon(285,300)(325,300){2}{6}
\GCirc(305,300){5}{0.7}
\ArrowLine(287,283)(323,283)
 
\put(305,270){\makebox(0,0)[c]{$p-k_{1}$}}
 
\put(380,292){\makebox(0,0)[c]
{$ \left\{ \begin{array}{c}  \\ \\ \\ \\ \\ \\ \\ \\ \\ 
\end{array} \right.$}}
 

\Line(400,325)(430,325)
\ArrowLine(430,325)(430,355)
\GlueArc(400,325)(17,  0, 90){2}{4}
\Photon(430,305)(430,325){2}{3}
\put(415,295){\makebox(0,0)[c]{(i)}}
 
\ArrowLine(400,250)(430,250)
\Line(430,250)(430,280)
\Gluon(400,265)(430,265){2}{4}
\Photon(430,230)(430,250){2}{3}
\put(415,220){\makebox(0,0)[c]{(j)}}

\put(-5,-25){\makebox(0,0)[l]{\small
Fig.\ 14. a--f: The two-loop QCD corrections to 
the process $q\gamma \rightarrow q\gamma$
consisting of unrenormalized one- }}
 
\put(-5,-40){\makebox(0,0)[l]{\small
loop gluon self-energy insertions, represented by the grey blobs, 
in the one-loop QCD corrections.}}

\end{picture}
\end{center}

We now contract the longitudinal factors
$k_{1}^{\rho}$, $k_{1}^{\rho'}$ which appear explicitly in the
transverse tensor $t^{\rho\rho'\!}(k_{1})$ 
in (\ref{fig14ad2}),
triggering the Ward identity (\ref{G4wid})
obeyed by the four-point functions
$G_{\rho\mu}^{r}$,
$G_{\rho'\!\nu}^{r}$.
Using this Ward identity, together with the fact that
$S^{-1}(p_{2}) = S^{-1}(p_{4}) = 0$ for the on-shell external fermions,
one obtains
\bea
{\rm Figs.\,14a\!-\!d}|_{\xi =  1}
&=&
-Q^{2}g^{2}\int[dk_{1}][dk_{2}]\,
\frac{1}{k_{1}^{4}} \biggl\{
\Bigl(  \hat{\Pi}^{\prime(1)\rho\rho'\!}(k_{1},k_{2})
-k_{1}^{2}g^{\rho\rho'\!}\Delta\Pi^{\prime(1)}(k_{1},k_{2},1)  \Bigr)
\nn\\
& &
\times
G_{\rho\mu}^{r}(k_{1},-p_{3},p-k_{1},p_{4})
S(p-k_{1})
G_{\rho'\!\nu}^{r}(-k_{1},p_{1},p_{2},p-k_{1}) \nn \\
& &
+\,C_{F}\Delta\Pi^{\prime(1)}(k_{1},k_{2},1)
\gamma_{\mu}S(p) \gamma_{\nu} 
\biggr\}\,\,.
\label{fig14ad3}
\eea

There remain the two external leg corrections shown in
Figs.~14e and 14f. 
The sum of these two diagrams for arbitrary $\xi$ is given by
\bea
\lefteqn{ {\rm Figs.\,14e + 14f}
\,\,=\,\,
-\frac{1}{2}Q^{2}g^{2}
\int[dk_{1}][dk_{2}]\,
\frac{1}{k_{1}^{4}}
\Pi^{\prime(1)\rho\rho'\!}(k_{1},k_{2},\xi)  } \nn\\
& &
\times \Bigl(
\gamma_{\mu}S(p)\gamma_{\nu} S(p_{2})
\gamma_{\rho}T^{r}S(p_{2}-k_{1})\gamma_{\rho'\!}T^{r} 
\,+\,
\gamma_{\rho}T^{r}S(p_{4}-k_{1})\gamma_{\rho'\!}T^{r}S(p_{4})
\gamma_{\mu}S(p)\gamma_{\nu} \Bigr) \,\,.\qquad\qquad
\label{fig14ef}
\eea
We again substitute the decomposition (\ref{Pidecomp})
for the gluon self-energy integrand,
and then the expression
(\ref{Piprimemunupinch1}) 
for the particular case of the Feynman gauge.
In an exactly similar way to the
analogous one-loop corrections in Figs.~2e and 2f,
the longitudinal factors $k_{1}^{\rho}$, $k_{1}^{\rho'}$
which appear in the transverse tensor $t^{\rho\rho'\!}(k_{1})$
for the pinch part of the gluon self-energy in (\ref{fig14ef})
trigger the elementary Ward identity (\ref{wid}), giving
\bea
\lefteqn{ {\rm Figs.\,14e + 14f}|_{\xi = 1}
\,\,=\,\,
-Q^{2}g^{2}
\int[dk_{1}][dk_{2}]\,
\frac{1}{k_{1}^{4}} \biggl\{
\Bigl(  \hat{\Pi}^{\prime(1)\rho\rho'\!}(k_{1},k_{2})
-k_{1}^{2}g^{\rho\rho'\!}\Delta\Pi^{\prime(1)}(k_{1},k_{2},1) 
\Bigr) } \nn\\
& &
\times \frac{1}{2}C_{F}\Bigl(
\gamma_{\mu}S(p)\gamma_{\nu} S(p_{2})
\gamma_{\rho}S(p_{2}-k_{1})\gamma_{\rho'\!} 
\,+\,
\gamma_{\rho}S(p_{4}-k_{1})\gamma_{\rho'\!}S(p_{4})
\gamma_{\mu}S(p)\gamma_{\nu} \Bigr) \nn \qquad\qquad \\
& &
-\,C_{F}\Delta\Pi^{\prime(1)}(k_{1},k_{2},1)\,
\gamma_{\mu}S(p)\gamma_{\nu}
\biggr\}\,\,.
\label{fig14ef2}
\eea

Combining the expressions (\ref{fig14ad3}) and (\ref{fig14ef2}),
the last terms cancel, giving
\bea
\lefteqn{ {\rm Figs.\,14a\!-\!f}|_{\xi =  1}
\,\,=\,\,
-Q^{2}g^{2}\int[dk_{1}][dk_{2}]\,
\frac{1}{k_{1}^{4}}
\Bigl(  \hat{\Pi}^{\prime(1)\rho\rho'\!}(k_{1},k_{2})
-k_{1}^{2}g^{\rho\rho'\!}\Delta\Pi^{\prime(1)}(k_{1},k_{2},1)  
\Bigr) } \nn\\
& &
\times\biggl\{
G_{\rho\mu}^{r}(k_{1},-p_{3},p-k_{1},p_{4})
S(p-k_{1})
G_{\rho'\!\nu}^{r}(-k_{1},p_{1},p_{2},p-k_{1}) \nn \\
& &
+ \,\frac{1}{2}C_{F}\Bigl(
\gamma_{\mu}S(p)\gamma_{\nu} S(p_{2})
\gamma_{\rho}S(p_{2}-k_{1})\gamma_{\rho'\!} 
\,+\,
\gamma_{\rho}S(p_{4}-k_{1})\gamma_{\rho'\!}S(p_{4})
\gamma_{\mu}S(p)\gamma_{\nu} \Bigr) \biggr\}\,\,.
\label{fig14af}
\eea
Having cancelled among the integrands in
Eqs.~(\ref{fig14ad}) and (\ref{fig14ef}) 
the factors of longitudinal
four-momentum $k_{1}^{\rho}$, $k_{1}^{\rho'}$
occurring in the pinch part
$-k_{1}^{2}t^{\rho\rho'\!}(k_{1})\Delta\Pi^{\prime(1)}(k_{1},k_{2},1)$
of the integrand for the one-loop gluon self-energy insertions,
the PT two-loop ``effective'' two-point component of the
diagrams 14a--f may now be identified from the 
fermion propagator
structure of the integrands in the resulting
expression (\ref{fig14af}):
\bea
{\rm Figs.\,14a\!-\!f}|_{\xi = 1,2-{\rm pt}}
&=&
-C_{F}Q^{2}g^{2}\int[dk_{1}][dk_{2}]\,
\frac{1}{k_{1}^{4}}
\Bigl(  \hat{\Pi}_{\rho\rho'\!}^{\prime(1)}(k_{1},k_{2})
-k_{1}^{2}g_{\rho\rho'\!}\Delta\Pi^{\prime(1)}(k_{1},k_{2},1) \Bigr)
\nn\\
& & \times
\gamma_{\mu}S(p)\gamma^{\rho}S(p-k_{1})\gamma^{\rho'\!}S(p)\gamma_{\nu}
\,\,. \label{fig14af2pt}
\eea
The first term in \eq{fig14af2pt} is the
two-loop integrand required for the contribution to 
$-i\hat{\Sigma}^{(2)}(p)$ of the diagram 10c,
involving the PT one-loop gluon self-energy insertion in the PT
one-loop fermion self-energy, here embedded in the process
$q\gamma \rightarrow q\gamma$.
The second term in \eq{fig14af2pt} is the $g_{\rho\rho'}$
component of the pinch part $-\Delta\Pi_{\rho\rho'}^{\prime(1)}$
of the conventional $\xi = 1$ self-energy,
the $k_{1\rho}k_{1\rho'}$ component having exactly cancelled
among the diagrams 14a--f as just described.
In the next subsection, it will be shown how this second term
is exactly cancelled by pinch contributions from the
two-loop diagrams involving a single triple gauge vertex.

\subsection{Two-loop corrections involving one triple gauge vertex}

We next consider the $\cO(\alpha_{s}^{2})$ QCD corrections to
$q\gamma \rightarrow q\gamma$ consisting of two-loop 
diagrams involving a single triple gauge vertex. 
Eight of the set of 
ten such corrections are shown in \mbox{Figs.~15a--h} 
(the remaining two are external leg corrections which, in the
Feynman gauge, make no contribution to the ``effective'' two-point
component of the interaction;
as before, there is also a corresponding set of diagrams
with crossed external photon legs).
As indicated in Fig.~15, the sums of the diagrams
15a--d, 15e+15f and 15g+15h may be written in terms of the
tree level four-point function defined in \eq{G4}.
It is for these two-loop diagrams that we shall use the
decomposition of the triple gauge vertex
given in \eq{GammaFFFP}.

We consider first the four diagrams 15a--d. In the Feynman gauge,
the sum of these four diagrams may be written
\bea
\lefteqn{ {\rm Figs.\,15a\!-\!d}|_{\xi = 1}
\,\,=\,\,
-Q^{2}g^{4}\int[dk_{1}][dk_{2}]\,
\frac{1}{k_{1}^{2}k_{2}^{2}k_{3}^{2}}\,
f^{rst}\Gamma^{\rho\sigma\tau}(k_{1},k_{2},k_{3}) } \nn \\
& &
\times 
G_{\sigma\mu}^{s}(-k_{2},-p_{3},p+k_{2},p_{4})
S(p+k_{2})\gamma_{\tau}T^{t}S(p-k_{1})
G_{\rho\nu}^{r}(-k_{1},p_{1},p_{2},p-k_{1}) \,\,,
\label{fig15ad}
\eea
where $\Gamma^{\rho\sigma\tau}(k_{1},k_{2},k_{3})$ is the
triple gauge vertex appearing in Fig.~15k, with
$k_{1} + k_{2} + k_{3} = 0$.
The decomposition (\ref{GammaFFFP}) substituted
directly into the expression (\ref{fig15ad}) for Figs.~15a--d
is represented diagrammatically in Fig.~16.
In particular, the four terms on the right hand side of 
\eq{GammaFFFP} are shown
in Figs.~16c--f, respectively. We now consider
the contraction of the diagrams 16a+16b and 16g+16h with
each of the four diagrams 16c--f in turn (the contractions
will be represented by the symbol $\otimes$).

The first term
$-2\Gamma^{P,\tau\rho\sigma}(k_{3};k_{1},k_{2})
= 2k_{1}^{\rho}g^{\sigma\tau} -2k_{2}^{\sigma}g^{\rho\tau}$
from the decomposition (\ref{GammaFFFP}) is shown in Fig.~16c.
Substituting this term into the expression (\ref{fig15ad}),
then using the Ward identity (\ref{G4wid}),
together with $S^{-1}(p_{2}) = S^{-1}(p_{4}) = 0$ for the on-shell 
external fermions, and also
$f^{rst}T^{s}T^{t}T^{r} = \frac{1}{2}iC_{\!A}C_{F}$,
one obtains
\bea
\lefteqn{
{\rm Figs.\,}(16a+16b)\otimes 16c \otimes (16g+16h)|_{\xi = 1} }\nn \\
&=&
iC_{\!A}C_{F}Q^{2}g^{4}\int[dk_{1}][dk_{2}]\,
\frac{1}{k_{1}^{2}k_{2}^{2}(k_{1}+k_{2})^{2}}
\biggl\{
\gamma_{\mu}S(p)\gamma_{\rho}
\Bigl(S(p-k_{1}) \,+\, S(p+k_{2})\Bigr)
\gamma^{\rho}S(p)\gamma_{\nu} \nn \\
& & +\, 
\gamma_{\mu}S(p)\gamma_{\rho}
S(p-k_{1})\gamma_{\nu}S(p_{2}-k_{1})\gamma^{\rho} 
\,+\,
\gamma_{\rho}S(p_{4}+k_{2})\gamma_{\mu}S(p+k_{2})
\gamma^{\rho}S(p)\gamma_{\nu} \biggr\}\,\,.
\label{fig16c} 
\eea
The four terms on the r.h.s.\ of \eq{fig16c} are represented in 
Figs.~16i, 16j, 16n and 16r, respectively.
Each of these four terms involves an
{\em internal one-loop gluon self-energy-like pinch part}:
the first two terms, shown in Figs.~16i and 16j, 
are internal pinch contributions to the conventional
one-loop gluon self-energy insertion in
the quark self-energy correction shown in Fig.~14a,
i.e.\ they are contributions to the PT two-loop
``effective'' two-point component of the interaction;
and the third and fourth terms, shown in Figs.~16n and 16r,
are internal pinch contributions to the conventional 
one-loop gluon self-energy insertion in
the vertex corrections shown in Figs.~14b and 14c, respectively.
Isolating just the first two terms from \eq{fig16c},
making the change of variables $k_{1} \leftrightarrow -k_{2}$ 
in the second and then using \eq{Piprimepinch1}
gives the ``effective'' two-point component of the integrand:

\pagebreak

\begin{center}
\begin{picture}(430,630)(0,-200)

 
\ArrowLine(  0,430)(  0,400)
\Line(  0,400)( 60,400)
\ArrowLine( 60,400)( 60,430)
\GlueArc( 30,400)(15,  0, 90){2}{3}
\GlueArc( 30,400)(15, 90,180){2}{3}
\Gluon( 31,400)(31,415){2}{2}
\Photon( 0,380)( 0,400){-2}{3}
\Photon(60,380)(60,400){2}{3}
\put( 30,370){\makebox(0,0)[c]{(a)}}
 
\Line(  0,355)(  0,325)
\Line(  0,325)( 60,325)
\ArrowLine( 60,325)( 60,355)
\GlueArc( 17,325)(17,  0, 90){2}{4}
\Gluon( 0,342)(17,342){2}{2}
\Gluon(17,325)(17,342){2}{2}
\Photon( 0,305)( 0,325){-2}{3}
\Photon(60,305)(60,325){2}{3}
\put( 30,295){\makebox(0,0)[c]{(b)}}
 
\ArrowLine(  0,280)(  0,250)
\Line(  0,250)( 60,250)
\Line( 60,250)( 60,280)
\GlueArc( 43,250)(17, 90,180){2}{4}
\Gluon(43,267)(60,267){2}{2}
\Gluon(43,250)(43,267){-2}{2}
\Photon( 0,230)( 0,250){-2}{3}
\Photon(60,230)(60,250){2}{3}
\put( 30,220){\makebox(0,0)[c]{(c)}}
 
\Line(  0,205)(  0,175)
\ArrowLine(  0,175)( 30,175)
\ArrowLine( 30,175)( 60,175)
\Line( 60,175)( 60,205)
\Gluon( 0,190)(30,190){2}{5}
\Gluon(30,190)(60,190){2}{5}
\Gluon(31,175)(31,190){2}{2}
\Photon( 0,155)( 0,175){-2}{3}
\Photon(60,155)(60,175){2}{3}
\put( 30,145){\makebox(0,0)[c]{(d)}}
 
 
\put( 72,292){\makebox(0,0)[c]
{$ \left. \begin{array}{c}  \\ \\ \\ \\ \\ \\ \\ \\ \\ 
\\ \\ \\ \\ \\ \\ \\ \\ \\ \\
\end{array} \right\}$}}
 
\put(  95,292){\makebox(0,0)[c]{
\Large $=$}}
 
\put(120,292){\makebox(0,0)[c]
{$ \left\{ \begin{array}{c}  \\ \\ \\ \\ \\ \\ \\ \\ \\ 
\end{array} \right.$}}
 
 
\ArrowLine(130,355)(130,325)
\Line(130,325)(160,325)
\GlueArc(160,325)(17, 90, 180){2}{4}
\Photon(130,305)(130,325){-2}{3}
\put(145,295){\makebox(0,0)[c]{(i)}}
 
\Line(130,280)(130,250)
\ArrowLine(130,250)(160,250)
\Gluon(130,265)(160,265){2}{4}
\Photon(130,230)(130,250){-2}{3}
\put(145,220){\makebox(0,0)[c]{(j)}}
 

\put(170,292){\makebox(0,0)[c]
{$ \left. \begin{array}{c}  \\ \\ \\ \\ \\ \\ \\ \\ \\ 
\end{array} \right\}$}}
 
\Gluon(200,302)(225,302){2}{4}
\ArrowLine(202,282)(223,282)
\put(213.5,320){\makebox(0,0)[c]{$k_{1}$}}
\put(208.5,310){\vector(1, 0){10}}
\put(213.5,270){\makebox(0,0)[c]{$p-k_{1}$}}
 
\put(255,292){\makebox(0,0)[c]
{$ \left\{ \begin{array}{c}  \\ \\ \\
\end{array} \right.$}}
 
\Gluon(264,305)(280,305){2}{2}
\Gluon(280,305)(296,305){2}{2}
\Gluon(280,280)(280,305){2}{4}
\Line(265,280)(295,280)
\put(280,265){\makebox(0,0)[c]{(k)}}
 
\put(305,292){\makebox(0,0)[c]
{$ \left. \begin{array}{c}  \\ \\ \\
\end{array} \right\}$}}
 
\Gluon(335,302)(360,302){2}{4}
\ArrowLine(337,282)(358,282)
\put(348.5,320){\makebox(0,0)[c]{$k_{2}$}}
\put(353.5,310){\vector(-1, 0){10}}
\put(348.5,270){\makebox(0,0)[c]{$p+k_{2}$}}
 
\put(390,292){\makebox(0,0)[c]
{$ \left\{ \begin{array}{c}  \\ \\ \\ \\ \\ \\ \\ \\ \\ 
\end{array} \right.$}}
 

\Line(400,325)(430,325)
\ArrowLine(430,325)(430,355)
\GlueArc(400,325)(17,  0, 90){2}{4}
\Photon(430,305)(430,325){2}{3}
\put(415,295){\makebox(0,0)[c]{(l)}}
 
\ArrowLine(400,250)(430,250)
\Line(430,250)(430,280)
\Gluon(400,265)(430,265){2}{4}
\Photon(430,230)(430,250){2}{3}
\put(415,220){\makebox(0,0)[c]{(m)}}
 
 
\Line(  0,130)(  0, 90)
\ArrowLine(  0, 90)( 30, 90)
\ArrowLine( 30, 90)( 60, 90)
\ArrowLine( 60, 90)( 60,130)
\GlueArc(  0,110)(10,-90,  0){2}{2}
\GlueArc(  0,110)(10,  0, 90){2}{2}
\GlueArc(10, 90)(20, 0, 90){2}{4}
\Photon( 0, 80)( 0, 90){2}{1.5}
\Photon(60, 80)(60, 90){-2}{1.5}
\put( 30, 70){\makebox(0,0)[c]{(e)}}
 
\Line(  0, 55)(  0, 15)
\ArrowLine(  0, 15)( 60, 15)
\Line( 60, 15)( 60, 55)
\GlueArc(  0, 35)(10,-90,  0){2}{2}
\GlueArc(  0, 35)(10,  0, 90){2}{2}
\Gluon(10,35)(60,35){2}{8}
\Photon( 0,  5)( 0, 15){2}{1.5}
\Photon(60,  5)(60, 15){-2}{1.5}
\put( 30, -5){\makebox(0,0)[c]{(f)}}
 

\put(72,67){\makebox(0,0)[c]
{$ \left. \begin{array}{c}  \\ \\ \\ \\ \\ \\ \\ \\ \\
\end{array} \right\}$}}
 
\put( 120,67){\makebox(0,0)[c]{
\Large $=$}}
 
\put(170,67){\makebox(0,0)[c]
{$ \left\{ \begin{array}{c}  \\ \\ \\ \\ \\
\end{array} \right.$}}
 
\Line(180,100)(180,60)
\ArrowLine(180,60)(210,60)
\GlueArc(180, 80)(10,-90,  0){2}{2}
\GlueArc(180, 80)(10,  0, 90){2}{2}
\Gluon(190,80)(210,80){2}{3}
\Photon(180, 40)(180, 60){-2}{3}
\put(195, 30){\makebox(0,0)[c]{(n)}}
 
\put(220,67){\makebox(0,0)[c]
{$ \left. \begin{array}{c}  \\ \\ \\ \\ \\
\end{array} \right\}$}}
 
\Gluon(264, 77)(296, 77){2}{5}
\ArrowLine(265,57)(295,57)
\put(281, 95){\makebox(0,0)[c]{$k_{2}$}}
\put(286, 85){\vector(-1, 0){10}}
\put(281, 45){\makebox(0,0)[c]{$p+k_{2}$}}
 
\put(340,67){\makebox(0,0)[c]
{$ \left\{ \begin{array}{c}  \\ \\ \\ \\ \\ \\ \\ \\ \\ 
\end{array} \right.$}}
 
 
\Line(350,100)(380,100)
\ArrowLine(380,100)(380,130)
\GlueArc(350,100)(17,  0, 90){2}{4}
\Photon(380,80)(380,100){2}{3}
\put(365,70){\makebox(0,0)[c]{(o)}}
 
\ArrowLine(350,25)(380,25)
\Line(380,25)(380,55)
\Gluon(350,40)(380,40){2}{4}
\Photon(380, 5)(380,25){2}{3}
\put(365,-5){\makebox(0,0)[c]{(p)}}
 
 
\ArrowLine(  0,-20)(  0,-60)
\ArrowLine(  0,-60)( 30,-60)
\ArrowLine( 30,-60)( 60,-60)
\Line( 60,-60)( 60,-20)
\GlueArc( 60,-40)(10, 90,180){2}{2}
\GlueArc( 60,-40)(10,180,270){2}{2}
\GlueArc(50,-60)(20,90,180){2}{4}
\Photon( 0,-70)( 0,-60){2}{1.5}
\Photon(60,-70)(60,-60){-2}{1.5}
\put( 30,-80){\makebox(0,0)[c]{(g)}}
 
\Line(  0,-95)(  0,-135)
\ArrowLine(  0,-135)( 60,-135)
\Line( 60,-135)( 60,-95)
\GlueArc( 60,-115)(10, 90,180){2}{2}
\GlueArc( 60,-115)(10,180,270){2}{2}
\Gluon(0,-115)(50,-115){2}{8}
\Photon( 0,-145)( 0,-135){2}{1.5}
\Photon(60,-145)(60,-135){-2}{1.5}
\put( 30,-155){\makebox(0,0)[c]{(h)}}
 
 
\put(72,-83){\makebox(0,0)[c]
{$ \left. \begin{array}{c}  \\ \\ \\ \\ \\ \\ \\ \\ \\
\end{array} \right\}$}}
 
\put( 120,-83){\makebox(0,0)[c]{
\Large $=$}}
 
\put(170,-83){\makebox(0,0)[c]
{$ \left\{ \begin{array}{c}  \\ \\ \\ \\ \\ \\ \\ \\ \\
\end{array} \right.$}}
 
\Line(380,-50)(380,-90)
\ArrowLine(350,-90)(380,-90)
\GlueArc(380,-70)(10, 90,180){2}{2}
\GlueArc(380,-70)(10,180,270){2}{2}
\Gluon(350,-70)(370,-70){2}{3}
\Photon(380,-110)(380,-90){2}{3}
\put(365,-120){\makebox(0,0)[c]{(s)}}
 
\put(220,-83){\makebox(0,0)[c]
{$ \left. \begin{array}{c}  \\ \\ \\ \\ \\ \\ \\ \\ \\
\end{array} \right\}$}}
 
\Gluon(264,-73)(296,-73){2}{5}
\ArrowLine(265,-93)(295,-93)
\put(281, -55){\makebox(0,0)[c]{$k_{1}$}}
\put(276, -65){\vector(1, 0){10}}
\put(281,-105){\makebox(0,0)[c]{$p-k_{1}$}}
 
\put(340,-83){\makebox(0,0)[c]
{$ \left\{ \begin{array}{c}  \\ \\ \\ \\ \\ 
\end{array} \right.$}}
 
 
\Line(180,-50)(210,-50)
\ArrowLine(180,-20)(180,-50)
\GlueArc(210,-50)(17, 90,180){2}{4}
\Photon(180,-70)(180,-50){-2}{3}
\put(195,-80){\makebox(0,0)[c]{(q)}}
 
\ArrowLine(180,-125)(210,-125)
\Line(180,-125)(180,-95)
\Gluon(180,-110)(210,-110){2}{4}
\Photon(180,-145)(180,-125){-2}{3}
\put(195,-155){\makebox(0,0)[c]{(r)}}

\put(-5,-175){\makebox(0,0)[l]{\small
Fig.\ 15. a--h: The two-loop QCD corrections to the process
$q\gamma \rightarrow q\gamma$ consisting of diagrams with one }}

\put(-5,-190){\makebox(0,0)[l]{\small
triple gauge vertex 
(the two such diagrams representing external leg corrections are
not shown).}}

\end{picture}
\end{center}

\pagebreak

\begin{center}
\begin{picture}(440,625)(55,-35)
 

\ArrowLine(130,555)(130,525)
\Line(130,525)(160,525)
\GlueArc(160,525)(17, 90, 180){2}{4}
\Photon(130,505)(130,525){-2}{3}
\put(145,495){\makebox(0,0)[c]{(a)}}
 
\Line(130,480)(130,450)
\ArrowLine(130,450)(160,450)
\Gluon(130,465)(160,465){2}{4}
\Photon(130,430)(130,450){-2}{3}
\put(145,420){\makebox(0,0)[c]{(b)}}
 
 
\put(170,492){\makebox(0,0)[c]
{$ \left. \begin{array}{c}  \\ \\ \\ \\ \\ \\ \\ \\ \\ 
\end{array} \right\}$}}
 
\Gluon(200,502)(225,502){2}{4}
\ArrowLine(202,482)(223,482)
\put(213.5,520){\makebox(0,0)[c]{$k_{1}$}}
\put(208.5,510){\vector(1, 0){10}}
\put(213.5,470){\makebox(0,0)[c]{$p-k_{1}$}}

\put(255,492){\makebox(0,0)[c]
{$ \left\{ \begin{array}{c}  \\ \\ \\ \\ \\ \\ \\ \\ \\ \\ \\ \\ \\ 
\end{array} \right.$}}
 

\put(280,590){\vector(-1, 0){11}}
\put(280,590){\vector( 1, 0){11}}
\put(260,570){\makebox(0,0)[c]{\large $-2$}}
\Gluon(264,580)(280,580){2}{2}
\Gluon(280,580)(296,580){2}{2}
\Gluon(280,555)(280,580){2}{4}
\Line(265,555)(295,555)
\put(280,545){\makebox(0,0)[c]{(c)}}
 
\put(260,520){\makebox(0,0)[c]{\large $+$}}
\Line(294,525)(286,525)
\Line(286,525)(286,517)
\Gluon(264,530)(280,530){2}{2}
\Gluon(280,530)(296,530){2}{2}
\Gluon(280,505)(280,530){2}{4}
\Line(265,505)(295,505)
\put(280,495){\makebox(0,0)[c]{(d)}}
 
\put(260,470){\makebox(0,0)[c]{\large $+$}}
\Line(266,475)(274,475)
\Line(274,475)(274,467)
\Gluon(264,480)(280,480){2}{2}
\Gluon(280,480)(296,480){2}{2}
\Gluon(280,455)(280,480){2}{4}
\Line(265,455)(295,455)
\put(280,445){\makebox(0,0)[c]{(e)}}
 
\put(260,415){\makebox(0,0)[c]{\large $-$}}
\Line(270,431)(290,431)
\Gluon(264,425)(280,425){2}{2}
\Gluon(280,425)(296,425){2}{2}
\Gluon(280,400)(280,425){2}{4}
\Line(265,400)(295,400)
\put(280,390){\makebox(0,0)[c]{(f)}}
 
\put(305,492){\makebox(0,0)[c]
{$ \left. \begin{array}{c}  \\ \\ \\ \\ \\ \\\ \\ \\ \\ \\ \\ \\ \\
\end{array} \right\}$}}
 
\Gluon(335,502)(360,502){2}{4}
\ArrowLine(337,482)(358,482)
\put(348.5,520){\makebox(0,0)[c]{$k_{2}$}}
\put(353.5,510){\vector(-1, 0){10}}
\put(348.5,470){\makebox(0,0)[c]{$p+k_{2}$}}

\put(390,492){\makebox(0,0)[c]
{$ \left\{ \begin{array}{c}  \\ \\ \\ \\ \\ \\ \\ \\ \\ 
\end{array} \right.$}}
 
 
\Line(400,525)(430,525)
\ArrowLine(430,525)(430,555)
\GlueArc(400,525)(17,  0, 90){2}{4}
\Photon(430,505)(430,525){2}{3}
\put(415,495){\makebox(0,0)[c]{(g)}}
 
\ArrowLine(400,450)(430,450)
\Line(430,450)(430,480)
\Gluon(400,465)(430,465){2}{4}
\Photon(430,430)(430,450){2}{3}
\put(415,420){\makebox(0,0)[c]{(h)}}
 
 
\put(440,492){\makebox(0,0)[c]
{$ \left. \begin{array}{c}  \\ \\ \\ \\ \\ \\ \\ \\ \\ 
\end{array} \right\}$}}
 
\put( 55,327){\makebox(0,0)[c]{\Large $=$}}
 
\ArrowLine( 70,355)( 70,325)
\Line( 70,325)(130,325)
\ArrowLine(130,325)(130,355)
\Photon( 70,305)( 70,325){-2}{3}
\Photon(130,305)(130,325){2}{3}
\GlueArc(100,325)(18,  0, 90){2}{3}
\GlueArc(100,325)(18, 90, 180){2}{4}
\GlueArc(120,345)(18, 180,270){2}{3}
\put(100,295){\makebox(0,0)[c]{(i)}}
 
\put(145,327){\makebox(0,0)[c]{\Large $+$}}
 
\ArrowLine(160,355)(160,325)
\Line(160,325)(220,325)
\ArrowLine(220,325)(220,355)
\Photon(160,305)(160,325){-2}{3}
\Photon(220,305)(220,325){2}{3}
\GlueArc(190,325)(18,  0, 90){2}{4}
\GlueArc(190,325)(18, 90,180){2}{3}
\GlueArc(170,345)(18,-90, 0){2}{3}
\put(190,295){\makebox(0,0)[c]{(j)}}
 
\put(235,327){\makebox(0,0)[c]{\Large $+$}}
 
\ArrowLine(250,355)(250,325)
\Line(250,325)(310,325)
\ArrowLine(310,325)(310,355)
\Photon(250,305)(250,325){-2}{3}
\Photon(310,305)(310,325){2}{3}
\GlueArc(280,325)(18,  0, 90){2}{4}
\GlueArc(280,325)(18, 90, 180){2}{4}
\Gluon(280,325)(280,343){2}{3}
\CArc(280,326)(12,20,63)
\Line(285,337)(285,328)
\put(280,295){\makebox(0,0)[c]{(k)}}
 
\put(325,327){\makebox(0,0)[c]{\Large $+$}}
 
\ArrowLine(340,355)(340,325)
\Line(340,325)(400,325)
\ArrowLine(400,325)(400,355)
\Photon(340,305)(340,325){-2}{3}
\Photon(400,305)(400,325){2}{3}
\GlueArc(370,325)(18,  0, 90){2}{4}
\GlueArc(370,325)(18, 90, 180){2}{4}
\Gluon(370,325)(370,343){2}{3}
\CArc(370,326)(12,117,160)
\Line(365,337)(365,328)
\put(370,295){\makebox(0,0)[c]{(l)}}
 
\put(415,327){\makebox(0,0)[c]{\Large $-$}}

\ArrowLine(430,355)(430,325)
\Line(430,325)(490,325)
\ArrowLine(490,325)(490,355)
\Photon(430,305)(430,325){-2}{3}
\Photon(490,305)(490,325){2}{3}
\GlueArc(460,325)(18,  0, 90){2}{4}
\GlueArc(460,325)(18, 90, 180){2}{4}
\Gluon(460,325)(460,343){2}{3}
\Line(460,354)(472,346)
\Line(460,354)(448,346)
\put(460,295){\makebox(0,0)[c]{(m)}}
 
 
\put(100,252){\makebox(0,0)[c]{\Large $+$}}
 
\Line(115,280)(115,250)
\ArrowLine(115,250)(155,250)
\Line(155,250)(175,250)
\ArrowLine(175,250)(175,280)
\Photon(115,230)(115,250){-2}{3}
\Photon(175,230)(175,250){2}{3}
\Gluon(115,268)(135,268){2}{3}
\GlueArc(135,250)(18,  0, 90){2}{3}
\GlueArc(155,270)(18, 180,270){2}{3}
\put(145,220){\makebox(0,0)[c]{(n)}}
 
\put(190,252){\makebox(0,0)[c]{\Large $+$}}
 
\Line(205,280)(205,250)
\Line(205,250)(265,250)
\ArrowLine(265,250)(265,280)
\Photon(205,230)(205,250){-2}{3}
\Photon(265,230)(265,250){2}{3}
\Gluon(205,268)(225,268){2}{3}
\GlueArc(225,250)(18,  0, 90){2}{4}
\Gluon(225,250)(225,268){2}{3}
\CArc(225,251)(12,20,63)
\Line(230,262)(230,253)
\put(235,220){\makebox(0,0)[c]{(o)}}
 
\put(280,252){\makebox(0,0)[c]{\Large $+$}}

\Line(295,280)(295,250)
\Line(295,250)(355,250)
\ArrowLine(355,250)(355,280)
\Photon(295,230)(295,250){-2}{3}
\Photon(355,230)(355,250){2}{3}
\Gluon(295,268)(315,268){2}{3}
\GlueArc(315,250)(18,  0, 90){2}{4}
\Gluon(315,250)(315,268){2}{3}
\Line(301,263)(309,263)
\Line(309,263)(309,255)
\put(325,220){\makebox(0,0)[c]{(p)}}
 
\put(370,252){\makebox(0,0)[c]{\Large $-$}}
 
\Line(385,280)(385,250)
\Line(385,250)(445,250)
\ArrowLine(445,250)(445,280)
\Photon(385,230)(385,250){-2}{3}
\Photon(445,230)(445,250){2}{3}
\Gluon(385,268)(405,268){2}{3}
\GlueArc(405,250)(18,  0, 90){2}{4}
\Gluon(405,250)(405,268){2}{3}
\Line(398,274)(408,274)
\Line(408,274)(416,271)
\put(415,220){\makebox(0,0)[c]{(q)}}


\put(100,177){\makebox(0,0)[c]{\Large $+$}}
 
\ArrowLine(115,205)(115,175)
\Line(115,175)(135,175)
\ArrowLine(135,175)(175,175)
\Line(175,175)(175,205)
\Photon(115,155)(115,175){-2}{3}
\Photon(175,155)(175,175){2}{3}
\Gluon(155,193)(175,193){2}{3}
\GlueArc(155,175)(18, 90,180){2}{3}
\GlueArc(135,195)(18,-90, 0){2}{3}
\put(145,145){\makebox(0,0)[c]{(r)}}
 
\put(190,177){\makebox(0,0)[c]{\Large $+$}}
 
\ArrowLine(205,205)(205,175)
\Line(205,175)(265,175)
\Line(265,175)(265,205)
\Photon(205,155)(205,175){-2}{3}
\Photon(265,155)(265,175){2}{3}
\Gluon(245,193)(265,193){2}{3}
\GlueArc(245,175)(18, 90, 180){2}{4}
\Gluon(245,175)(245,193){2}{3}
\CArc(245,176)(12,117,160)
\Line(240,187)(240,178)
\put(235,145){\makebox(0,0)[c]{(s)}}
 
\put(280,177){\makebox(0,0)[c]{\Large $+$}}
 
\ArrowLine(295,205)(295,175)
\Line(295,175)(355,175)
\Line(355,175)(355,205)
\Photon(295,155)(295,175){-2}{3}
\Photon(355,155)(355,175){2}{3}
\Gluon(335,193)(355,193){2}{3}
\GlueArc(335,175)(18, 90, 180){2}{4}
\Gluon(335,175)(335,193){2}{3}
\Line(349,188)(341,188)
\Line(341,188)(341,180)
\put(325,145){\makebox(0,0)[c]{(t)}}
 
\put(370,177){\makebox(0,0)[c]{\Large $-$}}
 
\ArrowLine(385,205)(385,175)
\Line(385,175)(445,175)
\Line(445,175)(445,205)
\Photon(385,155)(385,175){-2}{3}
\Photon(445,155)(445,175){2}{3}
\Gluon(425,193)(445,193){2}{3}
\GlueArc(425,175)(18, 90, 180){2}{4}
\Gluon(425,175)(425,193){2}{3}
\Line(422,199)(432,199)
\Line(414,196)(422,199)
\put(415,145){\makebox(0,0)[c]{(u)}}
 
 
\put(145,102){\makebox(0,0)[c]{\Large $+$}}
 
\Line(160,130)(160,100)
\ArrowLine(160,100)(190,100)
\ArrowLine(190,100)(220,100)
\Line(220,100)(220,130)
\Photon(160, 80)(160,100){-2}{3}
\Photon(220, 80)(220,100){2}{3}
\Gluon(160,118)(190,118){2}{5}
\Gluon(190,118)(220,118){2}{5}
\Gluon(190,100)(190,118){2}{3}
\Line(204,113)(196,113)
\Line(196,113)(196,105)
\put(190, 70){\makebox(0,0)[c]{(v)}}
 
\put(235,102){\makebox(0,0)[c]{\Large $+$}}
 
\Line(250,130)(250,100)
\ArrowLine(250,100)(280,100)
\ArrowLine(280,100)(310,100)
\Line(310,100)(310,130)
\Photon(250, 80)(250,100){-2}{3}
\Photon(310, 80)(310,100){2}{3}
\Gluon(250,118)(280,118){2}{5}
\Gluon(280,118)(310,118){2}{5}
\Gluon(280,100)(280,118){2}{3}
\Line(266,113)(274,113)
\Line(274,113)(274,105)
\put(280, 70){\makebox(0,0)[c]{(w)}}
 
\put(325,102){\makebox(0,0)[c]{\Large $-$}}

\Line(340,130)(340,100)
\ArrowLine(340,100)(370,100)
\ArrowLine(370,100)(400,100)
\Line(400,100)(400,130)
\Photon(340, 80)(340,100){-2}{3}
\Photon(400, 80)(400,100){2}{3}
\Gluon(340,118)(370,118){2}{5}
\Gluon(370,118)(400,118){2}{5}
\Gluon(370,100)(370,118){2}{3}
\Line(360,124)(380,124)
\put(370, 70){\makebox(0,0)[c]{(x)}}

\put(55, 50){\makebox(0,0)[l]{\small
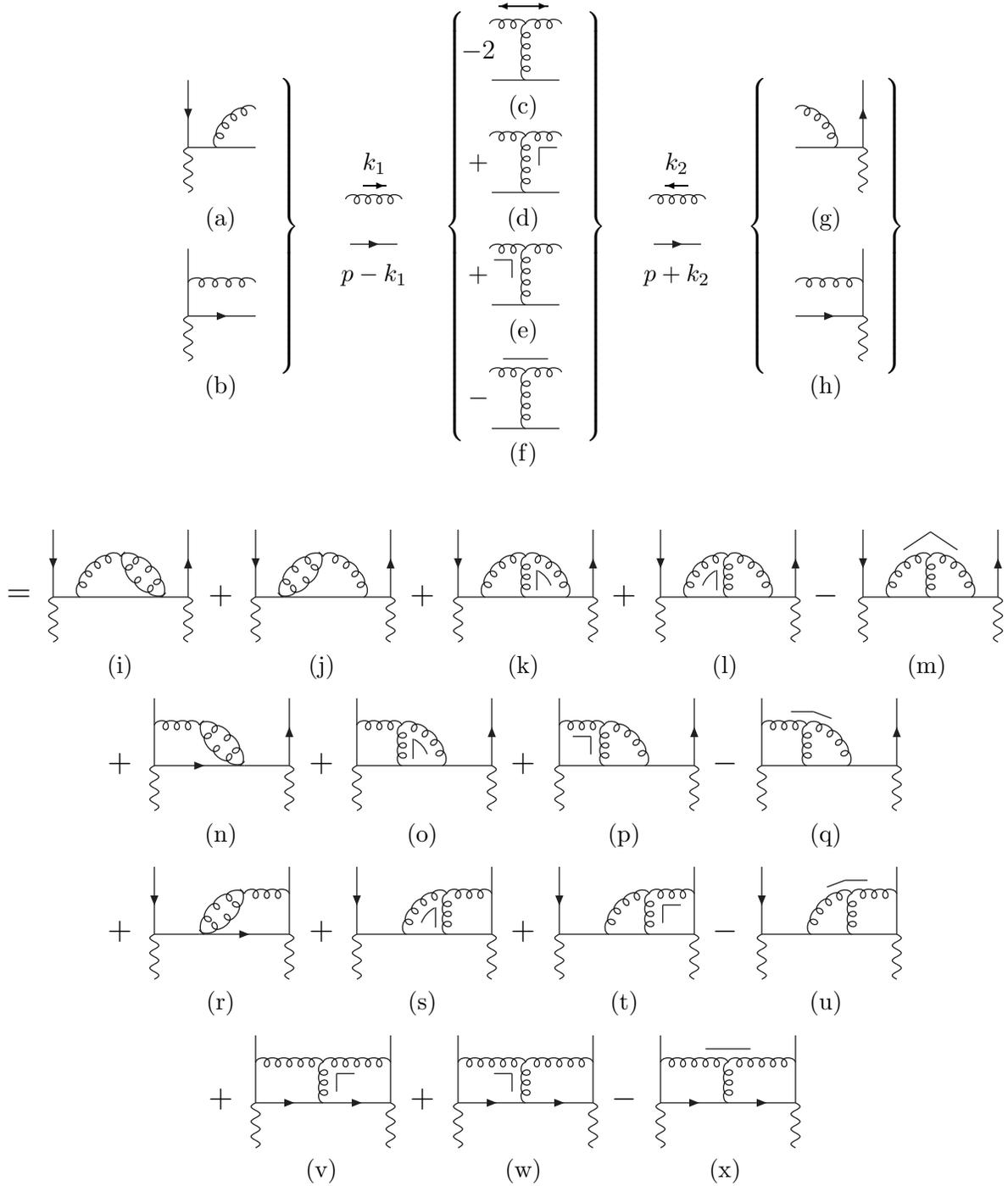
Fig.\ 16. a--h: The tree level diagrams shown in 
Figs.~15i--m, the contraction of which gives the two-}}

\put(55, 35){\makebox(0,0)[l]{\small
loop diagrams shown
in Figs.~15a--d, here with the decomposition (\ref{GammaFFFP}) 
of the triple gauge vertex}}

\put(55, 20){\makebox(0,0)[l]{\small
represented explicitly (16c--f). i--x: The contraction of these components.
The diagrams in the first }}

\put(55, 5){\makebox(0,0)[l]{\small
line (16i--m) are contributions to the PT self-energy
correction; those in the second and third lines}}

\put(55,-10){\makebox(0,0)[l]{\small
(16n--u) are contributions to the PT vertex corrections; 
and those in the fourth line (16v--x) are }}

\put(55,-25){\makebox(0,0)[l]{\small
contributions to the PT box correction.}}

\end{picture}
\end{center}

\bea
\lefteqn{
{\rm Figs.\,}(16a+16b)\otimes 16c \otimes 
(16g+16h)|_{\xi = 1,2-{\rm pt}} } \nn \\
&=&
-C_{F}Q^{2}g^{2}\int[dk_{1}][dk_{2}]\,
\frac{1}{k_{1}^{2}}\,\Delta\Pi^{\prime(1)}(k_{1},k_{2},1)
\gamma_{\mu}S(p)\gamma_{\rho}S(p-k_{1})
\gamma^{\rho}S(p)\gamma_{\nu} \,\,.
\label{fig16c2}
\eea
We immediately see that the contribution
\eq{fig16c2} exactly cancels the $g_{\rho\rho'}$
component of the pinch part $-\Delta\Pi_{\rho\rho'}^{\prime(1)}$
of the conventional Feynman gauge
one-loop gluon self-energy in \eq{fig14af2pt}.
This cancellation leaves in \eq{fig14af2pt}
precisely the required integrand 
$\hat{\Pi}_{\rho\rho'}^{\prime(1)}(k_{1},k_{2})$
for the unrenormalized
PT gauge-independent one-loop gluon self-energy insertion.
Adding the diagram 8i, with 
the PT one-loop renormalization constant 
$(Z_{3}-1)_{\rm PT}^{(1)}$
as the gluon self-energy counterterm insertion, 
we have therefore succeeded in obtaining the
contribution to the PT two-loop quark self-energy 
$-i\hat{\Sigma}^{(2)}(p)$ of the diagram 10c,
involving the renormalized PT one-loop gluon self-energy correction 
inserted into the PT one-loop quark self-energy,
embedded in the process $q\gamma\rightarrow q\gamma$.

The second and third terms
$\Gamma_{\rho\sigma\tau}^{F}(k_{1};k_{2},k_{3})$ and
$\Gamma_{\sigma\tau\rho}^{F}(k_{2};k_{3},k_{1})$
from the decomposition (\ref{GammaFFFP}) are shown in Figs.~16d and 16e.
The corrections to
$q\gamma \rightarrow q\gamma$ from these two components
of the triple gauge vertex in \eq{fig15ad}
are represented diagrammatically in
Figs.~16k, 16o, 16t and 16v
and Figs.~16l, 16p, 16s and 16w, respectively.
The contributions represented in Figs.~16k and 16l
are self-energy-like (``effective'' two-point)
corrections to the process.
We see immediately that they give precisely the contribution
of the diagram 14c to the PT one-loop quark-gluon
vertex insertions appearing in the 
diagrams 10d and 10e, respectively, 
for the PT two-loop quark self-energy $-i\hat{\Sigma}^{(2)}(p)$,
embedded in the process $q\gamma \rightarrow q\gamma$.
The contributions represented in 
Figs.~16o, 16p, 16s and 16t are vertex-like corrections to
$q\gamma \rightarrow q\gamma$,
while those in Figs.~16v and 16w are box-like corrections.

The last term
$-\Gamma_{\tau\rho\sigma}^{F}(k_{3};k_{1},k_{2})$
from the decomposition (\ref{GammaFFFP}) is shown in Fig.~16f.
The corrections to
$q\gamma \rightarrow q\gamma$ from this component
of the triple gauge vertex in \eq{fig15ad}
are represented in Figs.~16m, 16q, 16u and 16x, 
including explicitly in each case the associated minus sign.
The contribution in Fig.~16m is a self-energy-like correction;
those in Fig.~16q and 16u are vertex-like corrections;
and that in Fig.~16x is a box-like correction.
Having obtained, as just described, the required contributions 
from the diagrams of Fig.~15, thence Fig.~16, 
to the diagrams 10b--e for the PT two-loop fermion self-energy
$-i\hat{\Sigma}^{(2)}(p)$,
the self-energy-like contribution of Fig.~16m is a contribution
to the diagram 10f involving the PT
quark-gluon kernel $\hat{V}^{(0)}$, all
embedded in the process $q\gamma\rightarrow q\gamma$.
The contribution of Fig.~16m thus gives the component of the
PT kernel shown in Fig.~13b, involving just the
component $\Gamma^{F}$ of the triple gauge vertex.

There remain from Fig.~15 the diagrams 15e--h.
In the Feynman gauge, the sum of these diagrams may be written
\bea
\lefteqn{{\rm Figs.\,15e\!-\!h}|_{\xi = 1}
\,\,=\,\,
Q^{2}g^{4}\int[dk_{1}][dk_{2}]\,
\frac{1}{k_{1}^{2}k_{2}^{2}k_{3}^{2}}\,
f^{rst}\Gamma^{\rho\sigma\tau}(k_{1},k_{2},k_{3}) } \nn \\
& &
\times \biggl\{
G_{\sigma\mu}^{s}(-k_{2},-p_{3},p+k_{2},p_{4})
S(p+k_{2})\gamma_{\nu}S(p_{2} + k_{2})\gamma_{\tau}T^{t}
S(p_{2}-k_{1})\gamma_{\rho}T^{r} \nn \\
& & +\,
\gamma_{\sigma}T^{s}S(p_{4}+k_{2})\gamma_{\tau}T^{t}S(p_{4}-k_{1})
\gamma_{\mu}S(p-k_{1})
G_{\rho\nu}^{r}(-k_{1},p_{1},p_{2},p-k_{1}) \biggl\}
\label{fig15eh}
\eea
with $k_{1} + k_{2} + k_{3} = 0$.
For these diagrams, we again use the decomposition (\ref{GammaFFFP})
for the triple gauge vertex in \eq{fig15eh}.
It is found that, in the Feynman gauge, the diagrams \mbox{15e--h}
make no contribution to the PT self-energy-like
(``effective'' two-point) component of the process
$q\gamma \rightarrow q\gamma$.
Instead, the first term
$-2\Gamma^{P,\tau\rho\sigma}(k_{3};k_{1},k_{2})$
from \eq{GammaFFFP} results in a pair of two-loop vertex corrections
similar and
identical in magnitude to the diagrams 16n and 16r involving
internal one-loop gluon self-energy-like pinch parts,
together with a pair of two-loop external leg corrections,
also involving internal one-loop gluon self-energy-like pinch parts.
The remaining three terms from the decomposition \eq{GammaFFFP}
of the triple gauge vertex in \eq{fig15eh}
produce twelve diagrams, corresponding to the three 
different $\Gamma^{F}$ components from \eq{GammaFFFP}
substituted for the triple gauge vertices in the diagrams
15e--h (the orientations and signs should be clear).

\subsection{Two-loop gluonic corrections 
involving no triple gluon vertices}

Finally, we consider the $\cO(\alpha_{s}^{2})$ gluonic corrections to
$q\gamma\rightarrow q\gamma$ consisting both of
one-particle-irreducible and one-particle-reducible
two-loop diagrams with no triple gauge vertices.
The set of such QED-like diagrams are shown in Fig.~17
(the diagrams from this set which are purely
external leg corrections are not shown; 
nor, as always, are the corresponding diagrams 
involving crossed external photon legs).
In the Feynman gauge $\xi = 1$, there appear no
factors of longitudinal gluon four-momentum in the
corresponding two-loop integrands for these diagrams.
Thus, in this gauge,
no Ward identities are triggered (there is ``no pinching''),
and the diagrams in Fig.~17 contribute directly to the PT
two-loop self-energy, vertex and box corrections to
$q\gamma\rightarrow q\gamma$ as shown. In particular,
the only two contributions to the PT 1PI two-loop quark 
self-energy (``effective'' two-point) correction
are from the diagrams 17a and 17b
(the diagram 17a$^{\prime}$ gives the  
one-particle-reducible chain of two
PT one-loop quark self-energies).

{}From the diagram 17a,
adding the corresponding diagrams involving the
PT one-loop fermion self-energy counterterms
$(Z_{2}-1)_{\rm PT}^{(1)}$ and $(Z_{m}-1)_{\rm PT}^{(1)}$,
we immediately obtain the required contribution to the 
PT two-loop fermion self-energy $-i\hat{\Sigma}^{(2)}(p)$ of 
Fig.~10b, involving the 
renormalized PT one-loop quark self-energy as an internal
correction to, in turn, the PT one-loop quark self-energy,
all embedded in the process $q\gamma\rightarrow q\gamma$. 
This is as remarked at the beginning of this section.

{}From the diagram 17b, however, we first have to add and subtract
another identical such contribution.
In this way, we obtain from twice the diagram 17b
the required contribution 
of the diagram 12j to the PT one-loop quark-gluon vertex
insertions appearing in the diagrams 10d and 10e,
respectively, for $-i\hat{\Sigma}^{(2)}(p)$, 
all embedded in the process $q\gamma\rightarrow q\gamma$.
It then remains to subtract again the contribution
of the diagram 17b to avoid overcounting.
Having now obtained all of 
the required contributions to the diagrams 10b--e,
this subtracted contribution to $-i\hat{\Sigma}^{(2)}(p)$
is a contribution to the the diagram 10f involving
the PT quark-gluon kernel $\hat{V}^{(0)}$, 
again all embedded in the process $q\gamma\rightarrow q\gamma$. 
The contribution of minus Fig.~17b thus gives the component of the
PT kernel shown in Fig.~13c.

\pagebreak

\begin{center}
\begin{picture}(425,635)(0,-205)
 
 

\ArrowLine(  0,430)(  0,400)
\Line(  0,400)( 50,400)
\ArrowLine( 50,400)( 50,430)
\GlueArc( 25,400)(15,  0,180){2}{7}
\GlueArc( 25,400)( 8,  0,180){2}{4}
\Photon( 0,380)( 0,400){-2}{3}
\Photon(50,380)(50,400){2}{3}
\put( 25,370){\makebox(0,0)[c]{(a)}}
 
\ArrowLine( 75,430)( 75,400)
\Line( 75,400)(125,400)
\ArrowLine(125,400)(125,430)
\GlueArc( 95,400)(10,  0,180){2}{5}
\GlueArc(105,400)(10,180,360){2}{5}
\Photon( 75,380)( 75,400){-2}{3}
\Photon(125,380)(125,400){2}{3}
\put(100,370){\makebox(0,0)[c]{(b)}}
 
\Line(150,430)(150,400)
\Line(150,400)(200,400)
\ArrowLine(200,400)(200,430)
\Gluon(150,420)(175,400){2}{5}
\GlueArc(175,400)(10,180,360){2}{5}
\Photon(150,380)(150,400){-2}{3}
\Photon(200,380)(200,400){2}{3}
\put(175,370){\makebox(0,0)[c]{(c)}}
 
\ArrowLine(225,430)(225,400)
\Line(225,400)(275,400)
\Line(275,400)(275,430)
\Gluon(275,420)(250,400){-2}{5}
\GlueArc(250,400)(10,180,360){2}{5}
\Photon(225,380)(225,400){-2}{3}
\Photon(275,380)(275,400){2}{3}
\put(250,370){\makebox(0,0)[c]{(d)}}
 
\Line(300,430)(300,400)
\Line(300,400)(350,400)
\ArrowLine(350,400)(350,430)
\Gluon(300,420)(340,400){2}{7}
\GlueArc(318,400)(10,180,360){2}{5}
\Photon(300,380)(300,400){-2}{3}
\Photon(350,380)(350,400){2}{3}
\put(325,370){\makebox(0,0)[c]{(e)}}
 
\ArrowLine(375,430)(375,400)
\Line(375,400)(425,400)
\Line(425,400)(425,430)
\Gluon(425,420)(385,400){-2}{7}
\GlueArc(407,400)(10,180,360){2}{5}
\Photon(375,380)(375,400){-2}{3}
\Photon(425,380)(425,400){2}{3}
\put(400,370){\makebox(0,0)[c]{(f)}}
 
 
\Line(  0,355)(  0,325)
\ArrowLine(  0,325)( 50,325)
\Line( 50,325)( 50,355)
\Gluon(  0,340)( 50,340){2}{8}
\GlueArc( 25,325)(10,180,360){2}{5}
\Photon( 0,305)( 0,325){-2}{3}
\Photon(50,305)(50,325){2}{3}
\put( 25,295){\makebox(0,0)[c]{(g)}}
 
\Line( 75,355)( 75,325)
\ArrowLine( 75,325)(125,325)
\Line(125,325)(125,355)
\Gluon( 75,345)(115,325){2}{7}
\Gluon(125,345)(106,335){-2}{3}
\Gluon( 95,329)( 85,325){-2}{1}
\Photon( 75,305)( 75,325){-2}{3}
\Photon(125,305)(125,325){2}{3}
\put(100,295){\makebox(0,0)[c]{(h)}}
 
\Line(150,355)(150,325)
\Line(150,325)(200,325)
\ArrowLine(200,325)(200,355)
\Photon(150,305)(150,325){-2}{3}
\Photon(200,305)(200,325){2}{3}
\GlueArc(150,325)(11,  0, 90){2}{2}
\GlueArc(150,325)(20,  0, 90){2}{4}
\put(175,295){\makebox(0,0)[c]{(i)}}

\ArrowLine(225,355)(225,325)
\Line(225,325)(275,325)
\Line(275,325)(275,355)
\GlueArc(275,325)(11, 90,180){2}{2}
\GlueArc(275,325)(20, 90,180){2}{4}
\Photon(225,305)(225,325){-2}{3}
\Photon(275,305)(275,325){2}{3}
\put(250,295){\makebox(0,0)[c]{(j)}}
 
\Line(300,355)(300,325)
\Line(300,325)(350,325)
\ArrowLine(350,325)(350,355)
\Gluon(300,335)(307,335){2}{1}
\GlueArc(325,325)(10,  0, 90){2}{2}
\GlueArc(300,325)(20,  0, 90){2}{4}
\Photon(300,305)(300,325){-2}{3}
\Photon(350,305)(350,325){2}{3}
\put(325,295){\makebox(0,0)[c]{(k)}}
 
\ArrowLine(375,355)(375,325)
\Line(375,325)(425,325)
\Line(425,325)(425,355)
\Gluon(425,335)(418,335){-2}{1}
\GlueArc(400,325)(10, 90,180){2}{2}
\GlueArc(425,325)(20, 90,180){2}{4}
\Photon(375,305)(375,325){-2}{3}
\Photon(425,305)(425,325){2}{3}
\put(400,295){\makebox(0,0)[c]{(l)}}
 

\Line(  0,280)(  0,250)
\ArrowLine(  0,250)( 50,250)
\Line( 50,250)( 50,280)
\Gluon(  0,260)( 50,260){2}{8}
\Gluon(  0,270)( 50,270){2}{8}
\Photon( 0,230)(  0,250){-2}{3}
\Photon(50,230)( 50,250){2}{3}
\put( 25,220){\makebox(0,0)[c]{(m)}}
 
\Line( 75,280)( 75,250)
\ArrowLine( 75,250)(125,250)
\Line(125,250)(125,280)
\Gluon( 75,270)(125,260){2}{8}
\Gluon( 75,260)( 90,262){2}{2}
\Gluon(110,268)(125,270){2}{2}
\Photon( 75,230)( 75,250){-2}{3}
\Photon(125,230)(125,250){2}{3}
\put(100,220){\makebox(0,0)[c]{(n)}}
 
\Line(150,280)(150,250)
\ArrowLine(150,250)(200,250)
\Line(200,250)(200,280)
\Gluon(150,270)(200,270){2}{8}
\GlueArc(150,250)(11,  0, 90){2}{2}
\Photon(150,230)(150,250){-2}{3}
\Photon(200,230)(200,250){2}{3}
\put(175,220){\makebox(0,0)[c]{(o)}}
 
\Line(225,280)(225,250)
\ArrowLine(225,250)(275,250)
\Line(275,250)(275,280)
\Gluon(225,270)(275,270){2}{8}
\GlueArc(275,250)(11, 90,180){2}{2}
\Photon(225,230)(225,250){-2}{3}
\Photon(275,230)(275,250){2}{3}
\put(250,220){\makebox(0,0)[c]{(p)}}

\Line(300,280)(300,250)
\Line(300,250)(320,250)
\ArrowLine(320,250)(350,250)
\Line(350,250)(350,280)
\Gluon(324,260)(350,260){2}{4}
\Gluon(300,260)(310,260){2}{1}
\GlueArc(300,250)(20,  0, 90){2}{4}
\Photon(300,230)(300,250){-2}{3}
\Photon(350,230)(350,250){2}{3}
\put(325,220){\makebox(0,0)[c]{(q)}}
 
\Line(375,280)(375,250)
\ArrowLine(375,250)(405,250)
\Line(405,250)(425,250)
\Line(425,250)(425,280)
\Gluon(375,260)(401,260){2}{4}
\Gluon(415,260)(425,260){2}{1}
\GlueArc(425,250)(20, 90,180){2}{4}
\Photon(375,230)(375,250){-2}{3}
\Photon(425,230)(425,250){2}{3}
\put(400,220){\makebox(0,0)[c]{(r)}}
 

\Line( 75,205)( 75,165)
\ArrowLine( 75,165)(125,165)
\Line(125,165)(125,205)
\Gluon( 75,196)(125,196){2}{8}
\GlueArc( 75,180)( 8,-90, 90){2}{4}
\Photon( 75,155)( 75,165){2}{1.5}
\Photon(125,155)(125,165){-2}{1.5}
\put(100,145){\makebox(0,0)[c]{(s)}}
 
\Line(150,205)(150,165)
\Line(150,165)(200,165)
\ArrowLine(200,165)(200,205)
\GlueArc(150,180)( 8,-90, 90){2}{4}
\GlueArc(150,165)(31, 0, 90){2}{8}
\Photon(150,155)(150,165){2}{1.5}
\Photon(200,155)(200,165){-2}{1.5}
\put(175,145){\makebox(0,0)[c]{(t)}}
 
\Line(225,205)(225,165)
\ArrowLine(225,165)(275,165)
\Line(275,165)(275,205)
\GlueArc(225,185)(12,-90, 90){2}{5}
\Gluon(225,185)(232,185){2}{1}
\Gluon(242,185)(275,185){2}{5}
\Photon(225,155)(225,165){2}{1.5}
\Photon(275,155)(275,165){-2}{1.5}
\put(250,145){\makebox(0,0)[c]{(u)}}
 
\Line(300,205)(300,165)
\Line(300,165)(350,165)
\ArrowLine(350,165)(350,205)
\GlueArc(300,185)(12,-90, 90){2}{5}
\Gluon(300,185)(307,185){2}{1}
\GlueArc(317,165)(20,  0, 90){2}{5}
\Photon(300,155)(300,165){2}{1.5}
\Photon(350,155)(350,165){-2}{1.5}
\put(325,145){\makebox(0,0)[c]{(v)}}
 
 
\Line( 75,130)( 75, 90)
\ArrowLine( 75, 90)(125,90)
\Line(125, 90)(125,130)
\Gluon( 75,121)(125,121){2}{8}
\GlueArc(125,105)( 8, 90,270){2}{4}
\Photon( 75, 80)( 75, 90){2}{1.5}
\Photon(125, 80)(125, 90){-2}{1.5}
\put(100, 70){\makebox(0,0)[c]{(w)}}
 
\ArrowLine(150,130)(150,90)
\Line(150, 90)(200,90)
\Line(200, 90)(200,130)
\GlueArc(200,105)( 8, 90,270){2}{4}
\GlueArc(200, 90)(31, 90,180){2}{8}
\Photon(150, 80)(150, 90){2}{1.5}
\Photon(200, 80)(200, 90){-2}{1.5}
\put(175, 70){\makebox(0,0)[c]{(x)}}
 
\Line(225,130)(225,90)
\ArrowLine(225, 90)(275,90)
\Line(275, 90)(275,130)
\GlueArc(275,110)(12, 90,270){2}{5}
\Gluon(225,110)(258,110){2}{5}
\Gluon(268,110)(275,110){2}{1}
\Photon(225, 80)(225, 90){2}{1.5}
\Photon(275, 80)(275, 90){-2}{1.5}
\put(250, 70){\makebox(0,0)[c]{(y)}}
 
\ArrowLine(300,130)(300,90)
\Line(300, 90)(350,90)
\Line(350, 90)(350,130)
\GlueArc(350,110)(12, 90,270){2}{5}
\Gluon(343,110)(350,110){2}{1}
\GlueArc(333, 90)(20, 90,180){2}{5}
\Photon(300, 80)(300, 90){2}{1.5}
\Photon(350, 80)(350, 90){-2}{1.5}
\put(325, 70){\makebox(0,0)[c]{(z)}}
 

\ArrowLine( 75, 55)( 75, 25)
\Line( 75, 25)(125, 25)
\ArrowLine(125, 25)(125, 55)
\GlueArc( 88.5, 25)(8,  0,180){2}{4}
\GlueArc(111.5, 25)(8,  0,180){2}{4}
\Photon( 75,  5)( 75, 25){-2}{3}
\Photon(125,  5)(125, 25){2}{3}
\put(100, -5){\makebox(0,0)[c]{(a$^{\prime}$)}}
 
\Line(150, 55)(150, 25)
\Line(150, 25)(200, 25)
\ArrowLine(200, 25)(200, 55)
\GlueArc(150, 25)(11,  0, 90){2}{2}
\GlueArc(182, 25)(10,  0,180){2}{5}
\Photon(150,  5)(150, 25){-2}{3}
\Photon(200,  5)(200, 25){2}{3}
\put(175, -5){\makebox(0,0)[c]{(b$^{\prime}$)}}
 
\ArrowLine(225, 55)(225, 25)
\Line(225, 25)(275, 25)
\Line(275, 25)(275, 55)
\GlueArc(275, 25)(11, 90,180){2}{2}
\GlueArc(243, 25)(10,  0,180){2}{5}
\Photon(225,  5)(225, 25){-2}{3}
\Photon(275,  5)(275, 25){2}{3}
\put(250, -5){\makebox(0,0)[c]{(c$^{\prime}$)}}

\Line(300, 55)(300, 25)
\ArrowLine(300, 25)(350, 25)
\Line(350, 25)(350, 55)
\GlueArc(300, 25)(11,  0, 90){2}{2}
\GlueArc(350, 25)(11, 90,180){2}{2}
\Photon(300,  5)(300, 25){-2}{3}
\Photon(350,  5)(350, 25){2}{3}
\put(325, -5){\makebox(0,0)[c]{(d$^{\prime}$)}}
 

\Line( 75,-20)( 75,-60)
\Line( 75,-60)(125,-60)
\ArrowLine(125,-60)(125,-20)
\GlueArc( 75,-35.5)( 8.5,-90, 90){2}{4}
\GlueArc(100,-60)(10,  0,180){2}{5}
\Photon( 75,-70)( 75,-60){2}{1.5}
\Photon(125,-70)(125,-60){-2}{1.5}
\put(100,-80){\makebox(0,0)[c]{(e$^{\prime}$)}}
 
\Line(150,-20)(150,-60)
\Line(150,-60)(200,-60)
\ArrowLine(200,-60)(200,-20)
\GlueArc(150,-35)( 8,-90, 90){2}{4}
\GlueArc(150,-60)(11,  0, 90){2}{2}
\Photon(150,-70)(150,-60){2}{1.5}
\Photon(200,-70)(200,-60){-2}{1.5}
\put(175,-80){\makebox(0,0)[c]{(f\,$^{\prime}$)}}
 
\Line(225,-20)(225,-60)
\ArrowLine(225,-60)(275,-60)
\Line(275,-60)(275,-20)
\GlueArc(225,-35.5)( 8.5,-90, 90){2}{4}
\GlueArc(275,-60)(11, 90,180){2}{2}
\Photon(225,-70)(225,-60){2}{1.5}
\Photon(275,-70)(275,-60){-2}{1.5}
\put(250,-80){\makebox(0,0)[c]{(g$^{\prime}$)}}
 
\Line(300,-20)(300,-60)
\ArrowLine(300,-60)(350,-60)
\Line(350,-60)(350,-20)
\GlueArc(300,-35)( 8,-90, 90){2}{4}
\Gluon(300,-52)(350,-52){2}{8}
\Photon(300,-70)(300,-60){2}{1.5}
\Photon(350,-70)(350,-60){-2}{1.5}
\put(325,-80){\makebox(0,0)[c]{(h$^{\prime}$)}}
 
 
\ArrowLine( 75, -95)( 75,-135)
\Line( 75,-135)(125,-135)
\Line(125,-135)(125, -95)
\GlueArc(125,-110.5)( 8.5, 90,270){2}{4}
\GlueArc(100,-135)(10,  0,180){2}{5}
\Photon( 75,-145)( 75,-135){2}{1.5}
\Photon(125,-145)(125,-135){-2}{1.5}
\put(100,-155){\makebox(0,0)[c]{(i$^{\prime}$)}}
 
\ArrowLine(150,-95)(150,-135)
\Line(150,-135)(200,-135)
\Line(200,-135)(200, -95)
\GlueArc(200,-110)( 8, 90,270){2}{4}
\GlueArc(200,-135)(11, 90,180){2}{2}
\Photon(150,-145)(150,-135){2}{1.5}
\Photon(200,-145)(200,-135){-2}{1.5}
\put(175,-155){\makebox(0,0)[c]{(j$^{\prime}$)}}
 
\Line(225, -95)(225,-135)
\ArrowLine(225,-135)(275,-135)
\Line(275,-135)(275, -95)
\GlueArc(275,-110.5)( 8.5, 90,270){2}{4}
\GlueArc(225,-135)(11,  0, 90){2}{2}
\Photon(225,-145)(225,-135){2}{1.5}
\Photon(275,-145)(275,-135){-2}{1.5}
\put(250,-155){\makebox(0,0)[c]{(k$^{\prime}$)}}
 
\Line(300, -95)(300,-135)
\ArrowLine(300,-135)(350,-135)
\Line(350,-135)(350, -95)
\GlueArc(350,-110)( 8, 90,270){2}{4}
\Gluon(300,-127)(350,-127){2}{8}
\Photon(300,-145)(300,-135){2}{1.5}
\Photon(350,-145)(350,-135){-2}{1.5}
\put(325,-155){\makebox(0,0)[c]{(l$^{\prime}$)}}

\put(-8,-175){\makebox(0,0)[l]{\small
Fig.\ 17. The two-loop one-particle-irreducible
(17a--z) and one-particle-reducible 
(17a$^{\prime}$--l$^{\prime}$) gluonic }}

\put(-8,-190){\makebox(0,0)[l]{\small
corrections to the process
$q\gamma \rightarrow q\gamma$ consisting of diagrams with
no triple gauge vertices (the diagrams }}

\put(-8,-205){\makebox(0,0)[l]{\small
representing purely external leg corrections are not shown).}}
 
\end{picture}
\end{center}

\pagebreak

\subsection{Remarks}

The analysis presented in this section may be 
summarized and commented upon as follows:

(1) In section 4.1, the longitudinal $k_{1\rho}k_{1\rho'}$
component of the pinch part
$-\Delta\Pi_{\rho\rho'}^{\prime(1)}(k_{1},k_{2},1)$
of the Feynman gauge
one-loop gluon self-energy insertion exactly
cancelled among the integrands in 
Eqs.~(\ref{fig14ad}) and (\ref{fig14ef}) for the
diagrams 14a--f.
This algebraic cancellation is identical to the cancellation
of the longitudinal $k_{\rho}k_{\rho'}$
component of the lowest order gluon propagator
among the integrands in 
Eqs.~(\ref{fig3ad}) and (\ref{fig3ef})
for the diagrams 3a--f, 
described in section 2 for the construction
of the PT one-loop quark self-energy.

(2) In section 4.2, the elementary decomposition \eq{GammaFFFP} of the
triple gauge vertex, 
substituted into \eq{fig15ad} for the diagrams 15a--d,
provided {\em precisely}\, the required contributions to
$-i\hat{\Sigma}^{(2)}(p)$ of the diagrams 10c--e 
involving the PT one-loop gluon self-energy and quark-gluon vertex
internal corrections, and also a contribution to the diagram 10f
involving the PT quark-gluon kernel insertion:

\begin{itemize}

\item
The first term $-2\Gamma_{\tau\rho\sigma}^{P}(k_{3};k_{1},k_{2})$
provided the longitudinal factors which gave the
two-loop self-energy-like (``effective'' two-point)
contributions shown in Figs.~16i and 16j,
involving internal one-loop gluon self-energy-like pinch parts.
These contributions exactly cancelled
the $g_{\rho\rho'}$ component of the pinch part
$-\Delta\Pi_{\rho\rho'}^{\prime(1)}(k_{1},k_{2},1)$
of the Feynman gauge one-loop gluon self-energy
appearing in Fig.~14a.
This cancellation, together with that described in
section 4.1, left just the required 
contribution to $-i\hat{\Sigma}^{(2)}(p)$
of the diagram 10c, involving the PT one-loop
gluon self-energy insertion.

\item 
The second and third terms
$\Gamma_{\rho\sigma\tau}^{F}(k_{1};k_{2},k_{3})$,
$\Gamma_{\sigma\tau\rho}^{F}(k_{2};k_{3},k_{1})$
gave the two-loop self-energy-like
contributions shown in Figs.~16k and 16l.
These contributions provided just the required
triple gauge vertex component of the contribution
to $-i\hat{\Sigma}^{(2)}(p)$ of the diagrams 10d and 10e, 
involving the PT one-loop quark-gluon vertex insertions.

\item 
The last term $-\Gamma_{\tau\rho\sigma}^{F}(k_{3};k_{1},k_{2})$
gave the two-loop 
self-energy-like contribution shown in Fig.~16m.
Having obtained the required terms
involving triple gauge vertices in the contributions
to $-i\hat{\Sigma}^{(2)}(p)$ of the diagrams 10c--e,
this term provided a contribution to
the diagram 10f, involving the PT tree level
quark-gluon kernel insertion.
This contribution to the PT quark-gluon
kernel is shown in Fig.~13b.

\end{itemize}
It is remarkable that the elementary decomposition 
\eq{GammaFFFP} of the triple gauge vertex
not only results in the required
components of the diagrams 10c--e, 
but also that the remaining two-loop fermion
self-energy-like contribution, allocated to
the PT kernel insertion diagram 10f, involves
just the component $\Gamma^{F}$ of the triple gauge vertex,
and appears automatically with the appropriate minus sign.
Thus, as in the PT at one loop,
{\em all}\, tree level triple gauge vertices
which occur in the PT two-loop quark self-energy
$-i\hat{\Sigma}^{(2)}(p)$ involve only the
component $\Gamma^{F}$ of the triple gauge vertex,
given in \eq{GammaF}.

(3) In section 4.3, the diagrams
17a and 17b provided just 
the required QED-like components of the contributions to
$-i\hat{\Sigma}^{(2)}(p)$ of the diagrams 10b, 10d and 10e
involving the PT one-loop fermion 
self-energy and quark-gluon vertex
internal corrections, and also a second 
contribution to the diagram 10f
involving the PT quark-gluon kernel insertion.
This contribution to the PT quark-gluon kernel is
shown in Fig.~13c.
Given that in the Feynman gauge there occur
no longitudinal factors in the integrands for the 
diagrams of Fig.~17, this was trivial.

(4) The resulting PT tree level Bethe-Salpeter-type quark-gluon 
kernel thus corresponds, as stated in section~3, 
to the diagrams shown in Fig.~13:
\bea
{\rm Fig.\,13a}
\,\,=\,\,
i\hat{V}_{\rho\sigma}^{(0)rs}(p,k_{1},k_{2})
\!\!&=&\!\!
ig^{2}\Bigl\{ 
if^{mrs}T^{m}\Gamma_{\mu\rho\sigma}^{F}(-k_{1}-k_{2};k_{1},k_{2})
D^{\mu\nu}(k_{1}+k_{2},1) \gamma_{\nu} 
\nn \\
& &\!\!
-\,T^{r}T^{s}\gamma_{\rho}S(p-k_{1}+k_{2})\gamma_{\sigma}
\Bigr\}\,\,.
\label{fig13a}
\eea
It is emphasized that, having obtained the required 
contributions to 
$-i\hat{\Sigma}^{(2)}(p)$ shown in Figs.~10b--e,
the PT kernel insertion contribution shown in Fig.~10f emerged
immediately as the remaining
self-energy-like component of the two-loop diagrams,
without any further rearrangement.
It is seen from Fig.~13 that the PT kernel
indeed has no contributions involving
annihilation into a one-particle (quark) intermediate state.
Furthermore, and as already stated in section~3,
the fact that the diagram 13b involves only the component 
$\Gamma_{\mu\rho\sigma}^{F}$ of the triple gauge vertex, 
with the internal gluon propagator as in the Feynman gauge, 
means that {\em the PT tree level quark-gluon kernel  
$\hat{V}_{\rho\sigma}^{(0)rs}(p,k_{1},k_{2})$ 
provides no factors of longitudinal gluon four-momentum
$k_{1\rho}$, $k_{2\sigma}$ associated with the external gluon legs 
$A_{\rho}^{r}(k_{1})$, $A_{\sigma}^{s}(k_{2})$}.
When inserted in some diagram, 
e.g.\ as in Fig.~10f, the PT kernel therefore does not trigger 
any Ward identities, and so does not cause any
further PT rearrangement (it triggers no ``pinching'').

With hindsight, this is seen to be crucial for the consistency
of the PT algorithm at the two-loop level.
For if, having obtained the required contributions to 
$-i\hat{\Sigma}^{(2)}(p)$ shown in Figs.~10b--e,
thus accounting for the first three terms on 
the r.h.s.\ of the decomposition (\ref{GammaFFFP})
of the triple gauge vertex appearing in Fig.~15k,
the remaining term in (\ref{GammaFFFP}) {\em had}\, involved
further longitudinal factors 
$k_{1\rho}$, $k_{2\sigma}$, 
then these factors would have produced further internal
one-loop pinch parts.
The allocation of the contributions to
Figs.~10c--e would then have been ambiguous.

(5) The conventional tree level 
quark-gluon kernel shown in Figs.~9m--o 
can be written in terms of the PT kernel as
\bea
{\rm Fig.\,9m}
&=&
iV_{\rho\sigma}^{(0)rs}(p,k_{1},k_{2},\xi) \\
&=&
i\hat{V}_{\rho\sigma}^{(0)rs}(p,k_{1},k_{2})
\,+\, g^{2}f^{rsm}T^{m}\frac{1}{(k_{1}+k_{2})^{2}}
\biggl\{
-k_{1\rho}\gamma_{\sigma} +k_{2\sigma}\gamma_{\rho} 
\nn \\
& & 
-\, (1-\xi)
\Bigl( k_{1}^{2}t_{\rho\sigma}(k_{1})
      -k_{2}^{2}t_{\rho\sigma}(k_{2}) \Bigr) 
\frac{1}{(k_{1}+k_{2})^{2}}(\ks_{1}+\ks_{2})
\biggr\}\,\,.\qquad
\eea
For the case in which the gluon legs 
$A_{\rho}^{r}(k_{1})$, $A_{\sigma}^{s}(k_{2})$
are on-shell, contracting the above expression with
polarization vectors 
$\epsilon^{\rho}(k_{1})$, $\epsilon^{*\sigma}(-k_{2})$
for the incoming and outgoing gluon legs, and using
$k_{1}\cdot\epsilon(k_{1}) = 
k_{2}\cdot\epsilon^{*}(-k_{2}) =
k_{1}^{2} = k_{2}^{2} = 0$, we obtain
\be
V_{\rho\sigma}^{(0)rs}(p,k_{1},k_{2},\xi)
\epsilon^{\rho}(k_{1})\epsilon^{*\sigma}(-k_{2})
=
\hat{V}_{\rho\sigma}^{(0)rs}(p,k_{1},k_{2})
\epsilon^{\rho}(k_{1})\epsilon^{*\sigma}(-k_{2})\,\,.
\ee
Thus, for the case of tree level quark-gluon 
scattering in which both the
incoming and outgoing gluons are on-shell,
the amplitudes given by the PT and conventional kernels 
for the contributions not involving
a one-particle (quark) intermediate state exactly coincide.

(6) Although not explicitly considered here,
it requires only a little more effort to obtain the
two-loop QCD contributions to the PT quark-photon vertex
(``effective'' three-point function),
quark-photon box (``effective'' four-point function)
and quark external leg corrections to
$q\gamma \rightarrow q\gamma$.
In contrast to the case of the self-energy component,
some or all of the external legs of these functions are 
on-shell.
In the particular case of the QCD contribution to 
PT two-loop quark-photon vertex
$\hat{\Gamma}_{\mu}^{(2)}$,
this function is given by attaching 
an external photon leg in all possible ways 
to the internal fermion lines occurring in the diagrams for the
PT two-loop quark self-energy $-i\hat{\Sigma}^{(2)}$.
It is then straightforward to verify the simple
QED-like Ward identity for these two-loop functions:
\be
q^{\mu}\hat{\Gamma}^{(2)}_{\mu}(q,p_{1},p_{2})
=
\hat{\Sigma}^{(2)}(p_{1}) - \hat{\Sigma}^{(2)}(p_{2}) 
\ee
where $q+p_{1} = p_{2}$. Furthermore, the two-loop
external quark leg corrections are given by precisely
the same self-energy function $-i\hat{\Sigma}^{(2)}$
as that just obtained from the ``effective'' two-point
component of the process.
These two facts are also essential for the consistency of
the PT approach.

 
\setcounter{equation}{0}

\section{Gauge independence of $-i\hat{\Sigma}^{(2)}(p)$}


In this section, we again consider the first two steps
in the construction of $-i\hat{\Sigma}^{(2)}(p)$ outlined 
at the end of section 3, but now starting from an
arbitrary linear covariant gauge, i.e.\ arbitrary $\xi$. 
It is explicitly shown that the PT two-loop quark self-energy
constructed in the previous section starting from the Feynman
gauge $\xi = 1$ is in fact $\xi$-independent.
The strategy is to show that the additional contributions
to the PT self-energy-like
(``effective'' two-point) component of the interaction
$q\gamma \rightarrow q\gamma$ which occur when one moves
away from the Feynman gauge exactly cancel among themselves.
This cancellation takes place at the level of the two-loop 
integrands. 
In order to demonstrate this cancellation, it will again be 
convenient to deal separately with the three classes of 
two-loop QCD corrections to $q\gamma \rightarrow q\gamma$
considered in sections 4.1, 4.2 and 4.3, respectively.

\subsection{Two-loop corrections involving one-loop gluon self-energy
insertions}

The set of diagrams for the two-loop corrections to 
$q\gamma \rightarrow q\gamma$ involving unrenormalized
one-loop gluon self-energy
insertions are as shown already in Figs.~14a--f.
Using the usual covariant gauge Feynman rules
for the diagram shown in Fig.~9e, the integrand 
in \eq{Pixi} for the conventional one-loop
covariant gauge gluon self-energy may be written
\bea
i\Pi_{\mu\nu}^{\prime(1)}(q,k,\xi)
&=&
i\Pi_{\mu\nu}^{\prime(1)}(q,k,1)
\,+\, 
C_{\!A}g^{2}\frac{1}{k^{2}(k+q)^{2}}
\Gamma_{\mu\rho\sigma}(q,k,-k-q)
\Gamma_{\nu\rho'\!\sigma'\!}(-q,-k,k+q) \nn \\
& & \times
\biggl\{
(1-\xi)l^{\rho\rho'\!}(k)g^{\sigma\sigma'\!}
-\frac{1}{2}
(1-\xi)^{2}l^{\rho\rho'\!}(k)l^{\sigma\sigma'\!}(k+q)
\biggr\}\,\,,
\label{Piprimexi}
\eea
where $k$ and $k+q$ are the four-momenta of the gluons
propagating in the loop in Fig.~9e.
The contraction of the longitudinal terms
$l^{\rho\rho'\!}(k)$, $l^{\sigma\sigma'\!}(k+q)$ in \eq{Piprimexi}
is most easily carried out using 
the elementary Ward identities for the triple gauge vertex:
\bea
q_{1}^{\rho}\Gamma_{\rho\sigma\tau}(q_{1},q_{2},q_{3})
&=&
 q_{3}^{2}t_{\sigma\tau}(q_{3}) 
-q_{2}^{2}t_{\sigma\tau}(q_{2}) \,\,, \label{tgvwid1} \\
q_{1}^{\rho}q_{2}^{\sigma}\Gamma_{\rho\sigma\tau}(q_{1},q_{2},q_{3})
&=&
\frac{1}{2}(q_{2}-q_{1})^{\tau'\!}q_{3}^{2}t_{\tau'\!\tau}(q_{3})\,\,.
\label{tgvwid2}
\eea

The pinch part of the one-loop covariant gauge 
gluon self-energy integrand \eq{Piprimexi},
defined in \eq{Pidecomp} and given 
in Eqs.~(\ref{Piprimemunupinch1}) and (\ref{Piprimepinch1})
for $\xi = 1$, may be written for arbitrary $\xi$ as
\be
\Delta\Pi_{\mu\nu}^{\prime(1)}(q,k,\xi)
=
\Delta\Pi_{\mu\nu}^{\prime(1)}(q,k,1)
+
\sum_{i=1}^{3}\Delta\Pi_{i,\mu\nu}^{\prime(1)}(q,k,\xi)\,\,.
\label{Piprimepinchxi}
\ee
In \eq{Piprimepinchxi}, the contributions to
$\Pi_{\mu\nu}^{\prime(1)}(q,k,\xi)$, hence
$\Delta\Pi_{\mu\nu}^{\prime(1)}(q,k,\xi)$,
which occur for $\xi \neq 1$ have been written
as three distinct components.
In the construction of the PT one-loop gauge-independent
gluon self-energy starting from a four-fermion process,
these terms are distinguished as follows: the component
$\Delta\Pi_{1,\mu\nu}^{\prime(1)}(q,k,\xi)$
is that which is cancelled by the $\xi \neq 1$
self-energy-like pinch parts of the conventional 
one-loop vertex diagrams involving the triple gauge vertex;
the component
$\Delta\Pi_{2,\mu\nu}^{\prime(1)}(q,k,\xi)$
is that which is cancelled by the $\xi \neq 1$
self-energy-like pinch parts from the conventional 
one-loop box diagrams plus the conventional 
one-loop vertex diagrams not involving the triple gauge vertex
(i.e.\ the QED-like diagrams); and the remaining component
$\Delta\Pi_{3,\mu\nu}^{\prime(1)}(q,k,\xi)$,
purely longitudinal i.e.\ proportional to $q_{\mu}q_{\nu}$,
vanishes when the external fermions are on-shell,
as is the case in the $S$-matrix PT.
These three components are given by\footnote{
In \eq{DP3},
the dimensional regularization rule $\int[dk]\,k^{-2} = 0$
has been used to drop terms which vanish upon integration.}
\bea
\Delta\Pi_{1,\mu\nu}^{\prime(1)}(q,k,\xi)
&=& 
2A\Bigl\{q^{2}t_{\mu\nu}(q)- (k+q)^{2}t_{\mu\nu}(k+q)\Bigr\}
\,+\,B_{\rho\rho'}
\Bigl\{g_{\mu}^{\rho}t_{\nu}^{\rho'\!}(q)
+ t_{\mu}^{\rho}(q)g_{\nu}^{\rho'\!} \Bigr\} \,\,, \qquad
\label{DP1} \\
\Delta\Pi_{2,\mu\nu}^{\prime(1)}(q,k,\xi)
&=& 
A\Bigl\{(k+q)^{2} - q^{2}\Bigr\}g_{\mu\nu}
\,-\,B_{\mu\nu} \,\,, 
\label{DP2} \\
\Delta\Pi_{3,\mu\nu}^{\prime(1)}(q,k,\xi)
&=& 
A\Bigl\{
q^{2} - (k+q)^{2}[1 - 2l_{\rho\rho'\!}(q)t^{\rho\rho'\!}(k+q) ]
\Bigr\}l_{\mu\nu}(q) 
\,+\,B_{\rho\rho'}
l_{\mu}^{\rho}(q)l_{\nu}^{\rho'\!}(q) \,\,.
\label{DP3}
\eea
In the above expressions, for brevity we have introduced 
\bea
A
&=&
-i(1-\xi)C_{\!A}g^{2}\frac{1}{k^{4}(k+q)^{2}}\,q^{2}\,\,, 
\label {A} \\
B_{\rho\rho'}
&=&
i(1-\xi)^{2}C_{\!A}g^{2}
\frac{(2k+q)_{\rho}(2k+q)_{\rho'}}{8k^{4}(k+q)^{4}}\,q^{4}\,\,.
\label{B}
\eea

Substituting \eq{Piprimepinchxi} into \eq{fig14ad},
the sum of the four diagrams 14a--d for arbitrary $\xi$
can be written as
\bea
{\rm Figs.\,14a\!-\!d}
&=&
{\rm Figs.\,14a\!-\!d}|_{\xi =1} \,+\,
Q^{2}g^{2}\int[dk_{1}][dk_{2}]\,
\frac{1}{k_{1}^{4}}
\sum_{i=1}^{3}\Delta\Pi_{i}^{\prime(1)\rho\rho'\!}(k_{1},k_{2},\xi) \nn\\
& & \times 
G_{\rho\mu}^{r}(k_{1},-p_{3},p-k_{1},p_{4})
S(p-k_{1})
G_{\rho'\!\nu}^{r}(-k_{1},p_{1},p_{2},p-k_{1})\,\,.
\label{fig14adxi}
\eea

For the purely longitudinal pinch part term
$\Delta\Pi_{3}^{\prime(1)\rho\rho'\!}(k_{1},k_{2},\xi)$
in \eq{fig14adxi}, proportional to $k_{1}^{\rho}k_{1}^{\rho'}$,
the effect of the longitudinal factors is precisely
analogous to the effect of the longitudinal factors 
from the tree level gluon
propagator appearing in \eq{fig3ad} for Figs.~3a--d,
or the longitudinal factors from the pinch part 
$k_{1}^{2}t^{\rho\rho'\!}(k_{1})
\Delta\Pi^{\prime(1)}(k_{1},k_{2},1)$
of the one-loop-corrected gluon propagator
in \eq{fig14ad2} for Figs.~14a--d in the Feynman gauge.
Thus, using the Ward identity \eq{G4wid}, 
the $i=3$ term in
\eq{fig14adxi} results in a $p$-independent contribution
which is cancelled algebraically by two corresponding
contributions from the expression \eq{fig14ef}
for Figs.~14e and 14f. The term 
$\Delta\Pi_{3}^{\prime(1)\rho\rho'\!}(k_{1},k_{2},\xi)$
therefore makes no net contribution to the process
$q\gamma \rightarrow q\gamma$. 

The remaining $i=1,2$
pinch part terms in \eq{fig14adxi} are neither
purely longitudinal nor purely proportional
to $g^{\rho\rho'\!}$.
At this point, we could project out the
$k_{1}^{\rho}k_{1}^{\rho'}$ and $g^{\rho\rho'\!}$
components of 
$\Delta\Pi_{i}^{\prime(1)\rho\rho'\!}(k_{1},k_{2},\xi)$,
$i = 1,2$.
The purely longitudinal components would then each give
no net contribution, as in the case of 
$\Delta\Pi_{3}^{\rho\rho'\!}(k_{1},k_{2},\xi)$.
This would leave just the $g^{\rho\rho'\!}$ components
in \eq{fig14adxi}. 
In Fig.~14a, these latter
are then  a contribution to the PT self-energy-like
component of the process,
in Figs.~14b and 14c they are contributions 
to the PT vertex-like component of the process
and in Fig.~14d they are a contribution to the PT box-like
component of the process.
The task would then be to show that these 
$g^{\rho\rho'\!}$ components are cancelled by corresponding
terms from the remaining two-loop diagrams in
Figs.~15 and 17.
However, we shall show in the next two subsections
that the full functions 
$\Delta\Pi_{i}^{\prime(1)\rho\rho'\!}(k_{1},k_{2},\xi)$,
$i = 1,2$, in fact emerge naturally from the remaining
two-loop diagrams {\em in exactly the form of
\eq{fig14adxi}}\, but with the opposite sign.
The cancellation of the one-loop gluon self-energy pinch parts 
$\Delta\Pi_{i}^{\prime(1)\rho\rho'\!}(k_{1},k_{2},\xi)$,
$i = 1,2$, therefore occurs immediately,
and there is thus no need to do make this projection.

\subsection{Two-loop diagrams with one triple gauge vertex}

The subset of two-loop QCD corrections to
$q\gamma \rightarrow q\gamma$ involving one triple gauge
vertex are as shown already in Fig.~15
(purely external leg corrections are not shown).
As indicated in Fig.~15, 
the sum of these eight two-loop diagrams
may be written in terms of the four-point function (\ref{G4}):
\bea
\lefteqn{ {\rm Figs.\,15a\!-\!h}
\,\,=\,\,
Q^{2}g^{4}\int[dk_{1}][dk_{2}]\,
D^{\rho\rho'\!}(k_{1},\xi)
D^{\sigma\sigma'\!}(k_{2},\xi)
D^{\tau\tau'\!}(k_{3},\xi)
f^{rst}\Gamma_{\rho'\!\sigma'\!\tau'\!}(k_{1},k_{2},k_{3}) 
} \nn \\
& &
\times \biggl\{
G_{\sigma\mu}^{s}(-k_{2},-p_{3},p+k_{2},p_{4})
S(p+k_{2})\gamma_{\tau}T^{t}S(p-k_{1})
G_{\rho\nu}^{r}(-k_{1},p_{1},p_{2},p-k_{1}) \,\,\nn \\
& & -
G_{\sigma\mu}^{s}(-k_{2},-p_{3},p+k_{2},p_{4})
S(p+k_{2})\gamma_{\nu}S(p_{2} + k_{2})\gamma_{\tau}T^{t}
S(p_{2}-k_{1})\gamma_{\rho}T^{r} \nn \\
& & -
\gamma_{\sigma}T^{s}S(p_{4}+k_{2})\gamma_{\tau}T^{t}S(p_{4}-k_{1})
\gamma_{\mu}S(p-k_{1})
G_{\rho\nu}^{r}(-k_{1},p_{1},p_{2},p-k_{1}) \biggl\}\,\,,
\label{fig15ahxi}
\eea
with $k_{1} + k_{2} + k_{3} = 0$.
In the above expression, the first term on the r.h.s.\
corresponds to the sum of the diagrams 15a--d,
the second term to the diagrams 15e + 15f,
and the third term to the diagrams 15g + 15h.
\eq{fig15ahxi} is just the generalization to $\xi \neq 1$
of Eqs.~(\ref{fig15ad}) and (\ref{fig15eh}).

Using the expression (\ref{D}) together with the 
Ward identities (\ref{tgvwid1}) and (\ref{tgvwid2}),
the contraction of the three gluon propagators with
the triple gauge vertex in \eq{fig15ahxi} may be written
in the form

\pagebreak

\bea
\lefteqn{
D^{\rho\rho'\!}(k_{1},\xi)
D^{\sigma\sigma'\!}(k_{2},\xi)
D^{\tau\tau'\!}(k_{3},\xi)
\Gamma_{\rho'\!\sigma'\!\tau'\!}(k_{1},k_{2},k_{3}) } \nn \\
&=&
D^{\rho\rho'\!}(k_{1},1)
D^{\sigma\sigma'\!}(k_{2},1)
D^{\tau\tau'\!}(k_{3},1)
\Gamma_{\rho'\!\sigma'\!\tau'\!}(k_{1},k_{2},k_{3}) 
\,+\, (1-\xi)
\frac{1}{k_{1}^{2}k_{2}^{2}k_{3}^{2}} \nn \\
& & \times
\biggl\{
\biggl[ \frac{k_{1}^{\rho}}{k_{1}^{2}}
\biggl(k_{3}^{2}t^{\sigma\tau}(k_{3})- 
k_{2}^{2}t^{\sigma\tau}(k_{2}) \biggr) + {\rm c.p.} \biggr]
\,+\, (1-\xi)
\biggl[ 
\frac{k_{1}^{\rho}k_{2}^{\sigma}}{2k_{1}^{2}k_{2}^{2}}
(k_{1} - k_{2})^{\tau'}k_{3}^{2}t_{\tau'\!\tau}(k_{3})
\nn \\
& & -\,
\frac{k_{2}^{\sigma}}{4k_{2}^{2}k_{3}^{2}}
(k_{1} + (k_{2} - k_{3}))^{\tau}
(k_{2} - k_{3})^{\rho'}k_{1}^{2}t_{\rho'\!\rho}(k_{1})
\nn \\
& & -\,
\frac{k_{1}^{\rho}}{4k_{1}^{2}k_{3}^{2}}
(k_{2} - (k_{3}-k_{1}))^{\tau}
(k_{3} - k_{1})^{\sigma'}k_{2}^{2}t_{\sigma'\!\sigma}(k_{2})
\biggr]\biggr\}\,\,,
\label{DDDGamma}
\eea
where ``c.p.'' indicates cyclic permutations of
$\{k_{1},\rho\}$, $\{k_{2},\sigma\}$, $\{k_{3},\tau\}$.

Substituting the above expression into \eq{fig15ahxi}
and using the Ward identities (\ref{wid}) and (\ref{G4wid}),
together with $S^{-1}(p_{2}) = S^{-1}(p_{4}) = 0$
for the on-shell external fermions, it is a matter
of straightforward algebra to obtain
\bea
{\rm Figs.\,15a\!-\!h}
&=&
{\rm Figs.\,15a\!-\!h}|_{\xi = 1} 
\,-\, 
i(1-\xi)C_{\!A}Q^{2}g^{4}\int[dk_{1}][dk_{2}]\,
\frac{1}{k_{1}^{2}k_{2}^{2}k_{3}^{2}}   \nn \\
& &
\times \biggl\{ 
\biggl[
\frac{1}{2k_{2}^{2}} 
\Bigl( k_{3}^{2}t^{\rho\rho'\!}(k_{3}) 
 -     k_{1}^{2}t^{\rho\rho'\!}(k_{1}) \Bigr) 
+ 
\frac{1}{2k_{3}^{2}} 
\Bigl( k_{2}^{2}t^{\rho\rho'\!}(k_{2}) 
 -     k_{1}^{2}t^{\rho\rho'\!}(k_{1}) \Bigr)  \nn \\
& & \qquad
+\, 
(1-\xi) \frac{k_{1}^{2}}{8k_{2}^{2}k_{3}^{2}}
(k_{2}-k_{3})^{\rho'\!}(k_{2}-k_{3})^{\tau} 
t_{\tau}^{\rho}(k_{1}) \biggr] \nn \\
& & \qquad
\times G_{\rho'\!\mu}^{r}(k_{1},-p_{3},p-k_{1},p_{4})
S(p-k_{1})
G_{\rho\nu}^{r}(-k_{1},p_{1},p_{2},p-k_{1}) \nn \\ 
& &\,\,\,\,+
\biggl[
\frac{1}{2k_{1}^{2}} 
\Bigl( k_{3}^{2}t^{\sigma\sigma'\!}(k_{3}) 
 -     k_{2}^{2}t^{\sigma\sigma'\!}(k_{2}) \Bigr) 
+ 
\frac{1}{2k_{3}^{2}} 
\Bigl( k_{1}^{2}t^{\sigma\sigma'\!}(k_{1}) 
 -     k_{2}^{2}t^{\sigma\sigma'\!}(k_{2}) \Bigr)  \nn \\
& & \qquad
+ \,
(1-\xi) \frac{k_{2}^{2}}{8k_{1}^{2}k_{3}^{2}}
t_{\tau}^{\sigma}(k_{2}) 
(k_{1}-k_{3})^{\tau}(k_{1}-k_{3})^{\sigma'\!}
\biggr] \nn \\
& & \qquad
\times G_{\sigma\mu}^{s}(-k_{2},-p_{3},p+k_{2},p_{4})
S(p+k_{2})
G_{\sigma'\!\nu}^{s}(k_{2},p_{1},p_{2},p+k_{2})  \biggr\} \nn \\ 
& & \,\,\,\,+\,
\gamma_{\mu}S(p)\gamma_{\nu}\Bigl[ \,\,\cdots\,\, \Bigr]
\,+\,
\Bigl[ \,\,\cdots\,\, \Bigr]\gamma_{\mu}S(p)\gamma_{\nu} \,\,.
\label{fig15ahxi2}
\eea
In the above expression, the ellipses in the last line
post- and pre-multiplying
$\gamma_{\mu}S(p)\gamma_{\nu}$ stand for complicated expressions
which are contributions to the external leg corrections.

Making the change of variables $k_{1} \leftrightarrow -k_{2}$
in the term in \eq{fig15ahxi2} involving $S(p+k_{2})$ 
and using the definition \eq{DP1} then gives
\bea
{\rm Figs.\,15a\!-\!h}
&=&
{\rm Figs.\,15a\!-\!h}|_{\xi = 1} 
\,-\, 
Q^{2}g^{2}\int[dk_{1}][dk_{2}]\,
\frac{1}{k_{1}^{4}} \Delta\Pi_{1}^{\prime(1)\rho\rho'\!}
(k_{1},k_{2},\xi)  
\nn \\
& &
\times
G_{\rho\mu}^{r}(k_{1},-p_{3},p-k_{1},p_{4})
S(p-k_{1})
G_{\rho'\!\nu}^{r}(-k_{1},p_{1},p_{2},p-k_{1}) \nn \\ 
& & +\,
\gamma_{\mu}S(p)\gamma_{\nu}\Bigl[ \,\,\cdots\,\, \Bigr]
\,+\,
\Bigl[ \,\,\cdots\,\, \Bigr]\gamma_{\mu}S(p)\gamma_{\nu} \,\,.
\label{fig15ahxi3}
\eea
Comparing \eq{fig15ahxi3} with \eq{fig14adxi},
we immediately see that the $i=1$
pinch part component
$\Delta\Pi_{1}^{\prime(1)\rho\rho'\!}(k_{1},k_{2},\xi)$
of the conventional one-loop gluon self-energy insertion is cancelled 
{\em individually}\, for each of the diagrams 14a--d.

\pagebreak

\begin{center}
\begin{picture}(400,115)(0,15)


\ArrowLine(200,120)(230,120)
\ArrowLine(200, 90)(200,120)
\ArrowLine(200, 60)(200, 90)
\ArrowLine(230, 60)(200, 60)

\Photon(170, 60)(200, 60){2}{5}
\Gluon(170, 90)(200, 90){2}{5}
\Gluon(170,120)(200,120){2}{5}

\put(140,120){\makebox(0,0)[c]{$q_{1},\rho,r$}}
\put(140, 90){\makebox(0,0)[c]{$q_{2},\sigma,s$}}
\put(140, 60){\makebox(0,0)[c]{$q_{3},\mu$}}

\put(250, 60){\makebox(0,0)[c]{$q_{4}$}}
\put(250,120){\makebox(0,0)[c]{$q_{5}$}}

\put(300, 90){\makebox(0,0)[c]{$\times$ six perms.}}


\put(-20,25){\makebox(0,0)[l]{\small
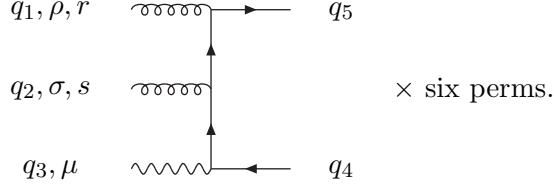
Fig.\ 18. The Feynman diagrams specifying the 
``abelian-like'' five-point function
$G_{\rho\sigma\mu}^{rs}(q_{1},q_{2},q_{3},q_{4},q_{5})$. }}

\end{picture}
\end{center}

\subsection{Two-loop diagrams involving no triple gauge vertices}

Finally, we consider the two-loop gluonic corrections to
$q\gamma \rightarrow q\gamma$ consisting of diagrams
with no triple gauge vertices.
The sets of one-particle-irreducible and one-particle-reducible
such diagrams are as shown already in Figs.~17a--z and 
17a$^{\prime}$--l$^{\prime}$, respectively.

In order to deal efficiently with the diagrams 17a--z,
it is very convenient to define the connected
``abelian-like'' five-point function
$G_{\rho\sigma\mu}^{rs}(q_{1},q_{2},q_{3},q_{4},q_{5})$,
specifying the tree level coupling of a pair of gluons
$A_{\rho}^{r}(q_{1})$, $A_{\sigma}^{s}(q_{2})$ and a photon
$A_{\mu}^{r}(q_{3})$
to a quark with electromagnetic charge $Q$
{\em via quark-gauge boson vertices only},
i.e.\ excluding triple gauge vertices.
The six relevant diagrams are as shown in Fig.~18:
\bea
{\rm Fig.\,18}
&=&
-Qg^{2}G_{\rho\sigma\mu}^{rs}(q_{1},q_{2},q_{3},q_{4},q_{5}) \\
&=&
-Qg^{2}\Bigl\{ 
i\gamma_{\rho}T^{r}S(q_{5}-q_{1})
G_{\sigma\mu}^{s}(q_{2},q_{3},q_{4},q_{5}-q_{1}) \nn \\
& &\qquad\,\,
+ i\gamma_{\sigma}T^{s}S(q_{5}-q_{2})
G_{\rho\mu}^{r}(q_{1},q_{3},q_{4},q_{5}-q_{2}) \nn \\
& &\qquad\,\,
-i\gamma_{\mu}S(q_{5}-q_{3})
\Bigl( \gamma_{\sigma}T^{s}S(q_{4}+q_{1})\gamma_{\rho}T^{r}
+ \gamma_{\rho}T^{r}S(q_{4}+q_{2})\gamma_{\sigma}T^{s} \Bigr) \Bigr\}\,\,.
\qquad\,\,
\label{G5}
\eea
In \eq{G5}, advantage has been taken of the four-point function
(\ref{G4}) to simplify the expression.
This five-point function satisfies the following ``Ward'' identities:
\bea
q_{1}^{\rho}G_{\rho\sigma\mu}^{rs}(q_{1},q_{2},q_{3},q_{4},q_{5})
&=&
f^{rst}G_{\sigma\mu}^{t}(q_{1}+q_{2},q_{3},q_{4},q_{5}) \nn \\
& &
-iG_{\sigma\mu}^{s}(q_{2},q_{3},q_{4}+q_{1},q_{5})
S(q_{4}+q_{1})T^{r}S^{-1}(q_{4}) \nn \\
& &
+ iS^{-1}(q_{5})T^{r}S(q_{5}-q_{1})
G_{\sigma\mu}^{s}(q_{2},q_{3},q_{4},q_{5}-q_{1}) \,\,,
\label{G5wid1} \\
q_{1}^{\rho}q_{2}^{\sigma}
G_{\rho\sigma\mu}^{rs}(q_{1},q_{2},q_{3},q_{4},q_{5})
&=&
\frac{1}{2}f^{rst}(q_{2}-q_{1})^{\tau}
G_{\tau\mu}^{t}(q_{1}+q_{2},q_{3},q_{4},q_{5}) \nn \\
& &
-\frac{1}{2}i\{T^{r},T^{s}\}\gamma_{\mu}
S(q_{4}+q_{1}+q_{2})S^{-1}(q_{4}) \nn \\
& &
-\frac{1}{2}i\{T^{r},T^{s}\}
S^{-1}(q_{5})S(q_{5}-q_{1}-q_{2})\gamma_{\mu} \nn \\
& &
+ iT^{r}T^{s}S^{-1}(q_{5})S(q_{5}-q_{1})\gamma_{\mu}S(q_{4}+q_{2})
S^{-1}(q_{4}) \nn \\
& &
+ iT^{s}T^{r}S^{-1}(q_{5})S(q_{5}-q_{2})\gamma_{\mu}S(q_{4}+q_{1})
S^{-1}(q_{4})\,\,.
\label{G5wid2}
\eea

It is easy to see that the eighteen diagrams 17a--r
may be obtained simply by contracting together via
a fermion and two gluon propagators a pair of the above
five-point functions, with an overall factor of one half
in order to account for the symmetry under interchange of
the two gluons.
Furthermore, the two sets of four diagrams 17s--v and 17w--z 
may be written in terms of the four-point function (\ref{G4}).
Explicitly, we have
\bea
\lefteqn{ {\rm Figs.\,17a\!-\!z}
\,=\,
-iQ^{2}g^{4}\int[dk_{1}][dk_{2}]\,
D^{\rho\rho'\!}(k_{1},\xi)
D^{\sigma\sigma'\!}(k_{2},\xi) } \nn \\
& &
\times\biggl\{ \frac{1}{2}
G_{\rho\sigma\mu}^{rs}(k_{1},k_{2},-p_{3},p-k_{1}-k_{2},p_{4})
S(p-k_{1}-k_{2})
G_{\rho'\!\sigma'\!\nu}^{rs}(-k_{1},-k_{2},p_{1},p_{2},p-k_{1}-k_{2})
\nn \\
& & \,\,\,\,
+ G_{\rho\mu}^{r}(k_{1},-p_{3},p-k_{1},p_{4})
S(p-k_{1})\gamma_{\nu}S(p_{2}-k_{1})
\gamma_{\sigma}T^{s}S(p_{2}-k_{1}-k_{2}) \nn \\
& & \,\,\,\,
\times\Bigl(
\gamma_{\sigma'\!}T^{s}S(p_{2}-k_{1})\gamma_{\rho'\!}T^{r}
\,+\, 
\gamma_{\rho'\!}T^{r}S(p_{2}-k_{2})\gamma_{\sigma'\!}T^{s} \Bigr) \nn \\
& & \,\,\,\,
+ \Bigl(
\gamma_{\rho'\!}T^{r}S(p_{4}-k_{1})\gamma_{\sigma'\!}T^{s}
\,+\, 
\gamma_{\sigma'\!}T^{s}S(p_{4}-k_{2})\gamma_{\rho'\!}T^{r} \Bigr)  \nn \\
& & \,\,\,\,
\times
S(p_{4}-k_{1}-k_{2})\gamma_{\sigma}T^{s}
S(p_{4}-k_{1})\gamma_{\mu}S(p-k_{1})
G_{\rho\nu}^{r}(-k_{1},p_{1},p_{2},p-k_{1})    \biggr\}\,\,.
\label{fig17az}
\eea
In the above expression, the term involving $S(p-k_{1}-k_{2})$
accounts for the diagrams 17a--r, the term involving
$G_{\rho\mu}^{r}(k_{1},-p_{3},p-k_{1},p_{4})$
accounts for the diagrams 17s--v and the term involving
$G_{\rho\nu}^{r}(-k_{1},p_{1},p_{2},p-k_{1})$
accounts for the diagrams 17w--z.

Using the Ward identities (\ref{wid}), (\ref{G4wid})
(\ref{G5wid1}) and (\ref{G5wid2}) for the factors
of longitudinal four-momentum originating from 
the gluon propagators in \eq{fig17az},
together with $S^{-1}(p_{2}) = S^{-1}(p_{4}) = 0$,
after some algebra
the resulting expression may be written
\bea
\lefteqn{{\rm Figs.\,17a\!-\!z}
\,=\,
{\rm Figs.\,17a\!-\!z}|_{\xi = 1} 
\,-\,
i(1-\xi)Q^{2}g^{4}\int[dk_{1}][dk_{2}]\,
\frac{1}{k_{1}^{2}k_{2}^{4}} }\nn \\
& & \times\biggl\{
G_{\tau\mu}^{t}(k_{1}+k_{2},-p_{3},p-k_{1}-k_{2},p_{4})
S(p-k_{1}-k_{2})
G_{\tau'\!\nu}^{t}(-k_{1}-k_{2},p_{1},p_{2},p-k_{1}-k_{2}) \nn \\
& & \times
C_{\!A} \biggl( g^{\tau\tau'\!} 
- (1-\xi)
\frac{(k_{1}-k_{2})^{\tau}(k_{1}-k_{2})^{\tau'}}{8k_{1}^{2}}
\biggr) \nn \\
& &
+\,(C_{F}-C_{\!A})
G_{\rho\mu}^{r}(k_{1},-p_{3},p-k_{1},p_{4})
S(p-k_{1})
G_{\nu}^{r,\rho}(-k_{1},p_{1},p_{2},p-k_{1}) \biggr\}
\nn \\
& & +\, 
\gamma_{\mu}S(p)\gamma_{\nu}\Bigl[ \,\,\cdots\,\, \Bigr]
\,+\,
\Bigl[ \,\,\cdots\,\, \Bigr]\gamma_{\mu}S(p)\gamma_{\nu} \,\,.
\label{fig17az2}
\eea
In the above expression, the ellipses post- and pre-multiplying
$\gamma_{\mu}S(p)\gamma_{\nu}$ again
stand for complicated expressions
which are contributions to the external leg corrections.

There remain the one-particle-reducible
diagrams 17a$^{\prime}$--l$^{\prime}$.
These diagrams may be written in terms of the 
connected four-point function defined in \eq{G4}:
\bea
\lefteqn{
{\rm Figs.\,17a'\!-\!l'}
\,=\,
iQ^{2}g^{4}\int[dk_{1}][dk_{2}]\,
D^{\rho\rho'\!}(k_{1})D^{\sigma\sigma'\!}(k_{2})  }\nn \\
& & 
\times\biggl\{
G_{\sigma\mu}^{s}(-k_{2},-p_{3},p+k_{2},p_{4})
S(p+k_{2})
\gamma_{\sigma'\!}T^{s}S(p)\gamma_{\rho'\!}T^{r}
S(p-k_{1})
G_{\rho\nu}^{r}(-k_{1},p_{1},p_{2},p-k_{1}) \nn \\
& & +\frac{1}{2}
G_{\rho\mu}^{r}(k_{1},-p_{3},p-k_{1},p_{4})
S(p-k_{1})
G_{\rho'\!\nu}^{r}(-k_{1},p_{1},p_{2},p-k_{1}) \,
S(p_{2})\gamma_{\sigma}T^{s}S(p_{2}-k_{2})\gamma_{\sigma'\!}T^{s}
\nn \\
& & +\frac{1}{2}
\gamma_{\sigma}T^{s}S(p_{4}-k_{2})\gamma_{\sigma'\!}T^{s}S(p_{4}) \,
G_{\rho\mu}^{r}(k_{1},-p_{3},p-k_{1},p_{4})
S(p-k_{1})
G_{\rho'\!\nu}^{r}(-k_{1},p_{1},p_{2},p-k_{1}) \biggr\} \,.\nn \\
\label{fig17aplp}
\eea
In the above expression,
the first term on the r.h.s.\ accounts for the diagrams
17a$^{\prime}$--d$^{\prime}$, the second term for the diagrams
17e$^{\prime}$--h$^{\prime}$ and the third term for the diagrams
17i$^{\prime}$--l$^{\prime}$.
The second and third terms are each just the expression (\ref{fig3ad})
for the one-loop
corrections shown in Figs.~3a--d, post- and pre-multiplied,
respectively, by the
expression for the one-loop external leg corrections.

Using the Ward identities (\ref{wid}) and (\ref{G4wid})
for the factors of longitudinal four-momentum from
the gluon propagators in \eq{fig17aplp},
together with $S^{-1}(p_{2}) = S^{-1}(p_{4}) = 0$, we obtain
\bea
{\rm Figs.\,17a'\!-\!l'}
&=&
{\rm Figs.\,17a'\!-\!l'}|_{\xi = 1} 
\,+\,
i(1-\xi)Q^{2}g^{4}\int[dk_{1}][dk_{2}]\,
\frac{1}{k_{1}^{2}k_{2}^{4}} \nn \\
& & \times
C_{F}G_{\rho\mu}^{r}(k_{1},-p_{3},p-k_{1},p_{4})
S(p-k_{1})
G_{\rho'\!\nu}^{r}(-k_{1},p_{1},p_{2},p-k_{1})   \nn \\
& & +\,
\gamma_{\mu}S(p)\gamma_{\nu}\Bigl[ \,\,\cdots\,\, \Bigr]
\,+\,
\Bigl[ \,\,\cdots\,\, \Bigr]\gamma_{\mu}S(p)\gamma_{\nu} \,\,.
\label{fig17aplp2}
\eea

Making the changes of variables $k_{2} \rightarrow -k_{2}$
followed by $k_{1} \rightarrow k_{1} + k_{2}$ in the
term in \eq{fig17az2} proportional to $S(p-k_{1}-k_{2})$,
then combining Eqs.~(\ref{fig17az2}) and (\ref{fig17aplp2})
and using the definition \eq{DP2},
we finally obtain for the sum of all of the diagrams in Fig.~17
\bea
{\rm Figs.\,17a\!-\!l'}
&=&
{\rm Figs.\,17a\!-\!l'}|_{\xi = 1} 
\,-\,
Q^{2}g^{2}\int[dk_{1}][dk_{2}]\,
\frac{1}{k_{1}^{4}} \Delta\Pi_{2}^{\rho\rho'\!}(k_{1},k_{2},\xi)  
\nn \\
& & \times
G_{\rho\mu}^{r}(k_{1},-p_{3},p-k_{1},p_{4})
S(p-k_{1})
G_{\rho'\!\nu}^{r}(-k_{1},p_{1},p_{2},p-k_{1})   \nn \\
& & +\,
\gamma_{\mu}S(p)\gamma_{\nu}\Bigl[ \,\,\cdots\,\, \Bigr]
\,+\,
\Bigl[ \,\,\cdots\,\, \Bigr]\gamma_{\mu}S(p)\gamma_{\nu} \,\,.
\label{fig17alp}
\eea
Comparing \eq{fig17alp} with \eq{fig14adxi},
we see that, just like the $i = 1$ pinch part 
component, the $i=2$ pinch part component
$\Delta\Pi_{2}^{\prime(1)\rho\rho'\!}(k_{1},k_{2},\xi)$
of the conventional one-loop gluon self-energy insertion 
is cancelled {\em individually}\, for each of the diagrams 14a--d.

We have thus succeeded in showing 
(i) that in the integrands for each of the two-loop 
QCD corrections to $q\gamma\rightarrow q\gamma$ 
shown in Figs.~14a--d, the additional pinch parts 
of the conventional one-loop gluon self-energy insertion
which occur for $\xi \neq 1$
are each exactly cancelled by corresponding 
pinch terms from the remaining two-loop corrections
shown in Figs.~15a--h and 17a--l$^{\prime}$, and 
(ii) that this cancellation exhausts 
the additional terms from the corrections shown 
in Figs.~15 and 17 which occur for $\xi \neq 1$
(for simplicitly, this latter was only shown explicitly
up to terms which are two-loop external leg corrections).

In the particular case of the PT self-energy-like
component of the process $q\gamma\rightarrow q\gamma$,
we have thus explicitly demonstrated the gauge independence
of the contributions of the two-loop diagrams in
Figs.~14, 15 and 17 to the PT two-loop quark self-energy
$-i\hat{\Sigma}^{(2)}(p)$.

\subsection{Remarks}

We finish this section with two remarks:

(1) At the one-loop level, the simple technique 
used in section~2 to implement the PT simultaneously
among subsets of diagrams, rather than individual
diagrams, lead to little saving of effort.
At the two-loop level, by contrast, 
with several dozen diagrams 
and more factors of longitudinal gluon four-momentum 
to deal with,
this technique enormously facilitated 
the implementation of the PT.
Furthermore, by grouping the diagrams for the 
two-loop QCD corrections to $q\gamma\rightarrow q\gamma$ 
as in the above three subsections,
the gauge cancellation mechanism 
which results in the gauge independence of the PT two-loop
quark self-energy
$-i\hat{\Sigma}^{(2)}(p)$
is seen to be essentially the same as that which results 
in the gauge independence of the PT one-loop gluon self-energy
$i\hat{\Pi}^{(1)}(p)$; only here, the PT one-loop
gluon self-energy occurs as internal corrections
in a two-loop process. 
This fact is consistent with the analysis of
Ref.~\cite{PTqcdeffch}, where it was shown
that the PT one-loop gluon self-energy 
may be obtained from one-loop processes in which the 
``external'' fields are explicitly off-shell,
i.e.\ the process does not constitute a one-loop
$S$-matrix element, as in the $S$-matrix PT \cite{PT1}.
Indeed, the diagrammatic technique used here is the
extension to two loops of that used in Ref.~\cite{PTqcdeffch}.

(2) In the above analysis, we did not consider the
$\cO(\alpha_{s}^{2})$ QCD corrections to 
$q \gamma \rightarrow q \gamma$ consisting of
one-loop diagrams with one-loop counterterm insertions.
These counterterm insertions 
are, of course, those obtained in the PT at one loop.
For the diagrams involving 
the one-loop quark-gluon vertex, quark wavefunction or 
quark mass renormalization constants
$(Z_{1}-1)_{\rm PT}^{(1)}$,
$(Z_{2}-1)_{\rm PT}^{(1)}$ and
$(Z_{m}-1)_{\rm PT}^{(1)}$, respectively.
the implementation of the PT algorithm,
in particular the gauge cancellation mechanism,
is precisely as for the $\cO(\alpha_{s})$ QCD corrections to 
$q \gamma \rightarrow q \gamma$ described in section~2. 
However, the counterterm corresponding to the one-loop
gluon wavefunction renormalization has the form
\be
-i(Z_{3}-1)_{\rm PT}^{(1)}k^{2}t_{\mu\nu}(k)
\label{Z3ct}
\ee
i.e.\ it is transverse. This follows from the fact that
the PT one-loop gluon self-energy is transverse.
When the counterterm (\ref{Z3ct}) is  
contracted with a pair of tree level gluon propagators,
the gauge-dependent longitudinal parts of the 
propagators thus vanish.\footnote{This statement does
not hold for the case in which one starts from a
non-covariant gauge. In this case, the entire
analysis of this section would be much more complicated,
although the results are expected to be the same.}
There remain, however,
the longitudinal factors from the counterterm itself.
If, as here, the ``induced'' longitudinal factors
from the PT one-loop gluon self-energy insertions
are not to be used to trigger the PT rearrangement,
then clearly the corresponding longitudinal factors from
the counterterm (\ref{Z3ct}) must not be used either;
for to do so would spoil the renormalizability of 
the PT two-loop quark self-energy (cf.\ section 6).

 
\setcounter{equation}{0}

\section{Renormalization of $-i\hat{\Sigma}^{(2)}(p)$}


Finally, we consider the third and last step outlined in 
section 3, viz.\ the calculation of the $\cO(\alpha_{s}^{2})$
renormalization constants 
$(Z_{2}-1)_{\rm PT}^{(2)}$ and $(Z_{m}-1)_{\rm PT}^{(2)}$
required to renormalize the PT two-loop quark self-energy.
For this task, we shall make use of
$Z_{1}$, $Z_{2}$, $Z_{3}$ and $Z_{m}$
to $\cO(\alpha_{s})$
in both the class of linear covariant gauges and the PT,
and also $Z_{2}$ and $Z_{m}$ to $\cO(\alpha_{s}^{2})$ in the
class of linear covariant gauges.
In the minimal subtraction (MS) scheme,
with $n_{f}$ active flavours of fermion, 
the $\cO(\alpha_{s})$ terms are given by \cite{PT1,itzzub}

\pagebreak

\be
\begin{array}{rclrcl}
\phantom{\Biggl|}\!\!\!\!
(Z_{1}-1)_{\xi}^{(1)} &=& 
\displaystyle{ \frac{\alpha_{s}}{4\pi}
\biggl(-\frac{3+\xi}{4}
C_{\!A}-\xi C_{F}\biggr)
\frac{1}{\epsilon}} \,\,, 
&
(Z_{1}-1)_{\rm PT}^{(1)} &=& 
\displaystyle{ \frac{\alpha_{s}}{4\pi} 
\biggl(-C_{F}\biggr)\frac{1}{\epsilon}} \,\,, \\
\phantom{\Biggl|}\!\!\!\!
(Z_{2}-1)_{\xi}^{(1)} &=& 
\displaystyle{ \frac{\alpha_{s}}{4\pi} 
\biggl(-\xi C_{F}\biggr)\frac{1}{\epsilon}} \,\,, 
&
(Z_{2}-1)_{\rm PT}^{(1)} &=& 
\displaystyle{ \frac{\alpha_{s}}{4\pi} 
\biggl(-C_{F}\biggr)\frac{1}{\epsilon}} \,\,, \\
\phantom{\Biggl|}\!\!\!\!
(Z_{3}-1)_{\xi}^{(1)} &=& 
\displaystyle{ \frac{\alpha_{s}}{4\pi}
\biggl(\frac{13-3\xi}{6}
C_{\!A}- \frac{4}{3}T_{F}n_{f}\biggr)
\frac{1}{\epsilon}} \,\,,
&
(Z_{3}-1)_{\rm PT}^{(1)} &=& 
\displaystyle{ \frac{\alpha_{s}}{4\pi}
\biggl(\frac{11}{3}C_{\!A}-\frac{4}{3}T_{F}n_{f}\biggr)
\frac{1}{\epsilon}} \,\,, \\
\phantom{\Biggl|}\!\!\!\!
(Z_{m}\!\!-1)_{\xi}^{(1)} &=& 
\displaystyle{ \frac{\alpha_{s}}{4\pi}
\biggl(-3C_{F}\biggr)\frac{1}{\epsilon}} \,\,,
&
(Z_{m}\!\!-1)_{\rm PT}^{(1)} &=& 
\displaystyle{ \frac{\alpha_{s}}{4\pi}
\biggl(-3C_{F}\biggr)\frac{1}{\epsilon}} \,\,,
\end{array}
\label{Z1s}
\ee
where $T_{F}$ is the normalization of the generators
for the fermion representation
($T_{F} = \frac{1}{2}$ for the fundamental representation).
The required $\cO(\alpha_{s}^{2})$ covariant gauge terms are given
by \cite{tarrach,nachtwet}
\bea 
(Z_{2}-1)_{\xi}^{(2)} 
\!\!\! &=& \!\!\!
\biggl(\frac{\alpha_{s}}{4\pi}\biggr)^{\!2}\!C_{F}\biggl\{ 
\biggl(\frac{\xi^{2}\!+\!3\xi}{4}C_{\!A}+\frac{\xi^{2}}{2}C_{F} 
\biggr)\frac{1}{\epsilon^{2}}
+ \biggl(-\frac{\xi^{2}\!+\!8\xi\!+\!25}{8}C_{A}+T_{F}n_{f} + \frac{3}{4}C_{F}
\biggr)\frac{1}{\epsilon}
\biggr\}\,\,,\nn \\ \\
(Z_{m}\!\!-1)_{\xi}^{(2)} 
\!\!\! &=& \!\!\!
\biggl(\frac{\alpha_{s}}{4\pi}\biggr)^{\!2}\!C_{F}\biggl\{ 
\biggl(\frac{11}{2}C_{A} - 2T_{F}n_{f} + \frac{9}{2}C_{F}\biggr)
\frac{1}{\epsilon^{2}}
+ \biggl(-\frac{97}{12}C_{A} + \frac{5}{3}T_{F}n_{f}-\frac{3}{4}C_{F}\biggr)
\frac{1}{\epsilon}
\biggr\} \,\,.\nn \\
\eea
Note that $Z_{m}$ $(=m_{0}/m)$ is gauge-independent.

In order to obtain  
$(Z_{2}-1)_{\rm PT}^{(2)}$ and $(Z_{m}-1)_{\rm PT}^{(2)}$, 
we first write the PT renormalized 
two-loop self-energy in terms of the
conventional self-energy as follows:
\be
\hat{\Sigma}^{(2)}(p)
=
\Sigma^{(2)}(p,\xi) + \Delta\Sigma^{(2)}(p,\xi)\,\,.
\label{SSDS}
\ee
The explicit expression for 
the conventional covariant gauge self-energy
$\Sigma^{(2)}(p,\xi)$ 
for arbitrary $\xi$ has recently been
given in Ref.~\cite{fleischeretal}.
The function $\Delta\Sigma^{(2)}(p,\xi)$ 
is the sum of the self-energy-like
pinch parts of the two-loop vertex and box corrections to
the given process involving the quark as virtual intermediate
state. However, having explicitly constructed 
$\hat{\Sigma}^{(2)}(p)$ in the previous sections
directly from subsets of diagrams, rather than via the
consideration of the pinch parts of individual vertex and
box diagrams, we will here obtain
$\Delta\Sigma^{(2)}(p)$ simply
as the difference between the PT and conventional self-energies.
Given the above renormalization constants,
together with the fact that the renormalized function
$\Sigma^{(2)}(p,\xi)$ is finite, 
it is then sufficient to determine the counterterm contributions
required to make finite
the function $\Delta\Sigma^{(2)}(p,\xi)$ in \eq{SSDS} 
in order to obtain 
$(Z_{2}-1)_{\rm PT}^{(2)}$ and $(Z_{m}-1)_{\rm PT}^{(2)}$.

The function $\Delta\Sigma^{(2)}(p,\xi)$ is conveniently
decomposed into four contributions, corresponding to
the difference between the diagrams 10b and 8b,
10c and 8c, 10d--f and 8d--f, and 10g and 8g, respectively:
\be
\Delta\Sigma^{(2)}(p,\xi)
=
\sum_{i=1}^{4}\Delta\Sigma_{i}^{(2)}(p,\xi)\,\,.
\ee
Given that the PT self-energy is gauge-independent,
we are at liberty to choose the value of $\xi$ for
which the function 
$\Delta\Sigma^{(2)}(p,\xi)$ is simplest, i.e.\ $\xi = 1$. We have:

\pagebreak

\bea
-i\Delta\Sigma_{1}^{(2)}(p,1)
&=&
     {\rm Fig.\,10b} 
\,-\,{\rm Fig.\,8b}|_{\xi=1}  {\phantom{\Bigl| }} \nn \\
&=&
C_{F}g^{2}\!\int[dk_{1}]\biggr\{
-\int[dk_{2}]\Delta\Pi^{\prime (1)}(k_{1},k_{2},1)   
+(Z_{3}-1)_{\rm PT}^{(1)}
-(Z_{3}-1)_{\xi =1}^{(1)} 
\biggr\}  \nn \\
& & \times
\frac{1}{k_{1}^{2}}t^{\rho\rho'}(k_{1})
\gamma_{\rho}S(p-k_{1})\gamma_{\rho'}\,\,;
\label{DS1} \\ \nn \\
-i\Delta\Sigma_{2}^{(2)}(p,1)
&=&
     {\rm Fig.\,10c} 
\,-\,{\rm Fig.\,8c}|_{\xi=1}  {\phantom{\Bigl| }} \nn \\
&=& 
0\,\,;
\label{DS2} \\ \nn \\
-i\Delta\Sigma_{3}^{(2)}(p,1) 
&=&
     {\rm Figs.\,10d\!-\!f} 
\,-\,{\rm Figs.\,8d\!-\!f}|_{\xi=1} {\phantom{\Bigl| }} \nn \\
&=&
-C_{\!A}C_{F}g^{4} 
\int[dk_{1}][dk_{2}]\,
\frac{i}{k_{1}^{2}k_{2}^{2}k_{3}^{2}}
\Gamma_{\tau\rho\sigma}^{P}(k_{3};k_{1},k_{2})\,
\gamma^{\sigma}
S(p+k_{2})\gamma^{\tau}S(p-k_{1})
\gamma^{\rho}
\nn \\ 
& &
- 2\Bigl\{
 (Z_{1}-1)_{\rm PT}^{(1)} 
-(Z_{1}-1)_{\xi =1}^{(1)} 
\Bigr\}
C_{F}g^{2}\!\int [dk_{1}]\frac{1}{k_{1}^{2}}
\gamma_{\rho}S(p-k_{1})\gamma^{\rho} \,\,;
\label{DS3} \\ \nn \\
-i\Delta\Sigma_{4}^{(2)}(p,1)
&=&
     {\rm Fig.\,10g} 
\,-\,{\rm Fig.\,8g}|_{\xi=1}  {\phantom{\Bigl| }} \nn \\
&=&
\Bigl\{
(Z_{2}-1)_{\rm PT}^{(2)} - (Z_{2}-1)_{\xi =1}^{(2)} 
\Bigr\}i(\ps - m) \nn \\
& &
-\,\Bigl\{
 (Z_{2}(Z_{m}-1))_{\rm PT}^{(2)} 
-(Z_{2}(Z_{m}-1))_{\xi =1}^{(2)} 
\Bigr\}im \,\,. 
\label{DS4}
\eea
In \eq{DS1}, we have used 
Eqs.~(\ref{Pidecomp})--(\ref{Piprimepinch1})
for the difference between the PT and Feynman gauge
one-loop gluon self-energy insertions.
In \eq{DS2}, we have used the fact that the
PT and Feynman gauge renormalized
one-loop quark self-energies coincide.
And in \eq{DS3}, we have used Eqs.~(\ref{GammaFFFP})
and (\ref{Gammatriv})
for the difference between the two sets of 
triple gauge vertices involved.

Substituting the expressions 
(\ref{Piprimepinch1}) and (\ref{GammaP})
for $\Delta\Pi^{\prime (1)}$ and
$\Gamma_{\tau\rho\sigma}^{P}$, respectively,
in Eqs.~(\ref{DS1}) and (\ref{DS2}),
and then using the elementary Ward identity (\ref{wid})
in the latter case,
the $g^{\rho\rho'}$ part of the 
$\Delta\Pi^{\prime (1)}$ term in \eq{DS1} is cancelled
by an internal pinch term in \eq{DS2}. This is
just the cancellation described in section 4.2.
Carrying out the $k_{2}$ integration in \eq{DS1},
and substituting the expresions (\ref{Z1s})
for the $\cO(\alpha_{s})$ renormalization constants,
the sum of the expressions (\ref{DS1})--(\ref{DS4})
may be written
\bea
\lefteqn{-i\Delta\Sigma^{(2)}(p,1)
\,\,=\,\,
C_{\!A}C_{F}g^{4}\biggr\{
\int[dk_{1}]\,
G(k_{1}^{2}/\mu^{2},\epsilon)
\frac{1}{k_{1}^{4}}\ks_{1} S(p-k_{1})\ks_{1}  }\nn \\
& & 
+\int[dk_{1}][dk_{2}]\,
\frac{i}{k_{1}^{2}k_{2}^{2}(k_{1}+k_{2})^{2}}
\Bigl(\gamma^{\mu}F_{\mu}(p,k_{1},k_{2})S^{-1}(p)
\,+\,S^{-1}(p)F_{\mu}(p,k_{1},k_{2})\gamma^{\mu}\Bigr)\biggr\}
\nn \\
& &
+\, \Bigl\{
(Z_{2}-1)_{\rm PT}^{(2)} - (Z_{2}-1)_{\xi =1}^{(2)} 
\Bigr\}i(\ps - m)
\,-\,\Bigl\{
 (Z_{m}-1)_{\rm PT}^{(2)} 
-(Z_{m}-1)_{\xi =1}^{(2)} 
\Bigr\}im   \label{DS}
\eea
In the above expression, for brevity we have defined
\bea
F_{\mu}(p,k_{1},k_{2})
&=& 
S(p+k_{2})\gamma_{\mu}S(p-k_{1})\,\,, \\
G(k_{1}^{2}/\mu^{2},\epsilon)
&=&
\frac{1}{8\pi^{2}\epsilon}
\biggl\{  
\frac{\Gamma(1+\epsilon)\Gamma^{2}(1-\epsilon)}{\Gamma(2-2\epsilon)} 
\biggl(\frac{-k_{1}^{2}}{4\pi\mu^{2}}\biggr)^{-\epsilon} - 1 \biggr\}
\,\,.
\eea

Carrying out the remaining integrations is \eq{DS}, we obtain
\bea
-i\Delta\Sigma^{(2)}(p,1)
&=&
\biggl\{\biggl(\frac{\alpha_{s}}{4\pi}\biggr)^{2}
C_{\!A}C_{F}
\biggl( \frac{1}{\epsilon^{2}} + \frac{1}{\epsilon}\biggr)
+ (Z_{2}-1)_{\rm PT}^{(2)} - (Z_{2}-1)_{\xi =1}^{(2)} 
\biggr\}
i(\ps - m)  \nn \\
& &
-\,\Bigl\{
 (Z_{m}\!\!-1)_{\rm PT}^{(2)} 
-(Z_{m}\!\!-1)_{\xi}^{(2)} 
\Bigr\}im \phantom{\biggl| }
\nn \\
& &
+\,\, {\rm terms\,\,which\,\,are\,\,finite\,\,as}\,\,
\epsilon\rightarrow 0 \,\,. 
\label{DSdiv}
\eea
Requiring that the function $\Delta\Sigma^{(2)}(p,1)$ be
finite as $\epsilon \rightarrow 0$
then determines the required $\cO(\alpha_{s}^{2})$
PT renormalization constants: in the MS scheme,
\bea 
(Z_{2}-1)_{\rm PT}^{(2)}
&=&
\biggl(\frac{\alpha_{s}}{4\pi}\biggr)^{2}C_{F}\biggl\{ 
\frac{1}{2}C_{F}\frac{1}{\epsilon^{2}}
+ \biggl(-\frac{21}{4}C_{A} + T_{F}n_{f} + \frac{3}{4}C_{F}\biggr)
\frac{1}{\epsilon}
\biggr\} \,\,,\\
\nn \\
(Z_{m}-1)_{\rm PT}^{(2)}
&=&
(Z_{m}-1)_{\xi}^{(2)}\,\,.
\eea

Using Eqs.~(\ref{SSDS}) and (\ref{DSdiv}), 
the PT renormalized gauge-independent
two-loop quark self-energy
$-i\hat{\Sigma}^{(2)}(p)$ in the MS scheme
may be written in terms of the corresponding
conventional Feynman gauge self-energy as
\bea
\lefteqn{-i\hat{\Sigma}^{(2)}(p)
\,\,=\,\,
-i\Sigma^{(2)}(p,\xi=1) \phantom{\Bigl|} }\nn \\
& &
+\, 
C_{\!A}C_{F}g^{4} \biggl\{
\int[dk_{1}]
G(k_{1}^{2}/\mu^{2},\epsilon)
\frac{1}{k_{1}^{4}}   
\Bigl( S^{-1}(p)S(p-k_{1})S^{-1}(p) \,-\, S^{-1}(p) \Bigr)
\nn \\
& &
+\int[dk_{1}][dk_{2}]
\frac{i}{k_{1}^{2}k_{2}^{2}(k_{1}+k_{2})^{2}}
\Bigl(\gamma^{\rho}F_{\rho}(p,k_{1},k_{2})S^{-1}(p)
\,+\,S^{-1}(p)F_{\rho}(p,k_{1},k_{2})\gamma^{\rho}\Bigr)
\nn \\
& &
 + \,\frac{1}{(4\pi)^{4}}
\biggl( -\frac{1}{\epsilon^{2}} - \frac{1}{\epsilon}\biggr)
iS^{-1}(p) \biggr\}\,\,. \label{Sfinal}
\eea
The integral expression \eq{Sfinal} is our final result.

We finish this section with three remarks:

(1) In this section, we have shown by explicit calculation
that the two-loop quark self-energy
obtained in the PT is multiplicatively 
renormalizable by local counterterms.
It is emphasized that this result was not obvious a priori. 
In particular, the worry was that the rearrangements
which gave the contributions to 
$-i\hat{\Sigma}^{(2)}(p)$ shown in Figs.~10b--f
would result in divergent terms of the form, e.g.,
$\epsilon^{-1}\ln (-p^{2}/\mu^{2})$
which then could not be cancelled by local
counterterms, shown in Fig.~10g.

(2) From \eq{Sfinal}, given that $S^{-1}(p)$ vanishes
at $\ps = m$, it is immediately clear that the PT
two-loop quark
self-energy does not shift the position of
the propagator pole (to $\cO(\alpha_{s}^{2})$):
\be
\hat{\Sigma}^{(2)}(p)\Bigl|_{p\!\!\!/ = M}
=
\Sigma^{(2)}(p,\xi=1)\Bigl|_{p\!\!\!/ = M}
\,+\, \cO(\alpha_{s}^{3})\,\,.
\ee
This is an essential requirement for the consistency 
of the PT approach, and is in fact expected on
rather general grounds \cite{PT2pt1}.

(3) The quark self-energy obtained in the BFM with quantum
gauge parameter $\xi_{Q}$ coincides with that obtained
in the class of linear covariant gauges with gauge 
parameter $\xi = \xi_{Q}$ to all orders\footnote{This 
statement is not entirely obvious
due to the question of the renormalization of the
quantum fields in the BFM beyond one loop. For a 
lucid discussion of this issue, see Ref.~\cite{capper}.}
in perturbation theory:
$-i\Sigma_{\rm BFM}^{(n)}(p,\xi_{Q})
=-i\Sigma^{(n)}(p,\xi=\xi_{Q})$.
At the two-loop level, we therefore obtain immediately from
\eq{Sfinal}
\be
-i\hat{\Sigma}^{(2)}(p)
\,\neq\,
-i\Sigma_{\rm BFM}^{(2)}(p,\xi_{Q}=1)\,\,.
\ee
We thus conclude that the correspondence between
the PT gauge-independent $n$-point functions 
and those obtained in the BFM in the Feynman quantum gauge
$\xi_{Q} = 1$ does not persist beyond one loop.

 
\setcounter{equation}{0}

\section{Summary and Conclusions}


The extension of the pinch technique (PT)
beyond the one-loop approximation requires 
the solution of the two problems described in the introduction:
(1) how to deal consistently with triple gauge vertices all three 
legs of which are associated with gauge fields propagating in loops;
and (2) whether or not to use the ``induced'' longitudinal 
factors from internal loop corrections to trigger further the PT 
rearrangement, and, if so, how.
In this paper, it has been shown how the first of these problems 
is consistently solved for the simplest non-trivial case, viz.\
the construction in the PT approach of the 1PI
two-loop fermion self-energy 
$-i\hat{\Sigma}^{(2)}(p)$ in QCD.

We began (section 2) by reviewing the construction of the
PT gauge-independent one-loop quark self-energy 
$-i\hat{\Sigma}^{(1)}(p)$ \cite{PTquark}.
This function was obtained from the ``effective'' two-point
component of the integrands for the one-loop QCD corrections
to the Compton scattering 
$q\gamma \rightarrow q\gamma$ of a photon off a quark.
The fact that gluons do not couple directly to the photons
meant that, for the construction of 
$-i\hat{\Sigma}^{(1)}(p)$,
this process was considerably simpler 
than, e.g., the Compton-like scattering 
$qg \rightarrow qg$ of a gluon off a quark.
Although the one-loop case was almost trivial, this review
was useful in order to introduce the simple technique
used subsequently at two loops 
to implement the PT simultaneously among
{\em subsets}\, of diagrams, rather than individual diagrams.

We then turned to the construction of the 1PI two-loop 
quark self-energy $-i\hat{\Sigma}^{(2)}(p)$ in the PT approach. 
By investigating here only the first of the above problems,
it was effectively assumed that the correct approach
to the second problem will turn out to be not to use the ``induced''
longitudinal factors to trigger further the PT rearrangement.
Thus, ``the PT'' here referred to the PT algorithm implemented
using longitudinal factors only from lowest order gauge field
propagators and triple gauge vertices.

The starting point for the construction (section 3)
was the general diagrammatic representation of the renormalized
two-loop quark self-energy in terms of renormalized
one-loop two- and three-point function and
tree level Bethe-Salpeter-type quark-gluon scattering kernel 
insertions in the one-loop quark self-energy,
shown in Figs.~8b--g for the case of the conventional self-energy.
This representation is precisely analogous to the
familiar representation \cite{bjdr} of the renormalized
two-loop QED photon self-energy (vacuum polarization)
in terms of renormalized one-loop two- and three-point function 
and tree level
electron-positron scattering kernel insertions in the
one-loop photon self-energy, shown in Figs.~6b--g.
The significance of these representations is two-fold:
first, they are explicitly 
in terms of renormalized one-loop $n$-point functions obtained in 
perturbation theory at the one-loop level; and second,
they are symmetric, in the sense that no orientation
of the overall sets of diagrams is preferred.
This latter property, made possible
by the introduction of the kernel insertion contributions,
would appear to be
essential for the consistent solution of the first problem
for the case of the two-loop quark self-energy
in the PT approach.

In order for the PT approach to be consistent, 
it was first of all required that the contributions to 
$-i\hat{\Sigma}^{(2)}(p)$ 
analogous to those shown in Figs.~8b--e for the conventional case
consist of the PT renormalized gauge-independent one-loop
$n$-point functions appearing as internal corrections
in the PT gauge-independent
one-loop quark self-energy $-i\hat{\Sigma}^{(1)}(p)$.
It then remained to determine the PT analogue of the
contribution in Fig.~8f involving the 
tree level quark-gluon kernel;
and finally the PT analogue of the
two-loop counterterm insertions shown in Fig.~8g.
The contributions to $-i\hat{\Sigma}^{(2)}(p)$ 
are thus as shown in Figs.~10b--g.

In order to carry out the explicit construction of 
$-i\hat{\Sigma}^{(2)}(p)$ (section 4),
we considered the two-loop QCD corrections to 
$q\gamma \rightarrow q\gamma$, starting from the Feynman gauge.
The key to the construction was the elementary decomposition
\eq{GammaFFFP} of the triple gauge vertex which
occurs in the two-loop diagrams shown in Figs.~15a--h.
It was shown that this decomposition not only results in
precisely the internal one-loop gluon self-energy-like
pinch parts and triple gauge vertex components $\Gamma^{F}$ 
required to obtain the contributions to
$-i\hat{\Sigma}^{(2)}(p)$ shown in Figs.~10c--e, 
involving the PT one-loop internal corrections;
but also the contribution to the PT quark-gluon kernel,
obtained from the remaining two-loop self-energy-like 
component of Figs.15a--h,
involves just the component $\Gamma^{F}$ 
of the triple gauge vertex which occurs in the
PT one-loop functions.
The resulting PT tree level kernel, shown in Fig.~13, 
thus provides no factors of longitudinal four-momentum
associated with the external gluon legs.
As explained at the end of the section (fourth remark),
this is crucial to the consistency of the PT construction.

In order to demonstrate the gauge independence of 
$-i\hat{\Sigma}^{(2)}(p)$ (section 5), 
we again considered the two-loop QCD corrections to
$q\gamma \rightarrow q\gamma$, but now starting
from arbitrary $\xi$. 
With $\xi \neq 1$, there then occurred many further
factors of longitudinal gluon four-momentum in the
corresponding Feynman integrands.
In the case of the forty-six
two-loop diagrams shown in Figs.~15 and 17, 
these factors were efficiently dealt with by 
writing the diagrams in terms 
of tree level four- and five-point functions 
contracted together via quark and gluon propagators.
Using the Ward identities for these tree level
$n$-point functions, we were then able
to implement the PT directly among entire
subsets of these diagrams, rather than diagram-by-diagram.
In this way, we were able to demonstrate the
algebraic cancellation of the additional contributions 
which occur for $\xi \neq 1$
not only to the 1PI two-loop 
self-energy-like component of the process,
but also the vertex-like and box-like components.
The gauge independence of 
$-i\hat{\Sigma}^{(2)}(p)$ was thus explicitly demonstrated 
in the class of linear covariant gauges.

In order to obtain the two-loop renormalization constants
required finally to renormalize
$-i\hat{\Sigma}^{(2)}(p)$ (section 6),
rather than evaluating the divergent part of 
the diagrams in Figs.~10b--f directly, we
computed only the difference between the divergent part
of these diagrams and that of the corresponding conventional
self-energy diagrams in Figs.~8b--f in the Feynman gauge.
Knowing the two-loop Feynman gauge counterterms,
we were then able to obtain the required two-loop
PT counterterms. In this way, it was explicitly 
shown that the PT two-loop quark self-energy
$-i\hat{\Sigma}^{(2)}(p)$ is renormalizable.

Before finishing, we briefly turn
to the second problem outlined in the introduction.
It is clear from the analysis in sections 2 and 4 that
if the integrand 
$\Pi^{\prime(1)\rho\rho'\!}(k_{1},k_{2},\xi)$
for the conventional one-loop gluon self-energy
insertions shown in Figs.~14a--f and given in 
Eqs.~(\ref{fig14ad}) and (\ref{fig14ef}) 
is explicitly written in terms of the transverse tensor 
$k_{1}^{2}g^{\rho\rho'\!} - k_{1}^{\rho}k_{1}^{\rho'\!}$,
then the entire longitudinal $k_{1}^{\rho}k_{1}^{\rho'\!}$
component exactly cancels among the integrands
for the six diagrams. 
This cancellation is identical to that of
the longitudinal component
of the tree level gluon propagators described in section~2,
and the longitudinal component of the pinch part of
the Feynman gauge one-loop gluon self-energy 
described in section~4.1.

Regarding the role of the longitudinal component of the
gauge field propagator, it was shown some years ago
in an elegant paper by Llewellyn Smith \cite{chris}
that the covariant gauge fixing parameter $\xi$ 
in QCD may legitimately be replaced in the lagrangian
by an operator $\xi(\partial^{2})$. 
This operator may then in principle be chosen
so that the radiatively-corrected gauge field propagator
$i\Delta_{\mu\nu}(q,\xi)$ is proportional to $g_{\mu\nu}$
to all orders in perturbation theory. 
In this way, it was shown that the longitudinal
component of the radiatively-corrected
covariant gauge field propagator in QCD
makes no contribution to $S$-matrix elements to all orders
(a result which is relatively obvious in QED).
The cancellation described above among the diagrams of Fig.~14 
is just an explicit two-loop example of this.

{}From the point of view of the PT, 
it therefore seems plausible that
the ``induced'' factors of longitudinal
gauge field four-momentum from internal corrections
should indeed be used to trigger further the PT rearrangement.
However, in order that the resulting $n$-point functions
continue to satisfy simple QED-like Ward identities,
it is necessary to use such factors not only from
the gauge field self-energies, but also the gauge field vertices.
For these latter, it is thus necessary
to use the associated invariant tensor decompositions 
(cf.\ e.g.\ Refs.~\cite{ball,osland})
to isolate the longitudinal factors. 
The way in which this works for the case of the 
$\cO(\alpha_{s}^{2}n_{f})$ i.e.\ mixed fermionic-bosonic 
contribution to the two-loop gluon self-energy 
has been described in Ref.~\cite{QCD98}.

There are two reasons, however, to believe that the correct
approach will turn out to be {\em not}\, to use
the ``induced'' longitudinal factors to trigger further
the PT rearrangement---as was the 
effective assumption in this paper.
First, it turns out that
the longitudinal factors from internal vertex corrections,
if used to trigger the PT rearrangement,
make non-vanishing contributions
to the $n$-point functions which, 
in contrast to the case at one loop, do not
have a simple interpretation in terms of ghosts
\cite{PT1,PTqcdeffch}.
The role of the ``induced'' longitudinal factors
is therefore qualitatively different from that of the
lowest order longitudinal factors which appear
in the PT at the one-loop level.
Second, it is hard to see how such an approach could ever be
obtained from a formulation at the level of the path 
integral---essential if the PT is to be put on a firm
field-theoretic footing.
In particular, the required tensor decompositions of the 
internal corrections are complicated for all but the
self-energy function; and such an approach clearly precludes
a set of Feynman rules.

To conclude, the analysis presented in this paper represents the
first, non-trivial step towards extending the pinch technique
beyond the one-loop level. 
It is emphasized that, although the analysis has been
complicated, the result---expressed essentially
in the diagrams shown in Figs.~10 to 13---is remarkably simple.
It remains to be seen if this extension can be generalized
beyond the case considered here.

\vspace{15pt}

{\bf Acknowledgements}

I am very grateful to the theory group at the 
Institut de Physique Nucl\'eaire, Orsay,
in particular Jan Stern, for their 
encouragement and support of this work.
I thank Chris Llewellyn Smith for bringing Ref.~\cite{chris}
to my attention.

\end{document}